\documentclass[fleqn,usenatbib]{mnras}

\usepackage[T1]{fontenc}

\DeclareRobustCommand{\VAN}[3]{#2}
\let\VANthebibliography\thebibliography
\def\thebibliography{\DeclareRobustCommand{\VAN}[3]{##3}\VANthebibliography}

\usepackage{ltxtable}
\usepackage{longtable}
\usepackage{supertabular}
\usepackage{rotating}
\usepackage{pdflscape}
\usepackage{color}
\usepackage{natbib}
\usepackage{lscape}
\usepackage{txfonts}

\usepackage{graphicx}	
\usepackage{amssymb}	
\usepackage{indentfirst}
\usepackage{enumitem}

\usepackage[normalem]{ulem}

\title[CoBiToM - II: Evolution close to period cut-off]{CoBiToM Project - II: Evolution of contact binary systems close to the orbital period cut-off}

\author[G. Loukaidou et al.]{
G. A. Loukaidou,$^{1}$\thanks{E-mail: georgialouk@phys.uoa.gr}
K. D. Gazeas,$^{1}$ 
S. Palafouta,$^{1}$
D. Athanasopoulos,$^{1}$
S. Zola,$^{2,3}$
\newauthor
A. Liakos,$^{4}$
P. G. Niarchos,$^{1}$
P. Hakala,$^{5}$
A. Essam$^{6}$
and D. Hatzidimitriou$^{1}$
\\
$^{1}$ Section of Astrophysics, Astronomy and Mechanics, Department of Physics, National and Kapodistrian University of Athens, \\ GR-15784 Zografos, Athens, Greece\\
$^{2}$ Astronomical Observatory, Jagiellonian University, ul. Orla 171, PL-30-244 Krakow, Poland\\
$^{3}$ Mt. Suhora Observatory, Pedagogical University, ul. Podchorazych 2, PL-30-084 Krakow, Poland\\
$^{4}$ IAASARS, National Observatory of Athens, GR-15236 Penteli, Athens, Greece\\
$^{5}$ Finnish Centre for Astronomy with ESO (FINCA), Quantum, University of Turku, FI-20014, Turku, Finland\\
$^{6}$ Department of Astronomy, National Research Institute of Astronomy and Geophysics (NRIAG), Helwan, Cairo 11421, Egypt
}

\date{Accepted XXX. Received YYY; in original form ZZZ}

\pubyear{2021}

\begin{document}
\label{firstpage}
\pagerange{\pageref{firstpage}--\pageref{lastpage}}
\maketitle

\begin{abstract}
Ultra-short orbital period contact binaries ($P_{orb}$ < 0.26~d) host some of the smallest and least massive stars. These systems are faint and rare, and it is believed that they have reached a contact configuration after several Gyrs of evolution via angular momentum loss, mass transfer and mass loss through stellar wind processes.
This study is conducted in the frame of \textit{Contact Binaries Towards Merging (CoBiToM) Project} and presents the results from light curve and orbital analysis of 30 ultra-short orbital period contact binaries, with the aim to investigate the possibility of them being red nova progenitors, eventually producing merger events.
Approximately half of the systems exhibit orbital period modulations, as a result of mass transfer or mass loss processes.
Although they are in contact, their fill-out factor is low (less than 30 per cent), while their mass ratio is larger than the one in longer period contact binaries. 
The present study investigates the orbital stability of these systems and examines their physical and orbital parameters in comparison to those of the entire sample of known and well-studied contact binaries, based on combined spectroscopic and photometric analysis.
It is found that ultra-short orbital period contact binaries have very stable orbits, while very often additional components are gravitationally bound in wide orbits around the central binary system.
We confirmed that the evolution of such systems is very slow, which explains why the components of ultra-short orbital period systems are still Main Sequence stars after several Gyrs of evolution.
\end{abstract}



\section{Introduction}
The present work is a continuation of \textit{CoBiToM Project I - Contact Binaries Towards Merging} \citep{Gazeas2021b}, which aims to investigate the stellar merging processes by utilising contact binary systems as probes of coalescence events. 
It follows the same rationale and scientific approach for deriving the physical parameters of the binary components. 
The present study focuses on contact binary systems with ultra-short orbital periods, with the main goal being the determination of their physical and orbital characteristics, providing indications about the evolution of the binary before merging.

Contact binary systems consist of Main Sequence (MS) stars, possibly formed from an initial detached configuration, after gradual angular momentum loss, which leads to orbital shrinking \citep{Stepien2012}. The stellar evolution of the components of a contact binary system is significantly different from that of single stars. The evolution is controlled by mass loss and hence angular momentum loss from the system, along with mass transfer between the components \citep[e.g.][]{Yakut2005,Stepien2006}, gradually brings the components closer to each other. The question of whether there is an evolutionary sequence among different types of contact binary systems is still open.
It is clear that energy transfer between the components in the contact configuration alters their evolution and therefore their parameters (such as mass, radius and temperature) depend on it. The high proximity between the components and the mass transfer affects the gravity vs radiation equilibrium. Consequently, components of close contact binary systems differ from their single stars counterparts. 

The continuous monitoring of the absolute physical and orbital parameters of the binary components in such systems, as well as their secular changes, provide the tools for understanding stellar evolution, under this environment. These parameters are constrained by Roche geometry, as the contact configuration has to be preserved, restricting the radius of the components, the orbital separation, and therefore the mass and orbital period. Correlation between orbital and physical parameters show that these parameters are highly correlated with each other, as a result of a common evolution scheme \citep{Hilditch1988,Gazeas2006}. 

The orbital period distribution of contact binary systems ranges between 0.22 and 1.1~d, while \cite{Rucinski1992} noticed a sharp period cut-off limit at a value of $\sim$0.22~d. More recent observational evidence has indicated that the period cut-off could be even shorter than 0.22~d. \cite{Davenport2013} presented the system SDSS~J001641-000925 with an orbital period of 0.19856~d. \cite{Nefs2012} detected nine eclipsing binary candidates with orbital period of less than 0.22~d (ranging between 0.15-0.21~d) in the Wide-Field Camera (WFCAM) Transit Survey, five of which were classified as contact binaries.
\cite{Drake2014} identified 231 contact binaries with an orbital period below the cut-off limit, utilizing data from the Catalina Sky Survey (CSS). \cite{Zhang2020} used a combination of observational and theoretical arguments to suggest a new orbital period cut-off limit, at approximately 0.15~d.

There have been several attempts to explain the existence of the observed orbital period cut-off. 
First, \cite{Rucinski1993} proposed a model, where stars with low surface temperature become fully convective. This ultimately constrains the range of orbital parameters of contact binaries and therefore a random combination of stars cannot always be placed into a contact configuration. An alternative approach was proposed by \cite{Stepien2006,Stepien2012}, who conducted detailed calculations and showed that the orbital period cut-off in low-mass contact binaries can be explained by the very large dynamical evolution timescale. The angular momentum and mass loss rate in these systems is so slow, that several billion years are needed for a close binary to reach a contact configuration from an initially detached one, which shall slowly present a period decrease. For example, for an equal mass component detached binary with an initial mass of 1 $M_{\odot}$, 7.5~Gyrs are approximately needed in order to start their contact phase. However, when the initial mass is 0.7 $M_{\odot}$, then more than 13~Gyrs are needed to reach a contact configuration and gradually decrease the orbital period to the values we observe today. 
On the other hand, \cite{Jiang2012} suggested that the period cut-off is a result of the instability of the mass transfer that occurs when the primaries of the initially detached binaries fill their Roche lobes. They suggested that when the initial mass of the primary is lower than a certain value (approximately 0.63~$M_{\odot}$), mass transfer occurs as soon as the primary component inflates and reaches its Roche lobe limit. 
The system is then dynamically unstable and quickly becomes a common envelope binary, ultimately leading to a coalescence event. Only stars with masses above this value, may form long-lived contact binaries with an orbital period longer than 0.22~d. 

The aforementioned possible explanations of the cut-off limit are still under debate and  observational constrains are necessary. This is the main goal of the present paper.The construction of accurate models and the correlation among orbital and physical parameters of ultra-short period contact binaries is the best way to test these theories and explain
the observed period cut-off limit. The key to achieve this goal is to have reliable results and improve significantly the accuracy of the determination of the physical parameters of the components of binary systems, as well as reduce the control systematics. This can be achieved by acquiring data with the same instrumentation, reaching sufficient signal to noise ratio and using the same well defined methodology and data reduction procedure. These criteria are fulfilled in the {\it CoBiToM Project} \citep{Gazeas2021b}, providing as accurate as possible solutions leading to important information about the nature of these systems and their environment. 

Thousands of contact binaries have been detected in recent major surveys and catalogues, such us AAVSO \citep{Watson2006}, ASAS \citep{Paczynski2006}, ASAS-SN \citep{Jayasinghe2019}, CoRoT \citep{Deleuil2018}, CSS \citep{Drake2014}, GCVS \citep{Samus2018}, Kepler \citep{Kirk2016}, LAMOST \citep{qian2020}, OGLE \citep{Soszynski2016}, SWASP \citep{Lohr2013}, WISE \citep{Petrosky2021}, and ZTF \citep{Chen2020}. The percentage of ultra-short contact binaries (with $P < 0.26~d$) among contact binaries in these surveys ranges between 1.8 and 13.8 per cent, with an average value of 6.6 per cent. These values depend on the detection and classification methods applied and vetting of false positives \citep[e.g. pulsating stars with short periods, i.e.][]{Rucinski2002, Drake2014var}. 
Hence, in the present study, a sample of 30 ultra-short orbital period contact binary systems ($P_{orb}$ < 0.26~d) is presented and examined under the scope of their physical and orbital properties. 
All systems were homogeneously observed and analysed, and only a few of them have been previously investigated in other studies using their light curves, $O-C$ diagrams and physical parameters at the same time. Therefore, we present the analysis of recent and unpublished multi-band photometric observations with well known and accurate techniques, combined with all the available spectroscopic data.

The paper is organized as follows: 
In Section 2 we describe the selection of ultra-short contact binaries, leading to a sample of 30 systems, as well as their new photometric observations. Section 3 presents the data analysis procedure used for obtaining the light curve models and interpreting the resulted uncertainties. The physical parameters of the targets are presented in Section 4, while an extended orbital period analysis is performed in Section 5. Finally, Section 6 includes an overview of the physical properties of the systems in the sample, followed by a discussion on their evolution status and dynamical stability. 


\section{Target Selection and Data Acquisition}
\subsection{Target Selection}
Our sample of ultra-short contact binaries were selected from the SWASP catalogue \citep{Norton2011}. The main criteria used to include a target in the working sample were a contact binary classification and an orbital period shorter than 0.26~d. Additionally, targets should have multi epoch data covering approximately 3-7 yr, thus ensuring a sufficient number of times of minimum light and hence a wide time span in the orbital period analysis through $O-C$ diagrams. Finally, all selected systems should be brighter than 16~mag due to the observational constraints of the collaborating observatories in the \textit{CoBiToM Project}. A sub-sample of our investigation, containing 30 ultra-short orbital period systems, will be shown in the present paper, as the rest of the sample will be discussed in forthcoming studies.

In Table \ref{Tab2.ObsLog} the IDs of our sample, along with the ephemerides and a detailed observing log are provided. During the definition of our sample, we have noticed in some cases that there was a confusion concerning the exact celestial coordinates, the amplitude of light variation and sometimes the orbital period in the literature. Close inspection revealed that seven of the selected 30 systems had wrong IDs in the literature, an issue that was also noted in some cases by \cite{Zhang2014} and \cite{Koen2016}. 
In Table \ref{Tab2.ObsLog} and throughout the paper, we use the updated IDs.

The amplitude of variability is also noticed to be larger when high angular resolution photometry is performed. For example, the eclipse depth reported by \cite{Norton2011} using the small aperture SWASP cameras was found to be significantly smaller in some cases, compared to our follow-up observations with much larger aperture telescopes. This issue is expected, when low-resolution instruments are used, when seeing conditions are poor or even when smearing effects play significant role in the observations. 
Low spatial resolution results in light contamination by close companions and the observed light curve (and therefore the eclipse depth) appears shallower. 

\subsection{Observations}
Publicly available photometric data for the targets under study provided a sufficient number of eclipse timings spanning over an adequate period of time. 
However, this is not long enough for the purpose of a detailed orbital period modulation analysis, as seen through the $O-C$ diagrams. 
In addition, photometric accuracy by small aperture telescopes is rather poor, limiting significantly the quality of the resulting models and the calculation of the absolute physical and orbital parameters. 
Therefore, we revisited our sample, by performing follow-up observations with the University of Athens Observatory (UOAO), and the telescopes of the National Observatory of Athens (NOA) at Helmos and Kryoneri Astronomical Stations. 
An analytical description of the astronomical equipment in each facility is described in the first paper of the series \citep{Gazeas2021b}.
Supplementary data of the southern targets were also utilized in this study for the systems that can not be observed from the northern latitudes. These data were collected with the 1~m telescope at South African Astronomical Observatory (SAAO), following the observing strategy described by \cite{Koen2016}.

The aim of the observations was the acquisition of multi-band photometric data (the filter set is mentioned in the seventh column in Table \ref{Tab2.ObsLog}) for all targets under study. Differential aperture photometry was performed in all cases. 
In order to collect uniform photometric data and construct the phased light curves, all observations were obtained using only one instrumental setup for each target, in order to avoid any systematic effects due to instrumental cross calibration or filter mismatch. 

For the purpose of extending the timescale of the $O-C$ diagrams, recent supplementary observations around the eclipse phase have been conducted in most targets in order to provide additional epochs. These data were not used in the light curve modeling process. 
Our goal during the entire observing period was to obtain a complete light curve within a few days,
in order to minimize the effect of any intrinsic variability (e.g. magnetic activity). 
The timescale of any intrinsic variability is usually significantly longer than the duration of our observations, since the orbital period of all systems is short enough to be covered within one night. 
Therefore, the light curves modeled in this paper represent a `snapshot' of a system's photometric behaviour over the duration of observations. 
A detailed observing log is presented in Table \ref{Tab2.ObsLog} and includes the observing dates, the total number of nights dedicated to each target and the observing telescope.

\begin{table*}
    \centering
	\caption{Targets observed with their linear ephemerides for phasing the observations. The standard errors for each value are expressed in parentheses, in units of last decimal digit quoted. The last three columns provide an observation log for all targets with the observing period, the number of data collecting nights and the observing site.}
	\label{Tab2.ObsLog}
	\setlength{\tabcolsep}{3pt}
	\begin{tabular}{lccccccccc} 
	\hline
System ID	                &	GCVS ID	    & $T_{0}$ (HJD)	    & $P_{orb}$~(d)	&	Observing Season    & Nights& Filters  & $Site^{*}$ &   \\
\hline																		
1SWASP J030749.87-365201.7	&	BL For	    & 2457010.41088(24)	& 0.2266707(4)	&	Dec	2014			&	1	&	BVRI & SAAO	        &	\\
1SWASP J040615.79-425002.3	&	AQ Hor	    & 2457013.33073(44)	& 0.22233739(3)	&	Dec	2014			&	1	&	BVRI & SAAO	        &	\\
1SWASP J044132.96+440613.7	&	V1110 Her 	& 2458039.60234(18)	& 0.2281521(1)	&	Sep	2017 - Apr 2021	&	4+1	&	BVRI & Helmos+UOAO	&	\\
1SWASP J050904.45-074144.4	&	OV Eri	    & 2457639.56377(25)	& 0.2295749(3)	&	Dec	2014 - Apr 2021	&	45+1&	BVRI & SAAO+UOAO    &	\\
1SWASP J052926.88+461147.5	&	V840 Aur	& 2458769.62318(21)	& 0.2266426(2)	&	Oct	2018 - Apr 2021	&	3+1	&	BVRI & Helmos+UOAO	&	\\
1SWASP J055416.98+442534.0	&	V853 Aur	& 2456353.38170(33)	& 0.21849667(2)	&	Feb	2013 - Apr 2021	&	8	&	BVRI & UOAO	        &	\\
1SWASP J080150.03+471433.8	&	LX Lyn	    & 2456778.35833(24)	& 0.21751919(3)	&	Apr	2014 - Apr 2021	&	8	&	BVRI & Kryoneri	    &	\\
1SWASP J092328.76+435044.8	&	-	        & 2457826.56109(21)	& 0.2348857(1)	&	Mar	2017 - Apr 2021	&	16	&	BVRI & UOAO	        &	\\
1SWASP J092754.99-391053.4	&	CO Ant	    & 2456765.37391(28)	& 0.22534530(4)	&	Apr	2014			&	1	&	UBVR & SAAO	        &	\\
1SWASP J093010.78+533859.5	&	V442 UMa	& 2456329.63930(20)	& 0.22771395(6)	&	Jan	2013 - Apr 2021	&	15	&	BVRI & UOAO	        &	\\
1SWASP J114929.22-423049.0	&	V1410 Cen	& 2456758.28401(22)	& 0.2273081(2)	&	Apr	2014			&	1	&	BVRI & SAAO	        &	\\
1SWASP J121906.35-240056.9	&	AE Crv	    & 2456768.38662(20)	& 0.22636763(3)	&	Apr	2014 - Apr 2021	&	1+1	&	BVRI & SAAO+UOAO	&	\\
1SWASP J133105.91+121538.0	&	-	        & 2456347.55370(15)	& 0.21801190(2)	&	Feb	2013 - Apr 2021	&	9	&	BVRI & UOAO	        &	\\
1SWASP J150822.80-054236.9	&	-	        & 2456352.62854(40)	& 0.26006086(5)	&	Feb	2013 - Apr 2021	&	12	&	BVRI & UOAO	        &	\\
2MASS J15165453+0048263	    &	V640 Ser	& 2457956.31802(43)	& 0.2107323(1)	&	Apr	2014 - Apr 2021	&	14+1&   UBVR & SAAO+UOAO	&	\\
1SWASP J161335.80-284722.2	&	V1677 Sco	& 2456877.30927(11)	& 0.2297735(1)	&	Aug	2014			&	1	&	BVRI & SAAO	        &	\\
1SWASP J170240.07+151123.5	&	-	        & 2457596.41084(18)	& 0.2614691(3)	&	Jul	2016 - Apr 2021	&	29	&	BVRI & UOAO	        &	\\
1SWASP J173003.21+344509.4	&	V1498 Her	& 2456832.36569(49)	& 0.2237088(1)	&	Jun	2014 - Apr 2021	&	4+1	&	BVRI & Kryoneri+UOAO&	\\
1SWASP J173828.46+111150.2	&	-	        & 2457568.55989(8)	& 0.2493487(3)	&	Aug	2014 - Apr 2021	&	27+1&	BVRI & SAAO+UOAO    &	\\
1SWASP J174310.98+432709.6	&	V1067 Her	& 2456778.49931(44)	& 0.2581081(1)	&	Apr	2014 - Apr 2021	&	9+1	&	BVRI & Kryoneri+UOAO&	\\
1SWASP J180947.64+490255.0	&	V1104 Her	& 2457629.48307(12)	& 0.2278766(1)	&	Jul	2016 - Apr 2021	&	40	&	BVRI & UOAO	        &	\\
1SWASP J195900.31-252723.1	&	-	        & 2456881.33791(38)	& 0.2381397(2)	&	Aug	2014 - Apr 2021	&	1+1	&	BVRI & SAAO+UOAO	&	\\
2MASS J21031997+0209339	    &	V496 Aqr	& 2457946.53396(43)	& 0.2285901(5)	&	Sep	2015 - Apr 2021	&	1+18 &	BVRI & SAAO+UOAO	&	\\
2MASS J21042404+0731381	    &	-	        & 2457656.36702(61)	& 0.2090908(2)	&	Sep	2015 - May 2021	&	18+2+1&	BVRI & SAAO+Helmos+UOAO&\\
1SWASP J212454.61+203030.8	&	-	        & 2457271.48793(30)	& 0.2278308(2)	&	Sep	2015 - Apr 2021	&	1+1	&	BVRI & SAAO+UOAO	&	\\
1SWASP J212808.86+151622.0	&	V694 Peg	& 2458014.28967(9)	& 0.22484157(9)	&	Sep 2017 - Apr 2021	&	1+1	&	BVRI & Helmos+UOAO	&	\\
1SWASP J220734.47+265528.6	&	V729 Peg	& 2457257.38289(14)	& 0.2312352(2)	&	Sep	2014 - Apr 2021	&	19+2&	BVRI & Helmos+UOAO	&	\\
1SWASP J221058.82+251123.4	&	V732 Peg	& 2458012.38783(26)	& 0.21372960(5)	&	Sep	2017 - Apr 2021	&	1+1	&	BVRI & Helmos+UOAO	&	\\
1SWASP J224747.20-351849.3	&	AS PsA	    & 2457279.30511(29)	& 0.2182159(1)	&	Sep	2015			&	1	&	BVRI & SAAO	        &	\\
1SWASP J232610.13-294146.6	&	DU Scl      & 2457274.54580(18)	& 0.2301173(4)	&	Sep	2015			&	1	&	BVRI & SAAO	        &	\\
\hline
\multicolumn{6}{l}{(*): \it{UOAO:  University of Athens Observatory, Kryoneri: Kryoneri Observatory of National Observatory of Athens}} \\
\multicolumn{6}{l} {\it{Helmos: Helmos Observatory of National Observatory of Athens, SAAO: South African Astronomical Observatory}} \\
\hline
\end{tabular}
\\
\end{table*}

\section{Data Analysis and Light Curve Modeling}
The data were reduced following standard procedures of aperture photometry and calibration. Differential photometry was subsequently performed, with a photometric accuracy between 5-10~mmag. Times of minimum light were used to derive precise ephemerides for the systems as described in section 3.1. The light curves were period folded and then modelled as described in section 3.3. This modeling requires the effective temperature of the primary component $T_{1}$ (the component that is eclipsed at phase 0) as a prior, as described in section 3.2.

\subsection{Times of minimum light and linear ephemerides}
Photometric times of minimum light were collected within the framework of the \textit{CoBiToM Project} over the last decades. The times of minimum light were calculated using the method of Kwee \& van Woerden \citep{Kwee1956}. From these timings, the linear astronomical ephemerides were calculated. Using the derived orbital period, the light curves were folded to show one single orbital period. Table \ref{Tab2.ObsLog} lists the linear ephemerides of all systems along with their uncertainties. 

Additional times of minimum light were also extracted from all the available online time-series photometric data (SWASP, ASAS-SN, ASAS, CSS, and NSVS surveys). 
In all cases, where the retrieved data are sparse (i.e. data from the catalogues NSVS, ASAS, ASAS-SN and CSS), we followed the same procedure of folding the data into "local" phase diagrams, according to the methodology described by \cite{Li2020}. 
Times of minimum light of some targets were also found in online minima databases \footnote{\url{http://var2.astro.cz/ocgate/}} \footnote{\url{http://www.oa.uj.edu.pl/ktt/krttk_dn.html}}. 

Early studies on individual targets provided an additional source of eclipse timing information.
We collected times of minimum light from the following publications: \cite{Li2020,Lu_2020,Fang2019,Zasche2019,Kjurkchieva2018,Haroon2018,Loukaidou2018,Darwish2017,Koen2016,Djurasevic2016,Saad2016,Dimitrov2015,Liu2015,Koo2014,Elkhateeb2014,Zhang2014,Terell2014}.

Utilising the information from all the above sources, $O-C$ diagrams for each target of our sample were constructed, following the procedure described in Section 5.

\begin{table*}
	\centering
	\caption{Astrometric and photometric parameters for the studied systems. The columns include parallax, distance, absolute magnitude in V filter and photometric depth in both eclipses, as well as the temperature values derived from photometric color index and spectroscopic observations. The adopted temperature for the modeling process is $T_{m}$.}
	\label{Tab3.Temperature}
	\begin{tabular}{lcccccccccccc} 
	\hline
System ID                   & Parallax      & d     &	$M_V$ 	 &	$min_{I}$       &	$min_{II}$ 	& $T_{BV}$	& $T_{gi}$	&	$T_{spec}$	&	$T_{m}$		\\
                            & (mas)         & (pc)  &	(mag)	 &	(mag)	            &	(mag)	        & (K)	    & (K)	    &	(K)	        &	(K)		\\
\hline 																											
1SWASP J030749.87-365201.7	&$	1.82	\pm	0.02	$&$	550	\pm	6	$&$	6.08	\pm	0.06	$&	0.78	&	0.64	&	4797	&	4591	&	-	        &	$	4700	$	\\
1SWASP J040615.79-425002.3	&$	2.59	\pm	0.02	$&$	387	\pm	3	$&$	6.18	\pm	0.03	$&	0.62	&	0.56	&	4865	&	4952	&	$5040^{a}$	&	$	4900	$	\\
1SWASP J044132.96+440613.7	&$	3.10	\pm	0.07	$&$	323	\pm	7	$&$	3.74	\pm	0.11	$&	0.95	&	0.78	&	5183	&	5565	&	-	        &	$	5350	$	\\
1SWASP J050904.45-074144.4	&$	3.87	\pm	0.02	$&$	258	\pm	1	$&$	6.00	\pm	0.15	$&	0.71	&	0.61	&	5030	&	5002	&	$5340^{a}$	&	$	5000	$	\\
1SWASP J052926.88+461147.5	&$	1.86	\pm	0.04	$&$	538	\pm	13	$&$	4.81	\pm	0.12	$&	0.69	&	0.59	&	5375	&	5425	&	-	        &	$	5400	$	\\
1SWASP J055416.98+442534.0	&$	4.94	\pm	0.03	$&$	203	\pm	1	$&$	5.48	\pm	0.20	$&	0.44	&	0.42	&	5189	&	5286	&	-	        &	$	5250	$	\\
1SWASP J080150.03+471433.8	&$	3.79	\pm	0.03	$&$	264	\pm	2	$&$	6.12	\pm	0.03	$&	0.71	&	0.69	&	4650	&	4656	&	$4690^{b}$	&	$	4650	$	\\
1SWASP J092328.76+435044.8	&$		-		        $&$		-		$&$		-		        $&	0.60	&	0.62	&	5619	&	5946	&	-	        &	$	5800	$	\\
1SWASP J092754.99-391053.4	&$	6.12	\pm	0.03	$&$	163	\pm	1	$&$	5.10	\pm	0.02	$&	0.50	&	0.42	&	5324	&	5469	&	-	        &	$	5400	$	\\
1SWASP J093010.78+533859.5	&$	14.29	\pm	0.06	$&$	70.0\pm	0.3	$&$	6.72	\pm	0.08	$&	0.18	&	0.17	&	4884	&	4347	&	$4700^{c}$	&	$	4700	$	\\
1SWASP J114929.22-423049.0	&$	4.87	\pm	0.02	$&$	205	\pm	1	$&$	7.29	\pm	0.04	$&	0.65	&	0.53	&	4194	&	4132	&	-	        &	$	4150	$	\\
1SWASP J121906.35-240056.9	&$	2.13	\pm	0.04	$&$	470	\pm	8	$&$	6.63	\pm	0.08	$&	0.61	&	0.51	&	4656	&	4675	&	-	        &	$	4650	$	\\
1SWASP J133105.91+121538.0	&$	14.00	\pm	0.07	$&$	71.4\pm	0.4	$&$	6.33	\pm	0.03	$&	0.71	&	0.54	&	5143	&	5050	&	-	        &	$	5150	$	\\
1SWASP J150822.80-054236.9	&$	4.30	\pm	0.04	$&$	232	\pm	2	$&$	5.40	\pm	0.05	$&	0.80	&	0.71	&	5202	&	5115	&	$4500^{d}$	&	$	5150	$	\\
2MASS J15165453+0048263	    &$	2.58	\pm	0.04	$&$	388	\pm	6	$&$	5.89	\pm	0.08	$&	0.58	&	0.49	&	5910	&	6417	&	-	        &	$	6150	$	\\
1SWASP J161335.80-284722.2	&$	8.62	\pm	0.04	$&$	116	\pm	1	$&$	6.66	\pm	0.02	$&	0.81	&	0.61	&	4661	&	4455	&	-	        &	$	4550	$	\\
1SWASP J170240.07+151123.5	&$	2.61	\pm	0.02	$&$	382	\pm	4	$&$	5.35	\pm	0.05	$&	0.70	&	0.61	&	4885	&	5090	&	-	        &	$	5000	$	\\
1SWASP J173003.21+344509.4	&$	2.89	\pm	0.01	$&$	347	\pm	2	$&$	6.00	\pm	0.03	$&	0.36	&	0.30	&	4838	&	4570	&	-	        &	$	4700	$	\\
1SWASP J173828.46+111150.2	&$		-		        $&$		-		$&$		-		        $&	0.41	&	0.33	&	5248	&	5275	&	$4940-5280^{a}$&$	5250	$	\\
1SWASP J174310.98+432709.6	&$	3.21	\pm	0.02	$&$	311	\pm	2	$&$	5.95	\pm	0.03	$&	0.70	&	0.52	&	5230	&	5353	&	-	        &	$	5300	$	\\
1SWASP J180947.64+490255.0	&$	5.40	\pm	0.01	$&$	185	\pm	1	$&$	7.02	\pm	0.01	$&	1.13	&	0.75	&	4083	&	4049	&	-	        &	$	4050	$	\\
1SWASP J195900.31-252723.1	&$		-		        $&$		-		$&$		-		        $&	0.72	&	0.70	&	5064	&	5752	&	-	        &	$	5400	$	\\
2MASS J21031997+0209339	    &$	2.07	\pm	0.05	$&$	483	\pm	11	$&$	6.65	\pm	0.12	$&	1.00	&	0.79	&	4647	&	4220	&	$4450^{a}$	&	$	4400	$	\\
2MASS J21042404+0731381	    &$	1.98	\pm	0.07	$&$	505	\pm	18	$&$	5.14	\pm	0.20	$&	0.55	&	0.49	&	4788	&	4865	&	$4450-5040^{a}$&$	4800	$	\\
1SWASP J212454.61+203030.8	&$		-		        $&$		-		$&$		-		        $&	0.64	&	0.56	&	5093	&	5359	&	-	        &	$	5250	$	\\
1SWASP J212808.86+151622.0	&$	2.53	\pm	0.05	$&$	395	\pm	7	$&$	6.29	\pm	0.09	$&	0.67	&	0.52	&	4621	&	4727	&	$4450-4840^{a}$&$	4700	$	\\
1SWASP J220734.47+265528.6	&$	1.65	\pm	0.03	$&$	607	\pm	12	$&$	5.14	\pm	0.10	$&	0.44	&	0.42	&	4933	&	4850	&	-	        &	$	4900	$	\\
1SWASP J221058.82+251123.4	&$		-		        $&$		-		$&$		-		        $&	0.66	&	0.51	&	4968	&	4937	&	-	        &	$	4950	$	\\
1SWASP J224747.20-351849.3	&$	2.97	\pm	0.02	$&$	337	\pm	3	$&$	6.21	\pm	0.04	$&	0.20	&	0.15	&	4369	&	4267	&	$4450-4620^{a}$&$	4300	$	\\
1SWASP J232610.13-294146.6	&$	4.04	\pm	0.04	$&$	248	\pm	3	$&$	6.55	\pm	0.05	$&	0.60	&	0.58	&	4820	&	4855	&	$4450-4840^{a}$&$	4850	$	\\
\hline 
\multicolumn{6}{l}{Median temperature in the last column is accounted for error approximately of 200~K} \\
\multicolumn{6}{l}{(a):\cite{Koen2016}, (b):\cite{Dimitrov2015}, (c):\cite{Lohr2015a}, (d):\cite{Lohr2014}}\\

    \end{tabular}
\end{table*}

\subsection{Temperature information and spectroscopy}
The effective temperature $T_{1}$ in each system was determined from $B-V$ and $g-i$ color indices, using the following procedure:
We calculated $T_{BV}$ as the temperature based on the $B-V$ color index, provided by AAVSO Photometric All Sky Survey (APASS) DR9; \citep{Henden2016}, following the conversion provided by \cite{Pecaut2013}. 
Also $T_{gi}$ is the calculated temperature based on the $g-i$ color index (also provided by APASS DR9), following the conversion provided by \cite{Covey2007}.
Both $B-V$ and $g-i$ color indices were corrected for reddening, using the extinction tables by \cite{Schlafly2011}, which are based on the IRAS \footnote{\url{https://irsa.ipac.caltech.edu/applications/DUST/}} photometric database. 
Effective temperature based on the infrared passbands ($JHK$) and their color indices were not sensitive enough in the range of interest (4000-5000~K), as the $B-V$ and $g-i$ are, and hence they were not included in our approach.
Additional spectroscopic data were retrieved from the literature \citep{Koen2016, Dimitrov2015, Lohr2015a, Lohr2014, Drake2014} for 11 systems of our sample, providing independent estimates of the effective temperature $T_{1}$ of the primary component. 

Table \ref{Tab3.Temperature} lists in the seventh and eight column the temperature values $T_{BV}$ and $T_{gi}$ obtained by photometric color indices and in the ninth column mentions the spectroscopic temperature $T_{sp}$. 
The rounded average ($T_{m}$) in the last column was assigned as the effective temperature of the primary component in each system, as it is needed for a prior value in the modeling process.
It was calculated by averaging the photometric values in columns $T_{BV}$ and $T_{gi}$ and rounding the value within 50~K, in order to match the closest spectral type.
The uncertainty on temperature determination with this method is estimated to be approximately 200~K. For very few cases, the photometric interpretation of the temperature deviated larger than the error. 
This occurs, due to the possibly spurious reddening determination, that can affect the temperature estimation and cause large uncertainty. 
As it can be seen from Table \ref{Tab3.Temperature}, the $T_{sp}$ values are within the error range of our final effective temperature, except for the cases of 1SWASP~J050904.45-074144.4 and 1SWASP~J150822.80-054236.9, where larger deviations are noted. 
This fact might be due to the low resolution spectra that were taken in these publications or due to the large uncertainty in reddening determination.

\cite{Casagrande2020} developed a method for determining the effective temperature of single Main Sequence stars using the Gaia color indices, metallicities and limb darkening coefficients. It is known that the majority of contact binaries (about 93 per cent, according to \cite{deJong2010}; and 96 per cent, according to \cite{Aumer2009}, respectively) are solar-metallicity objects, which are spread over the thin Galactic disk. Therefore solar metallicity is a plausible assumption to describe the current sample, a fact that is also confirmed spectroscopically by \cite{Rucinski2013}. Consequently, by assuming solar metallicity $[Fe/H] \sim0$ for our sample and surface gravity coefficient $logg \sim4$ for dwarf stars, it was found that the majority of the primary temperature values were within the error range of our approximation, confirming our initial hypothesis for the temperature determination and the metallicity of our sample.

Table \ref{Tab3.Temperature} also includes the apparent brightness decrease (depth in light curve) during the primary and secondary eclipses, as well as the parallax, distance and absolute magnitude information as derived from Gaia DR2 \citep{Gaia2018}. Taking into account the high precision astrometric observations by Gaia satellite for 25 targets of our sample, the distance is accurately determined. It is found that the current sample consists of the nearest binaries in our solar vicinity, all within a radius of $\sim$600 pc.

\subsection{Light curve modeling}
We used the Wilson-Devinney (W-D) code \citep{Wilson1979, Wilson1990} appended with the Monte Carlo (MC) algorithm as the search procedure, as described in detail in \citet{Zola2004,Gazeas2021a}. We took advantage of the fact that the applied method does not require initial values for the free parameters. Instead, it searches for the best solution within given ranges. We determined the uncertainties of the free parameters using $\chi^{2}$ minimization according to the method described in Numerical Recipes in Fortran \citep{Fortran}. 

Performing the light curve modeling, the albedo and gravity darkening coefficients are fixed at their theoretical values of $A = 0.5$ and $g = 0.32$, respectively \citep{Lucy1967,Rucinski1969}, since all binary members of the studied systems are low temperature stars ($T < 6500$~K) with convective envelopes. The limb darkening coefficients are taken from the tables of \cite{Claret2011}, according to the effective temperature of the components and the filters used. 

The parameters which are considered free are: the inclination (i), the phase shift, the effective temperature of the secondary component ($T_2$), the gravitational potential ($\Omega_{1,2}$), the luminosity of the primary ($L_{1}$), the third light ($l_{3}$) and the mass ratio (q). The relative luminosity of the secondary star ($L_{2}$) is not a free parameter, because the IPB control parameter was set to 0. In that case, $L_{2}$ is computed from geometrical parameters, the luminosity of the primary component, temperature values ($T_1$ and $T_2$) and the black body radiation law.

Cool photospheric spots had to be introduced when the light curves showed obvious asymmetries, different maximum brightness levels in the light curves, which are expressed as the O' Connell effect \citep{OConnell1951}. Light curve asymmetries are exhibited in 19 out of 30 studied systems.
In the cases that a cool spot was imposed, four additional parameters were added, in order to describe its location (latitude and longitude), size and temperature factor. It was found that in all cases a single spot was sufficient to explain the observed asymmetry. Large size photospheric spots are introduced, in order to explain asymmetries in light curves of some systems. These are not necessarily single large spots, but could be rather an extended spotted area covered with smaller spots.
It is expected that when introducing spots in a solution, the code results in a better fit with smaller residuals, as the degrees of freedom increase. However, the non-uniqueness of a spotted solution is a well known issue in stellar modeling, thus the number of spots is kept to a minimum and spots were added only to explain the observed asymmetries. The third light ($l_{3}$) parameter was also adjusted for systems which have confirmed or proposed additional companions (e.g. based on the $O-C$ diagram or on spectroscopic observations). 

An additional check on our modeling results was made for the systems 1SWASP~J093010.78+533859.5 and 1SWASP~J150822.80-054236.9, which were also studied previously by \cite{Lohr2015a} and \cite{Lohr2014}, respectively. These two systems were observed spectroscopically and their mass ratio is determined to be $q_{sp} = 0.397 \pm 0.006$ and $q_{sp} = 0.510 \pm 0.015$, respectively. Computations with the MC code converged to the photometric mass ratio, which was found to be $q_{ph} = 0.415 \pm 0.007$ and $q_{ph} = 0.578 \pm 0.032$, respectively. These values agree within 5 per cent for the totally eclipsing system 1SWASP~J093010.78+533859.5 and 13 per cent for the partial eclipsing system 1SWASP~J150822.80-054236.9. This gives confidence about the reliability of the mass ratio determination for other systems in the sample in the present study. It is also noted that for nine systems (about 30 per cent of the total sample) that show total eclipses, the photometric determination of the mass ratio is known to be accurately determined \citep{Pribulla2003,Terrell2005}.

\begin{figure*}
\includegraphics[width=4.1cm,scale=1.0,angle=270]{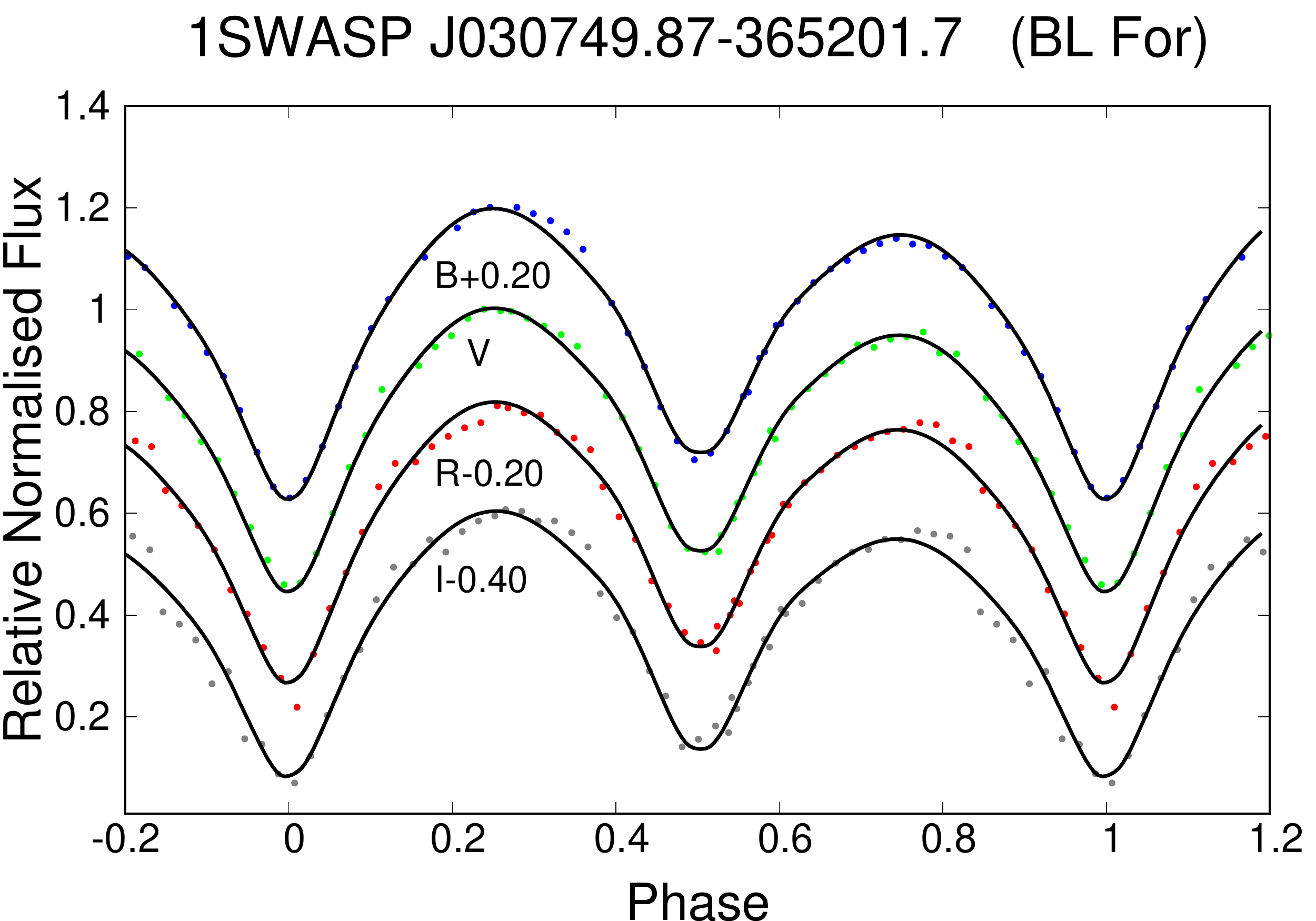}
\includegraphics[width=4.1cm,scale=1.0,angle=270]{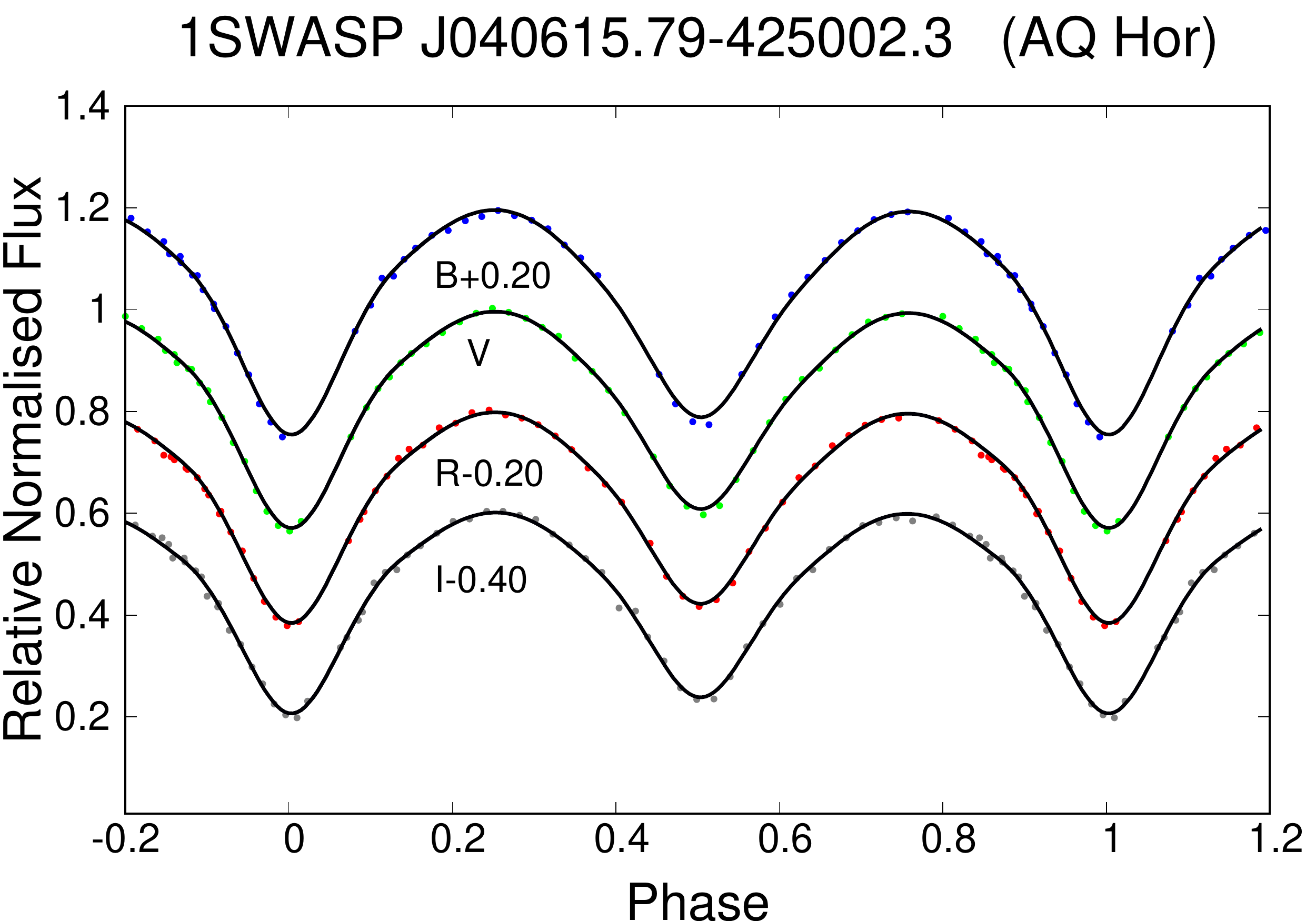}
\includegraphics[width=4.1cm,scale=1.0,angle=270]{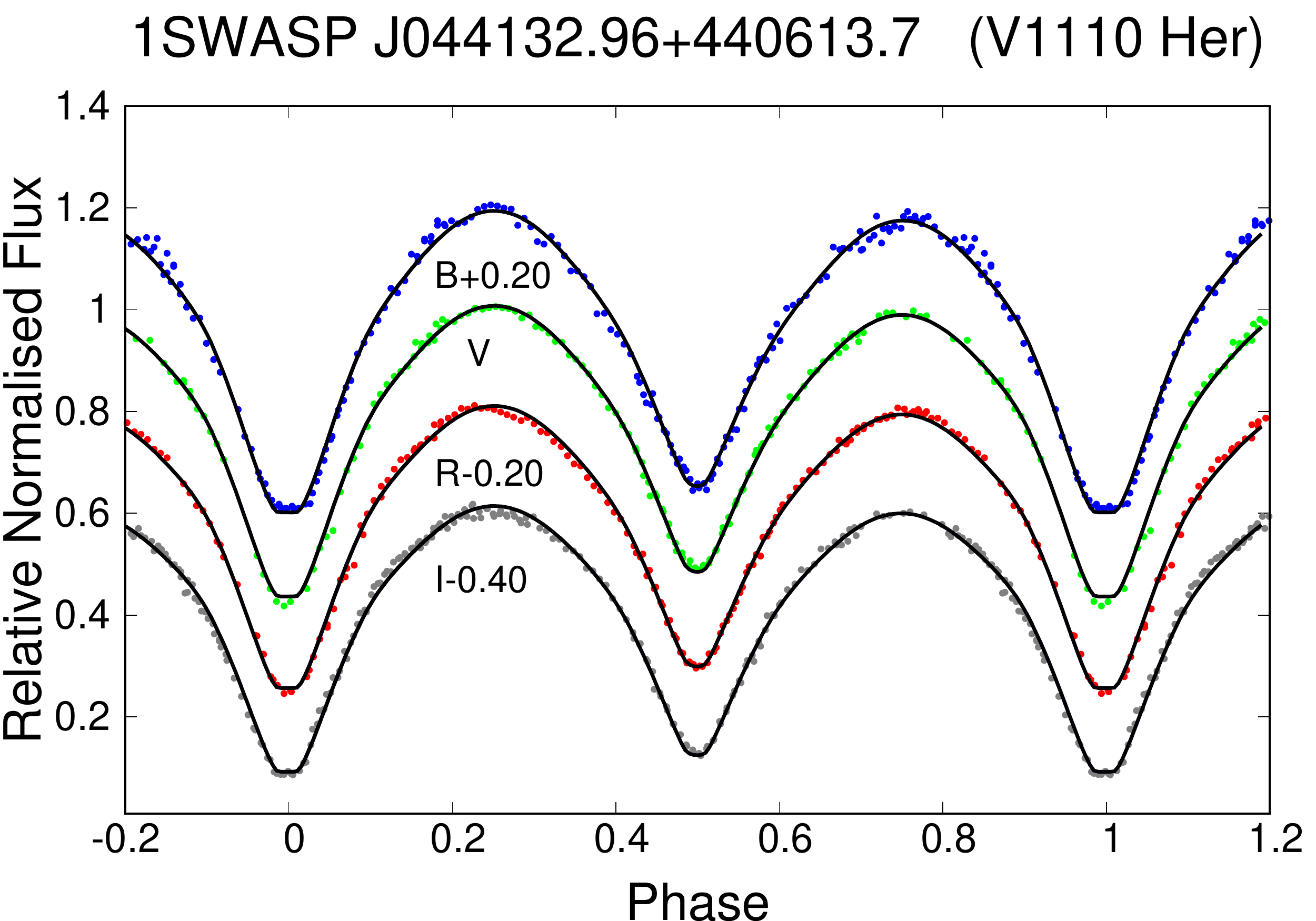}
\\
\vspace{10 pt}
\includegraphics[width=4.1cm,scale=1.0,angle=270]{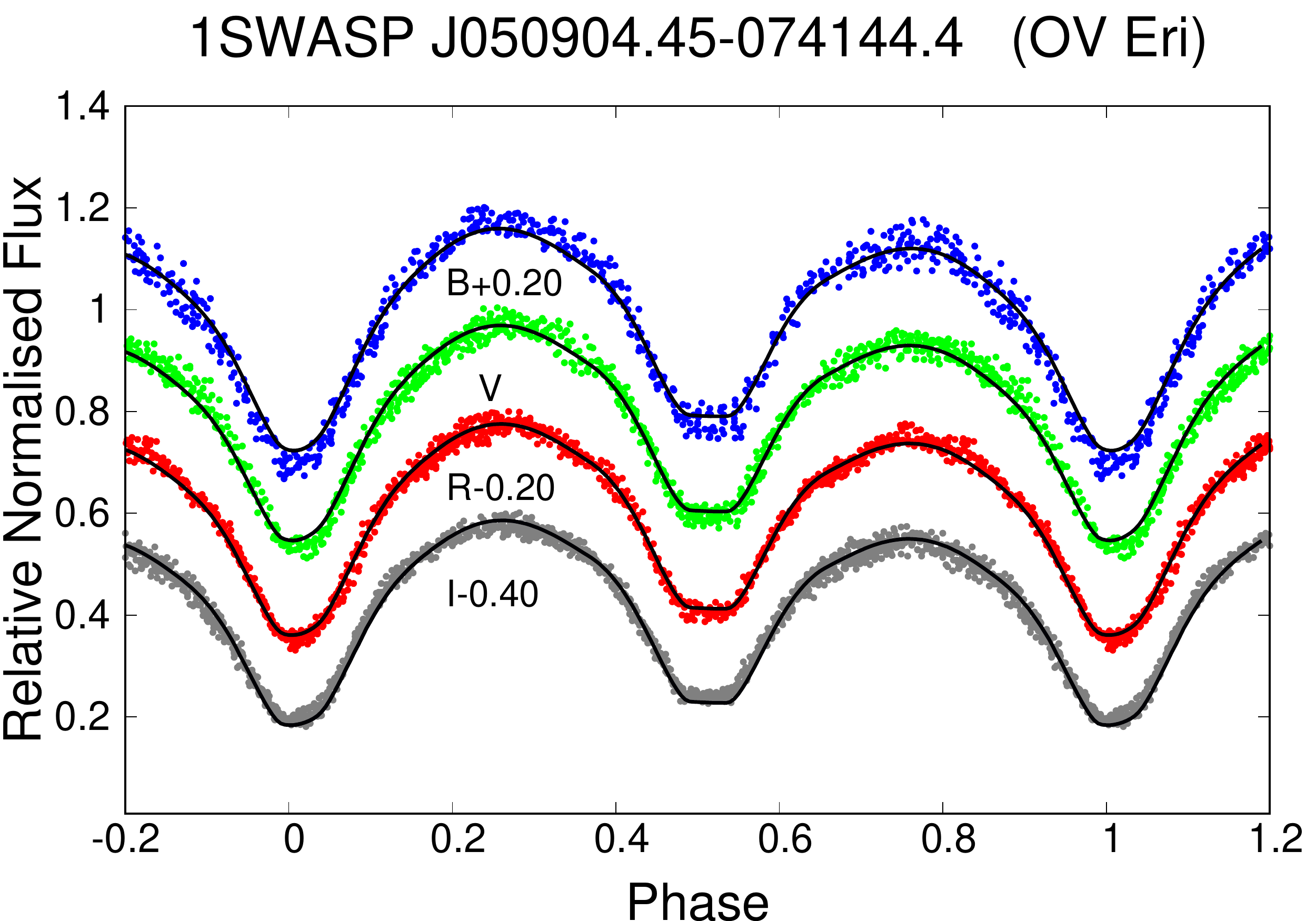}
\includegraphics[width=4.1cm,scale=1.0,angle=270]{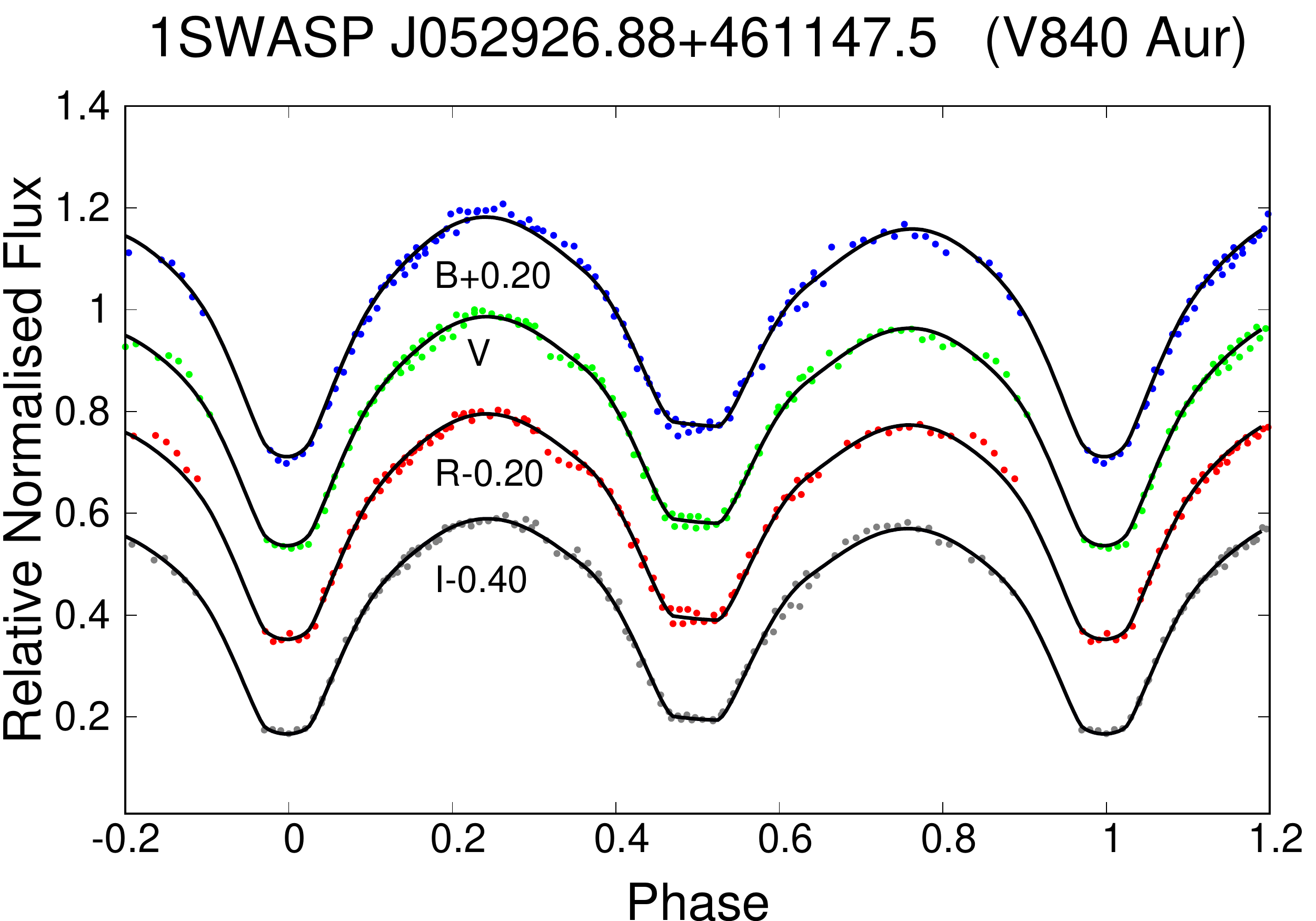}
\includegraphics[width=4.1cm,scale=1.0,angle=270]{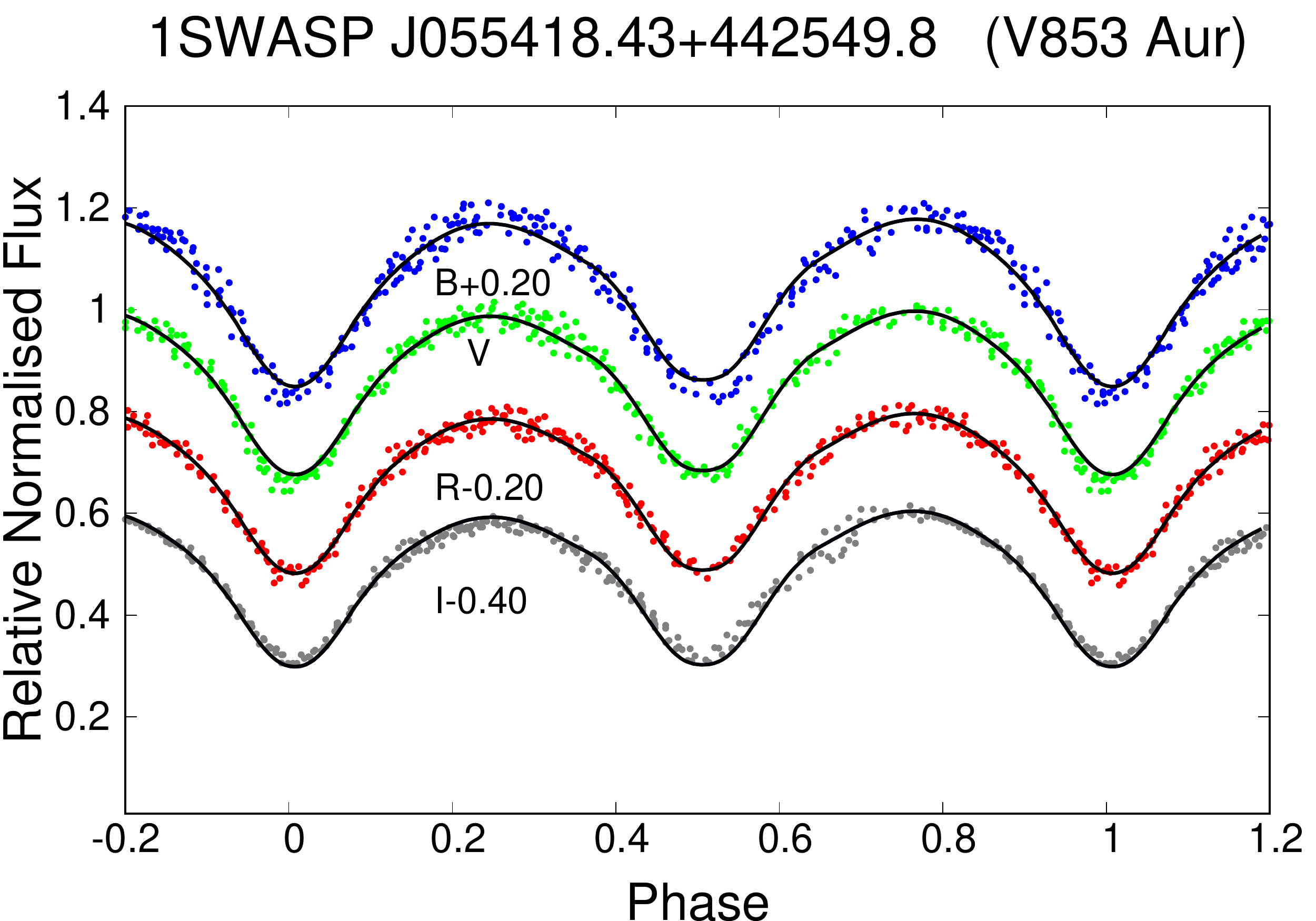}
\\
\vspace{10 pt}
\includegraphics[width=4.1cm,scale=1.0,angle=270]{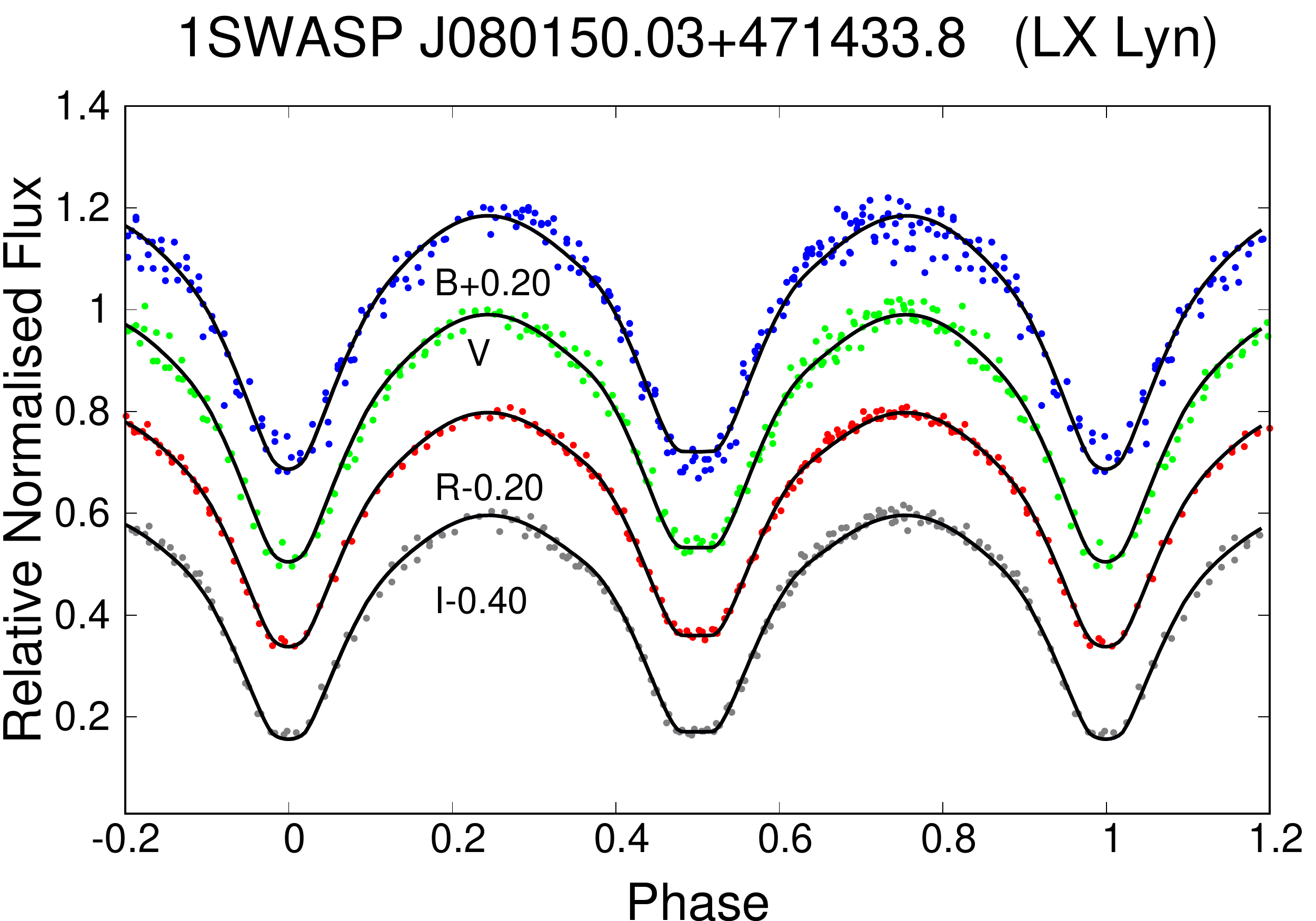}
\includegraphics[width=4.1cm,scale=1.0,angle=270]{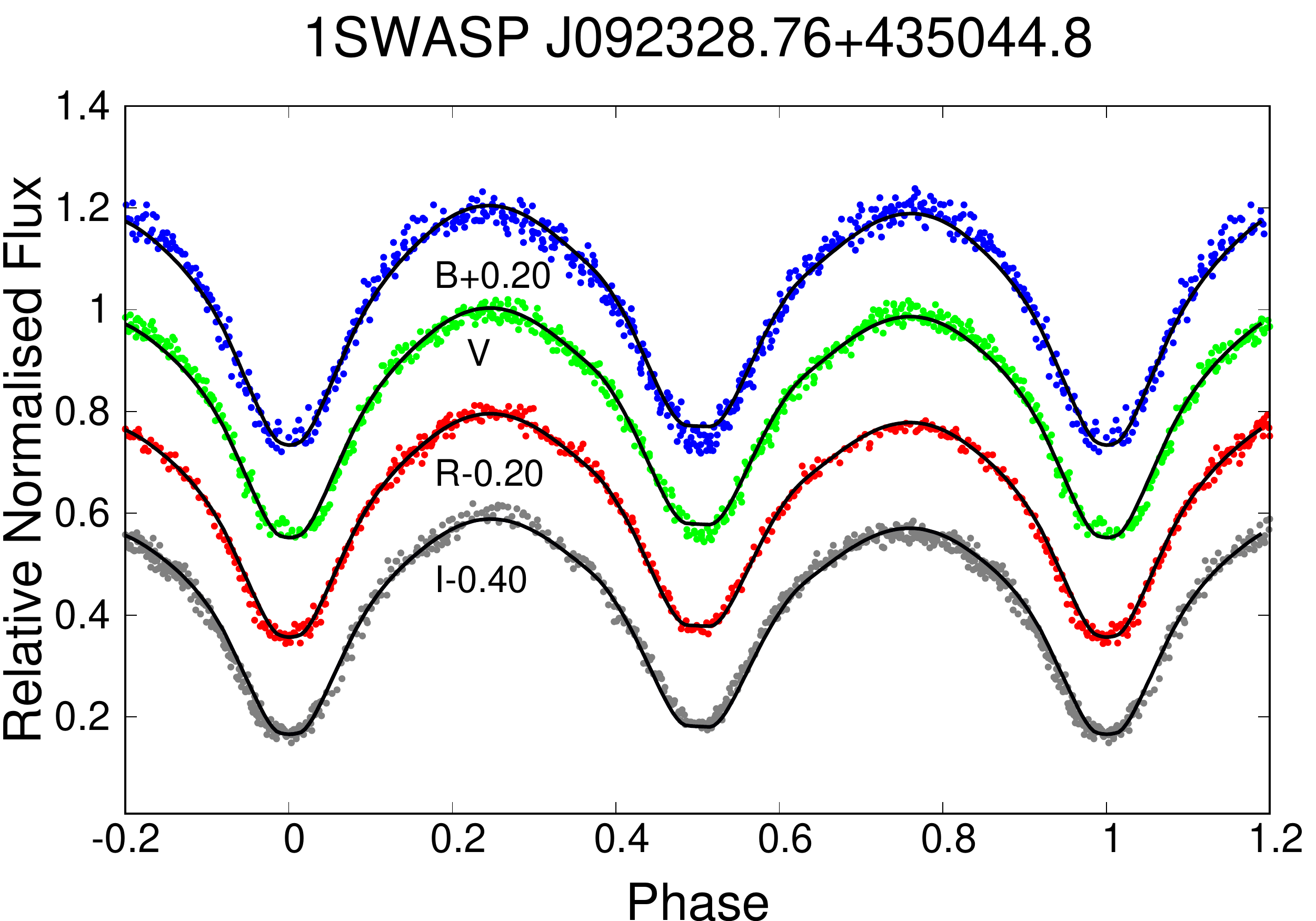}
\includegraphics[width=4.1cm,scale=1.0,angle=270]{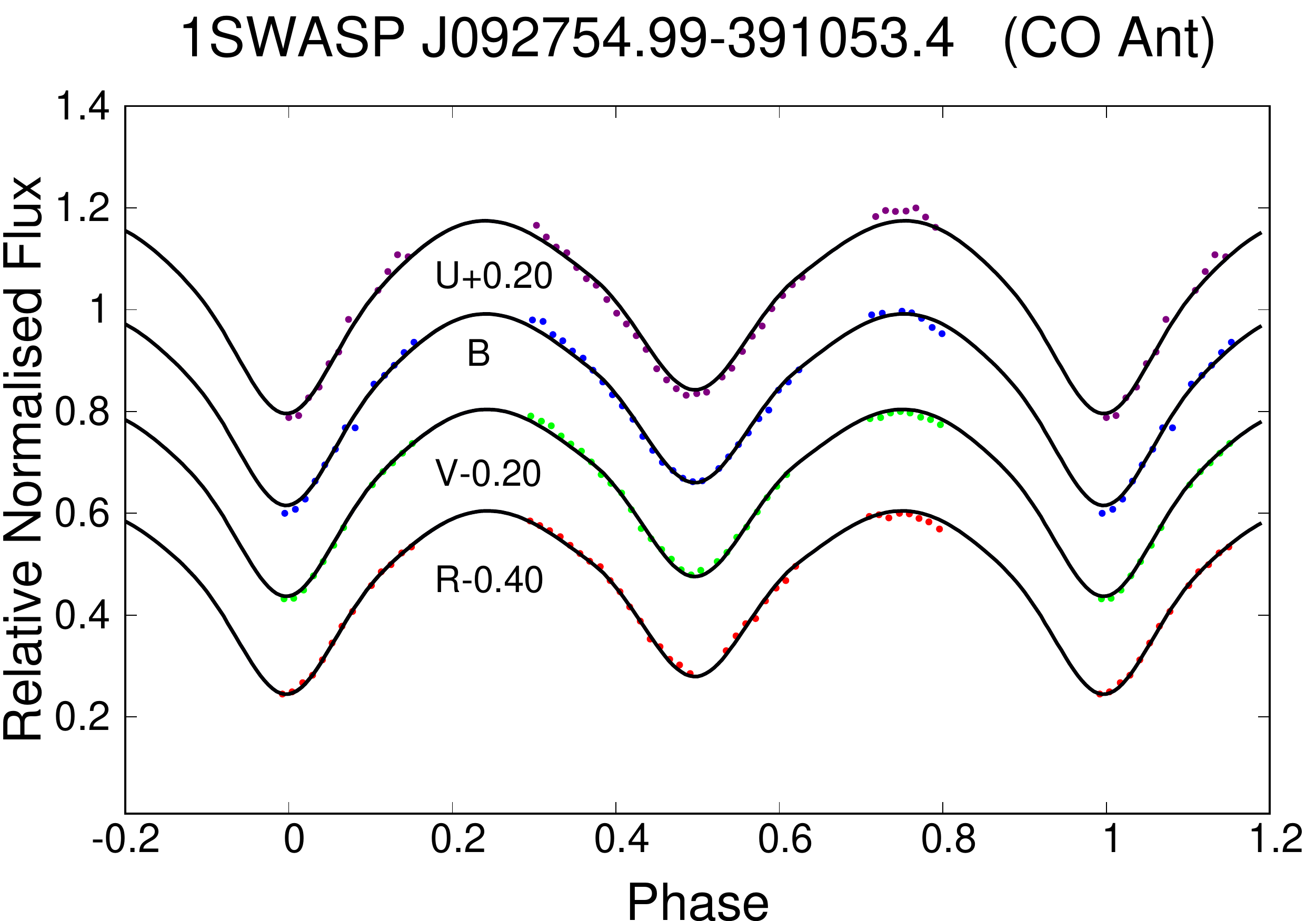}
\\
\vspace{10 pt}
\includegraphics[width=4.1cm,scale=1.0,angle=270]{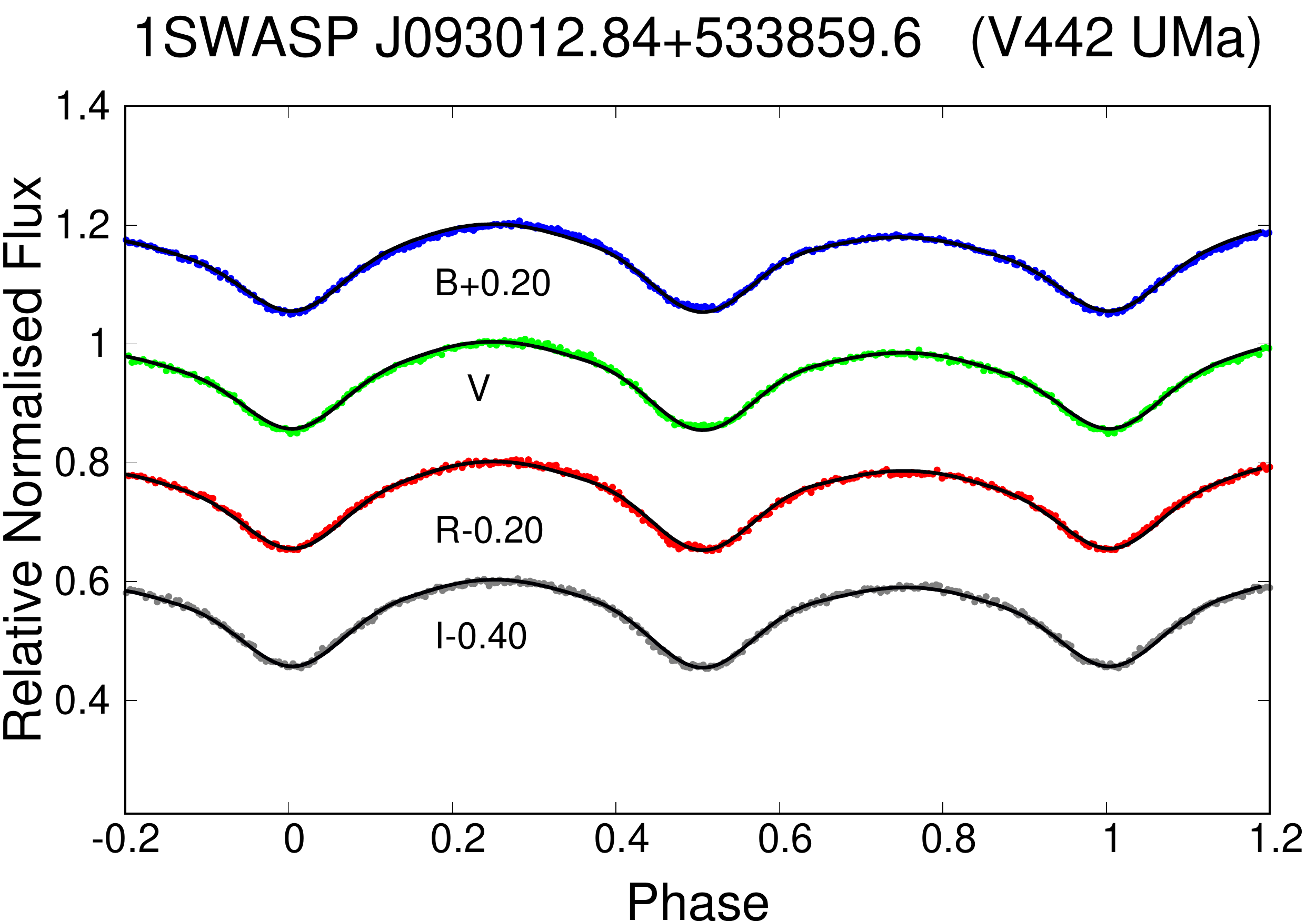}
\includegraphics[width=4.1cm,scale=1.0,angle=270]{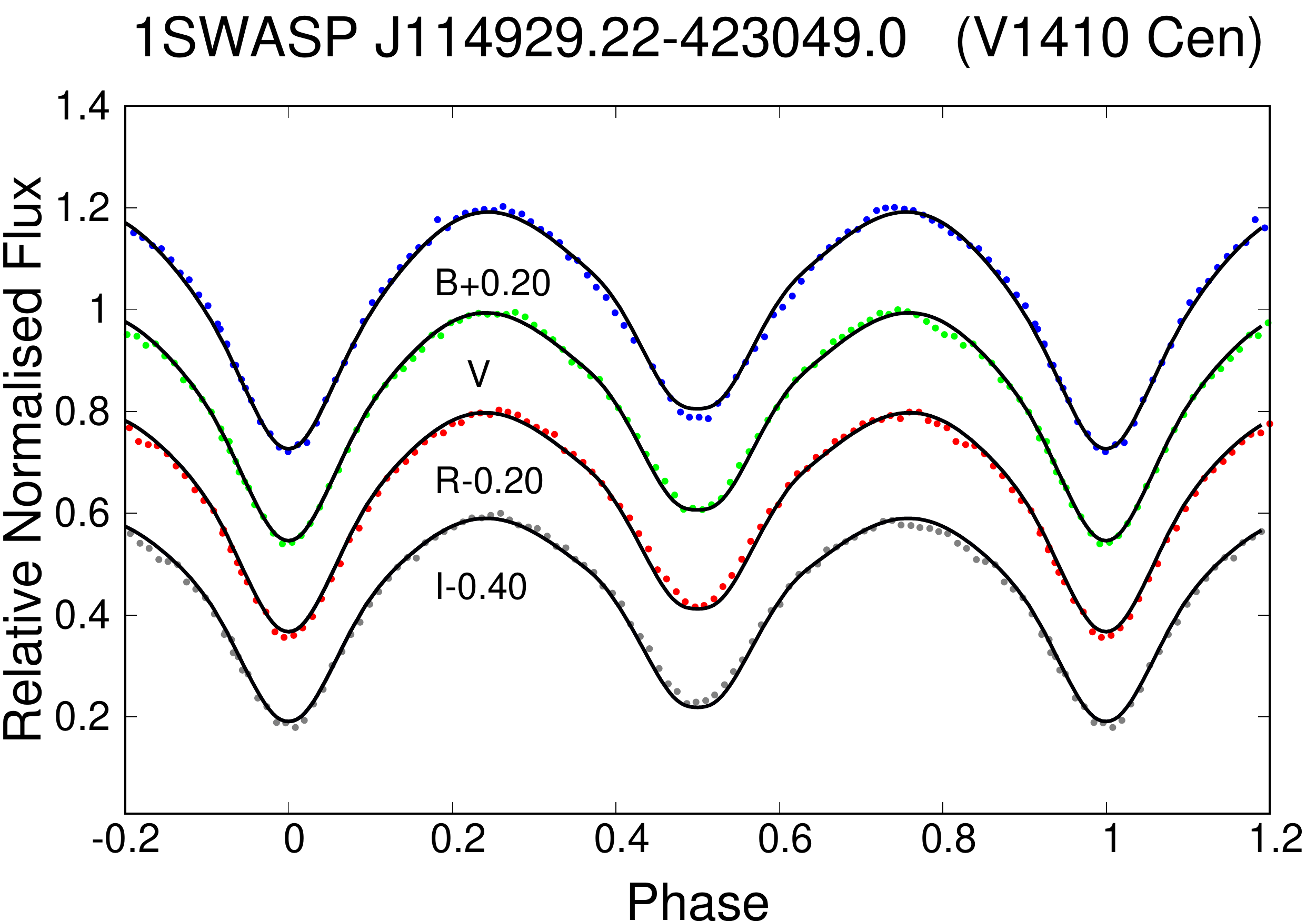}
\includegraphics[width=4.1cm,scale=1.0,angle=270]{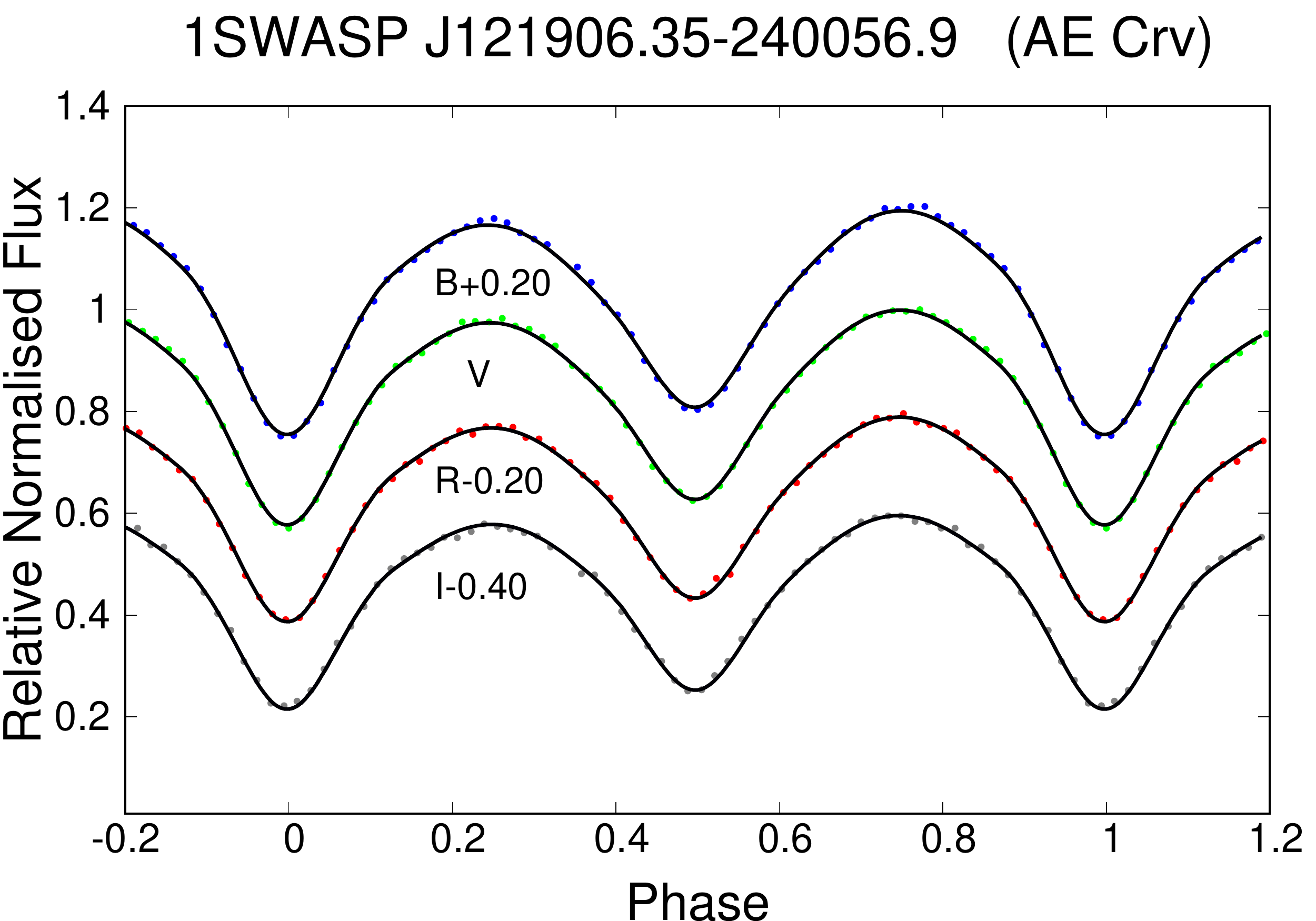}
\\
\vspace{10 pt}
\includegraphics[width=4.1cm,scale=1.0,angle=270]{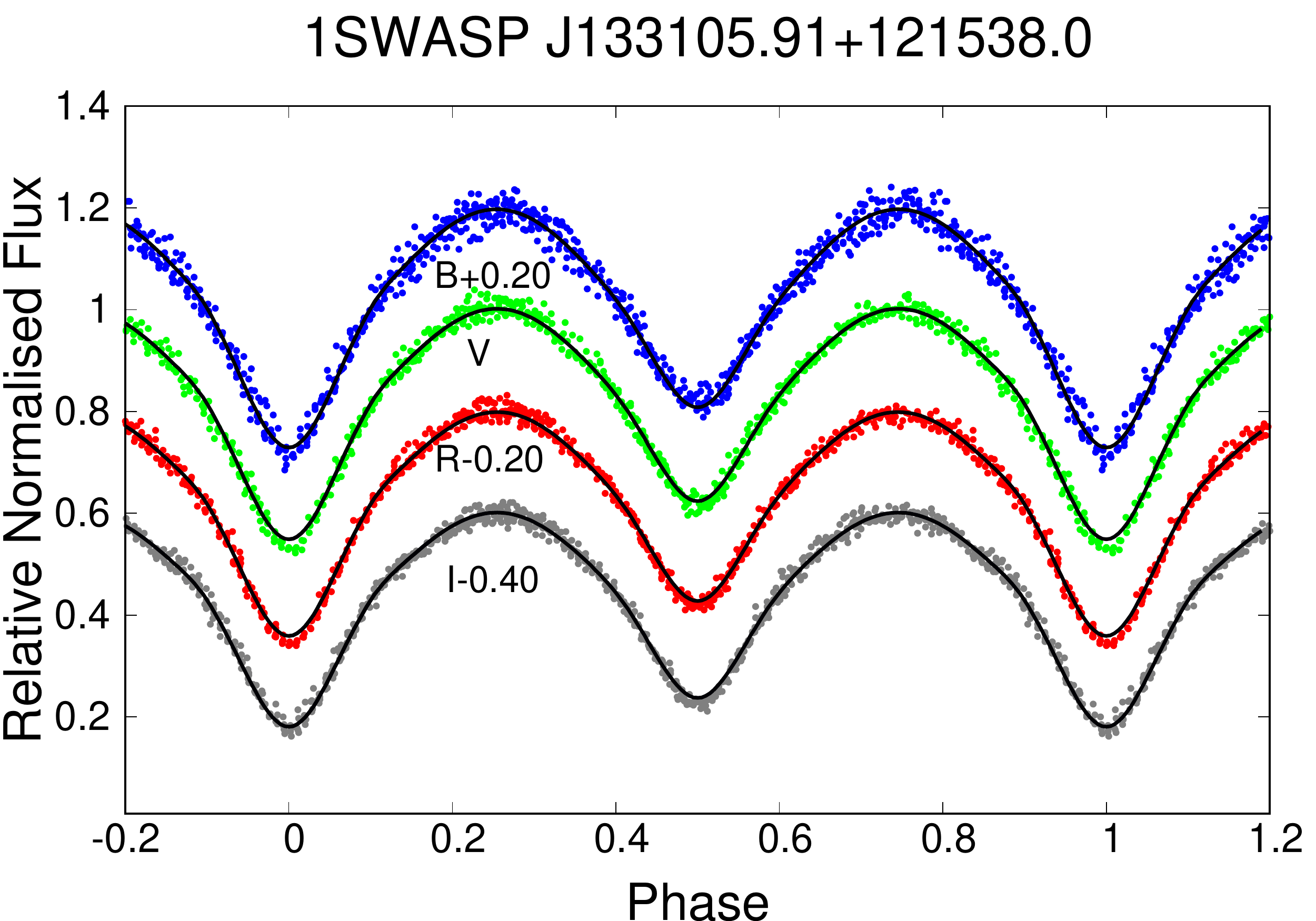}
\includegraphics[width=4.1cm,scale=1.0,angle=270]{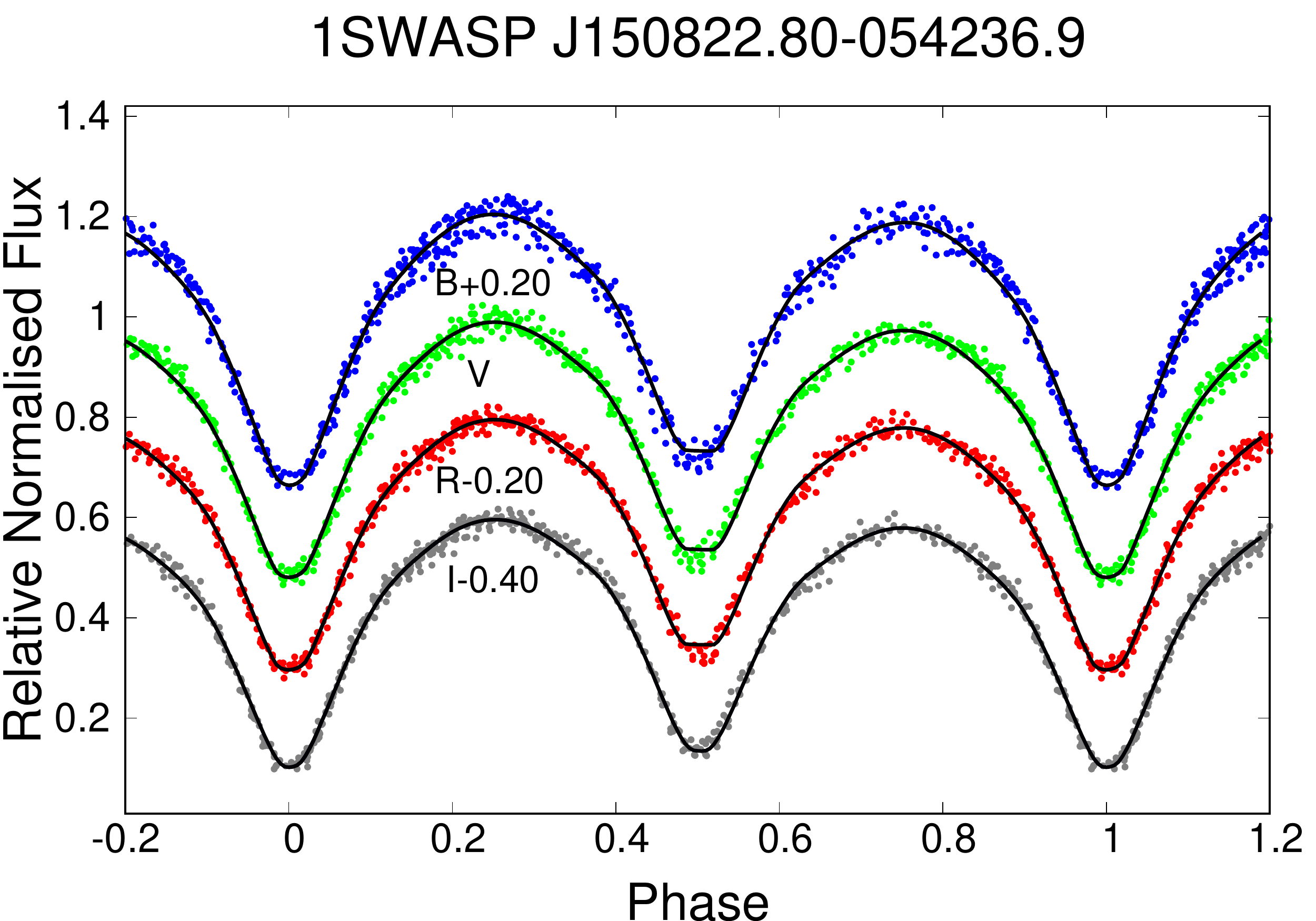}
\includegraphics[width=4.1cm,scale=1.0,angle=270]{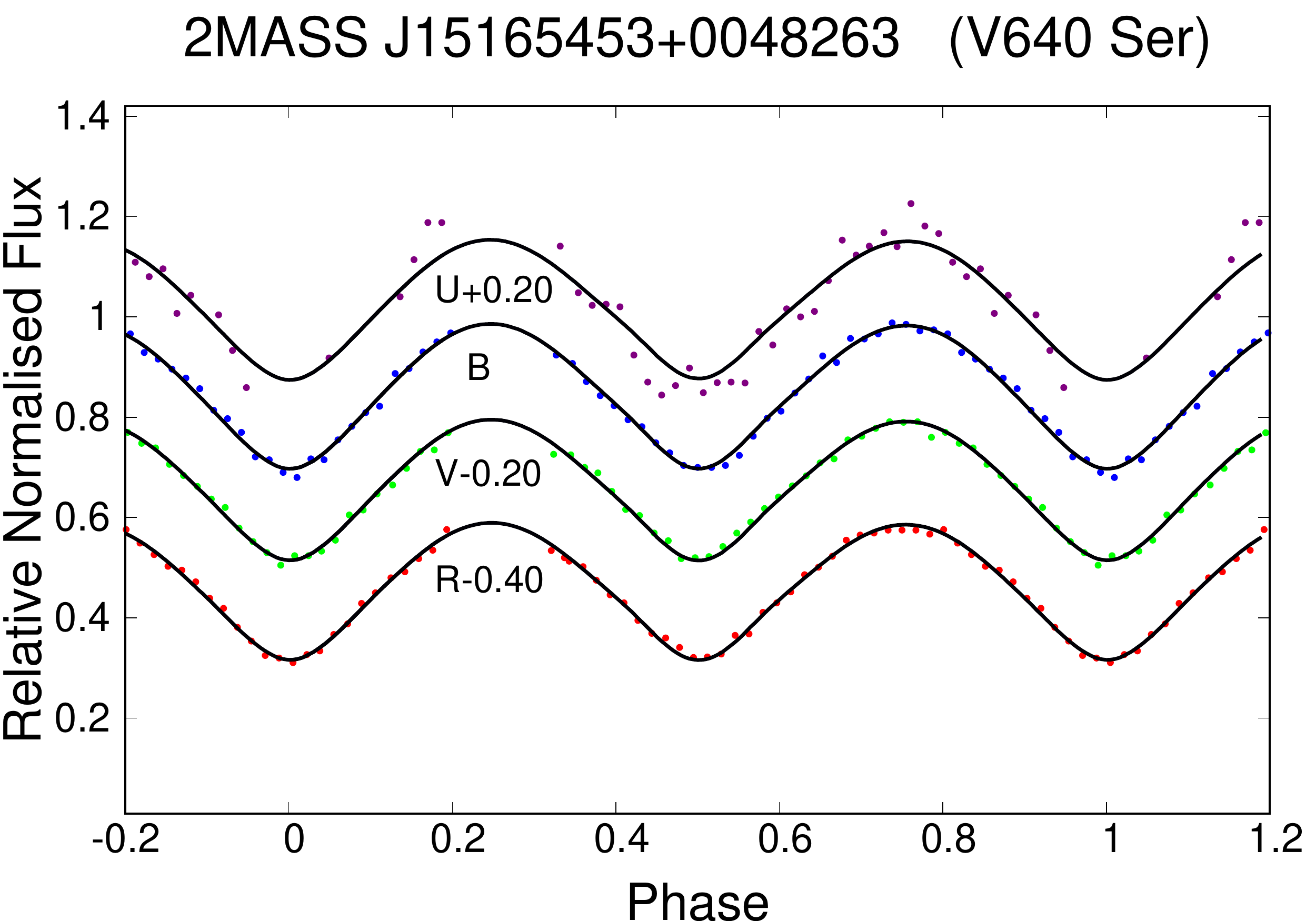}
\\
\vspace{10 pt}
\caption{Observed (points) and theoretical (lines) light curves for the 15 systems of our sample in four bands ($BVRI$ or $UBVR$ filters). The light curves are shifted vertically for clarity.}
\label{FigLC1}
\end{figure*}

\begin{figure*}
\includegraphics[width=4.1cm,scale=1.0,angle=270]{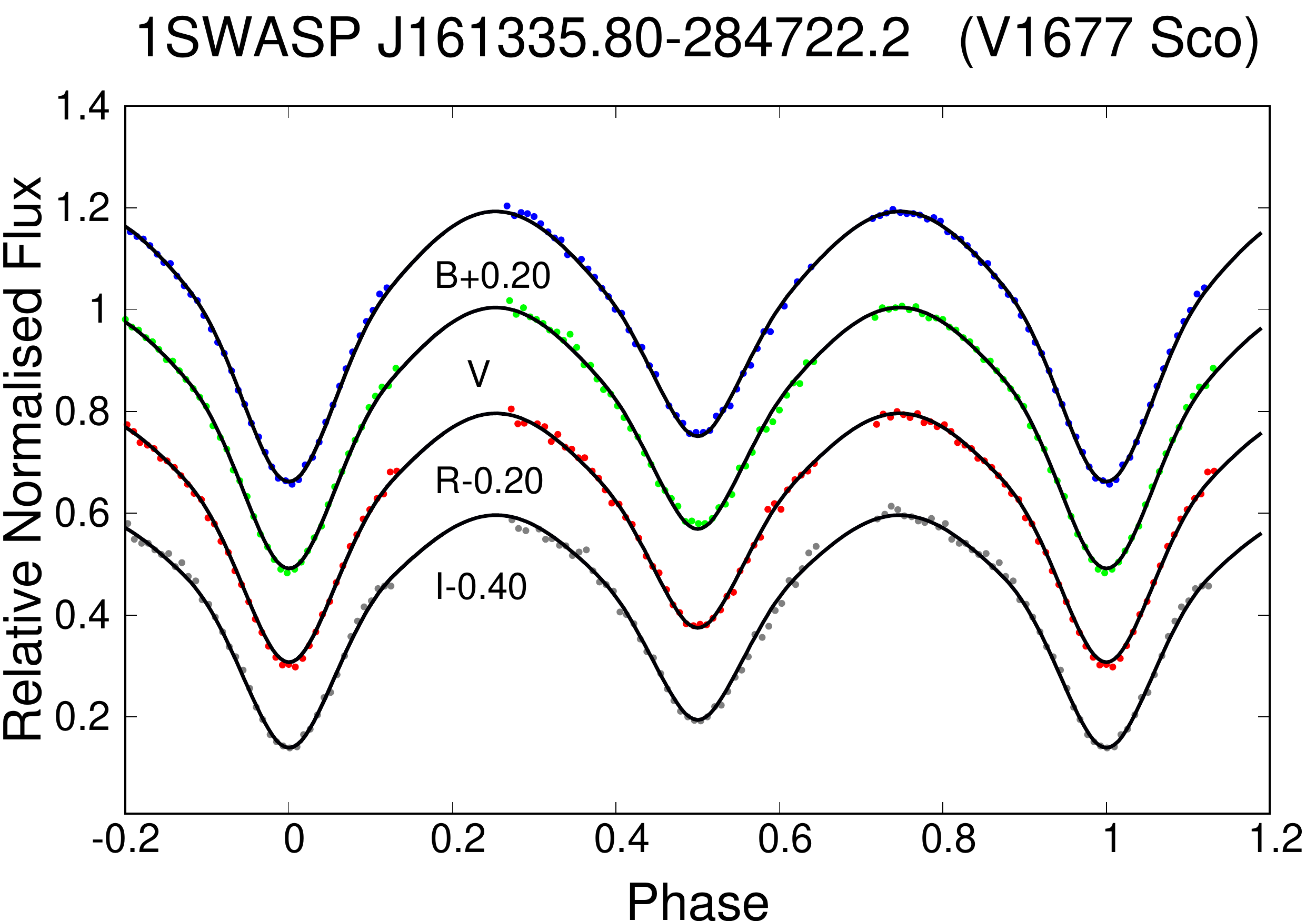}
\includegraphics[width=4.1cm,scale=1.0,angle=270]{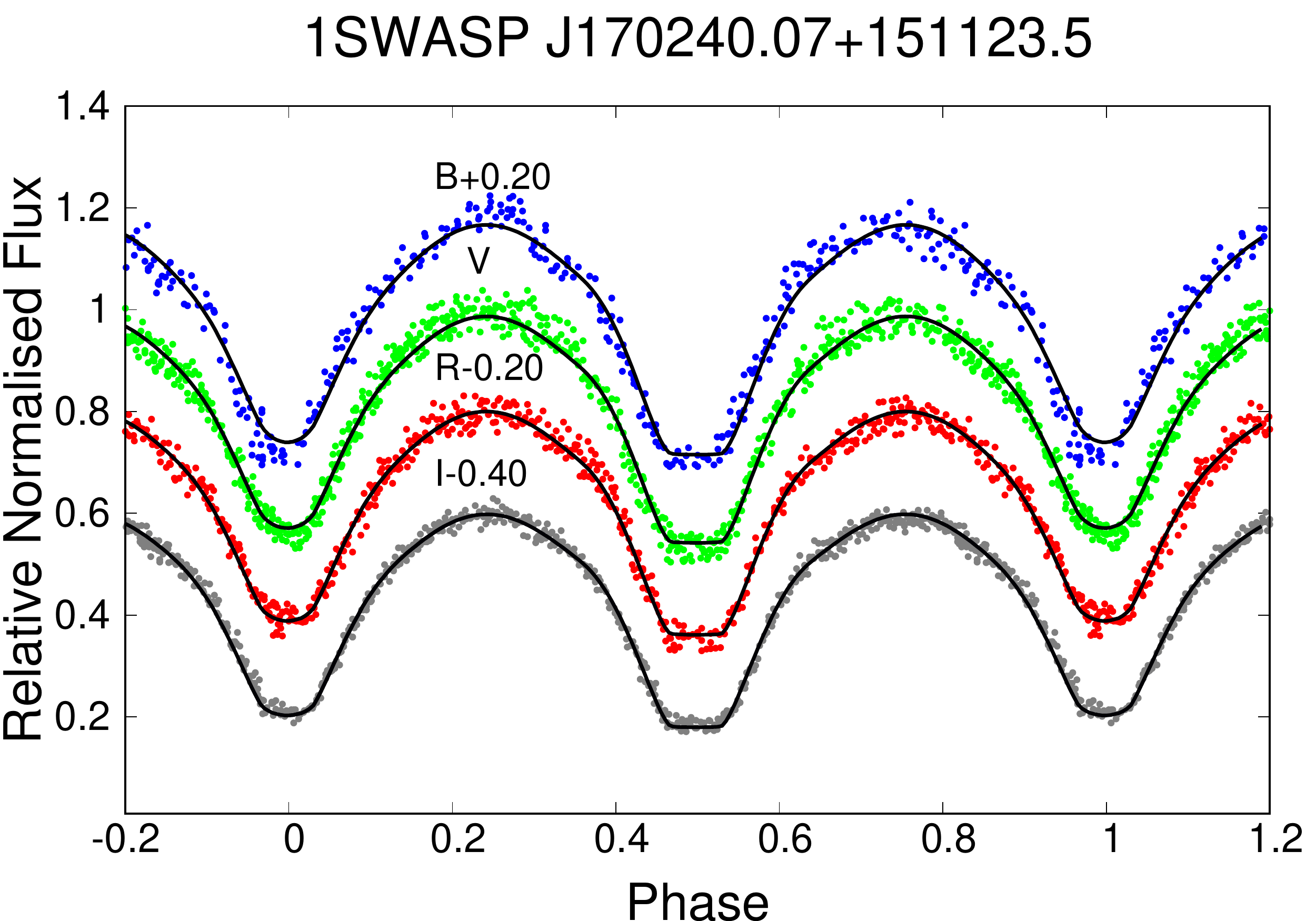}
\includegraphics[width=4.1cm,scale=1.0,angle=270]{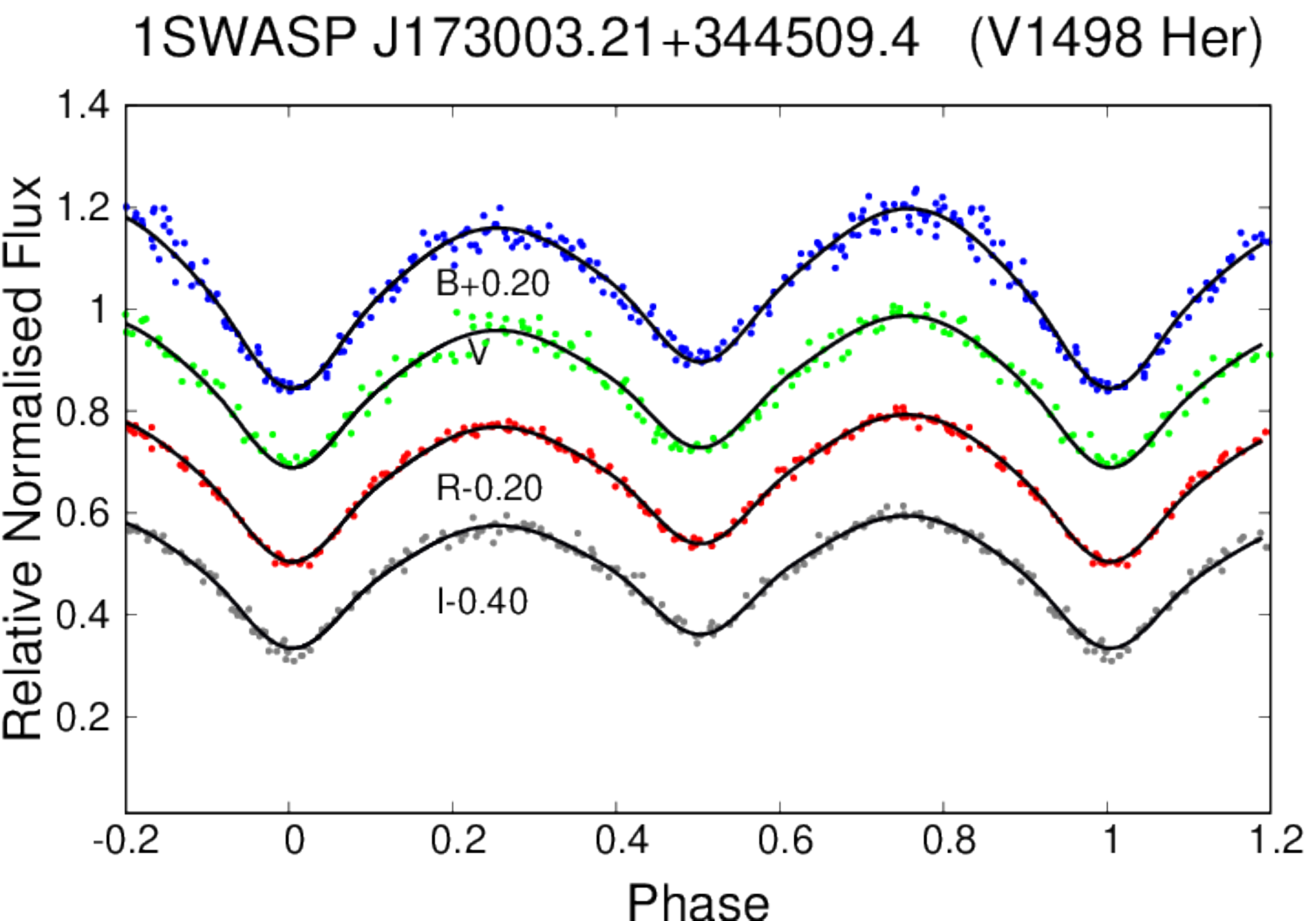}
\\
\vspace{10 pt}
\includegraphics[width=4.1cm,scale=1.0,angle=270]{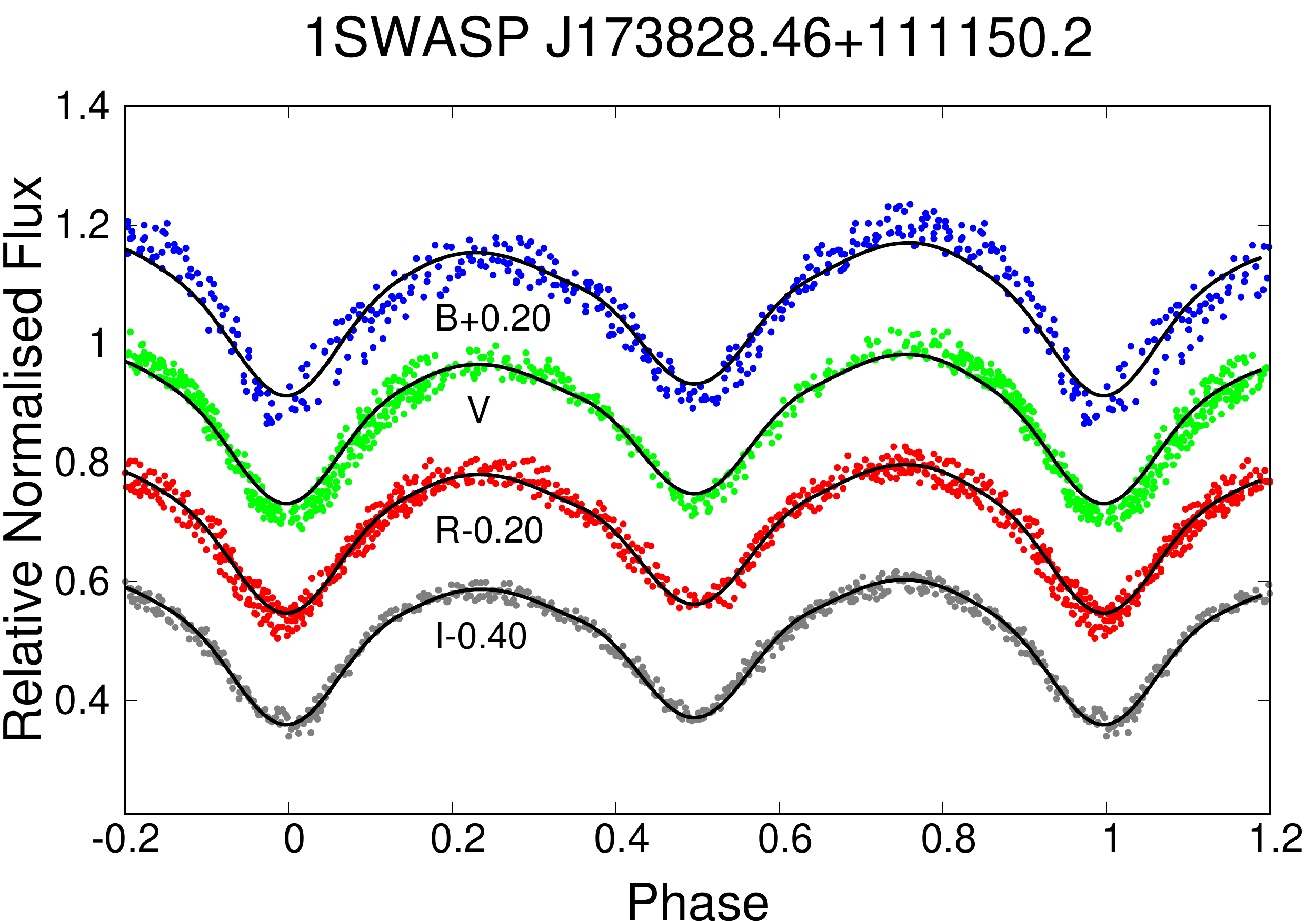}
\includegraphics[width=4.1cm,scale=1.0,angle=270]{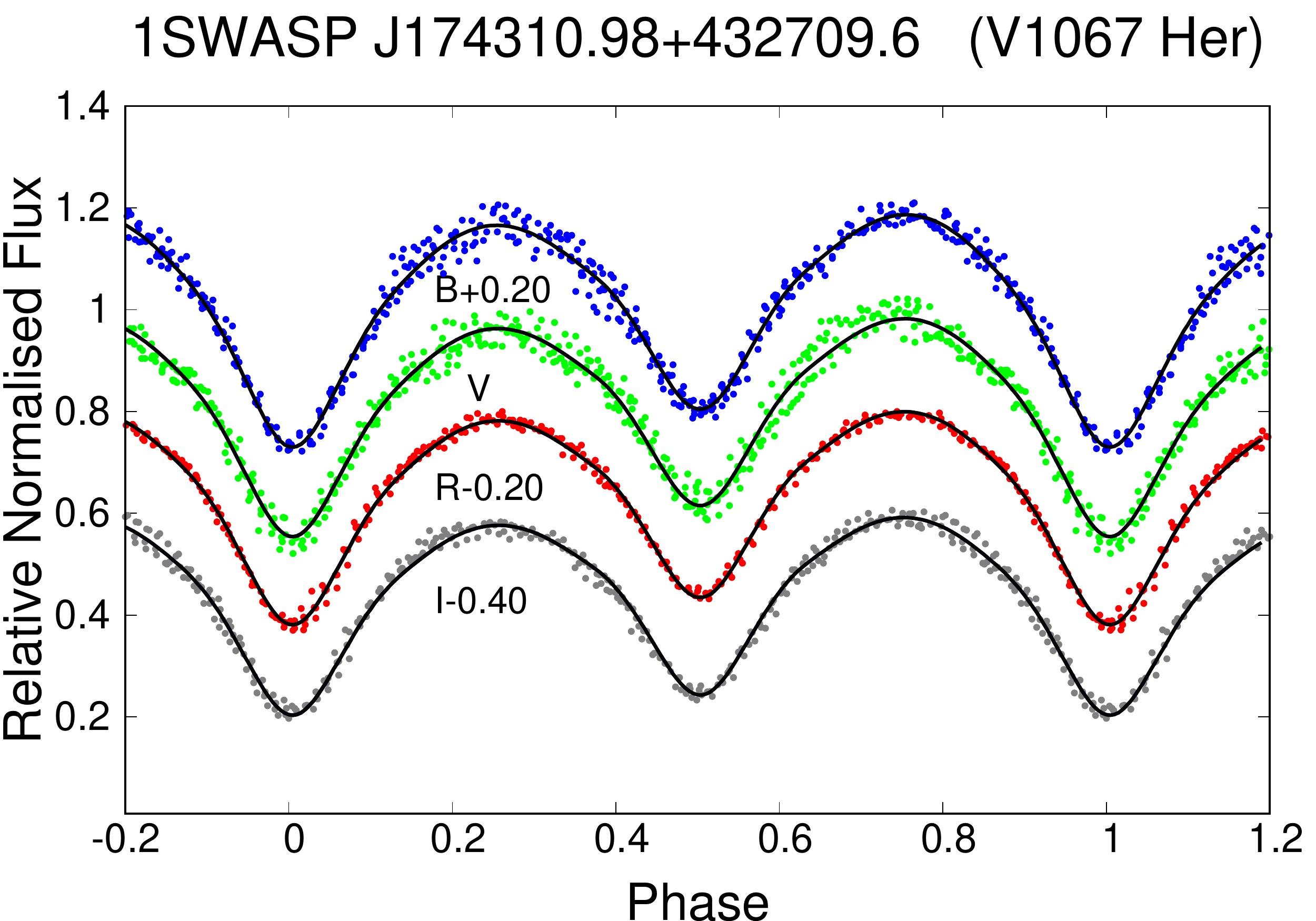}
\includegraphics[width=4.1cm,scale=1.0,angle=270]{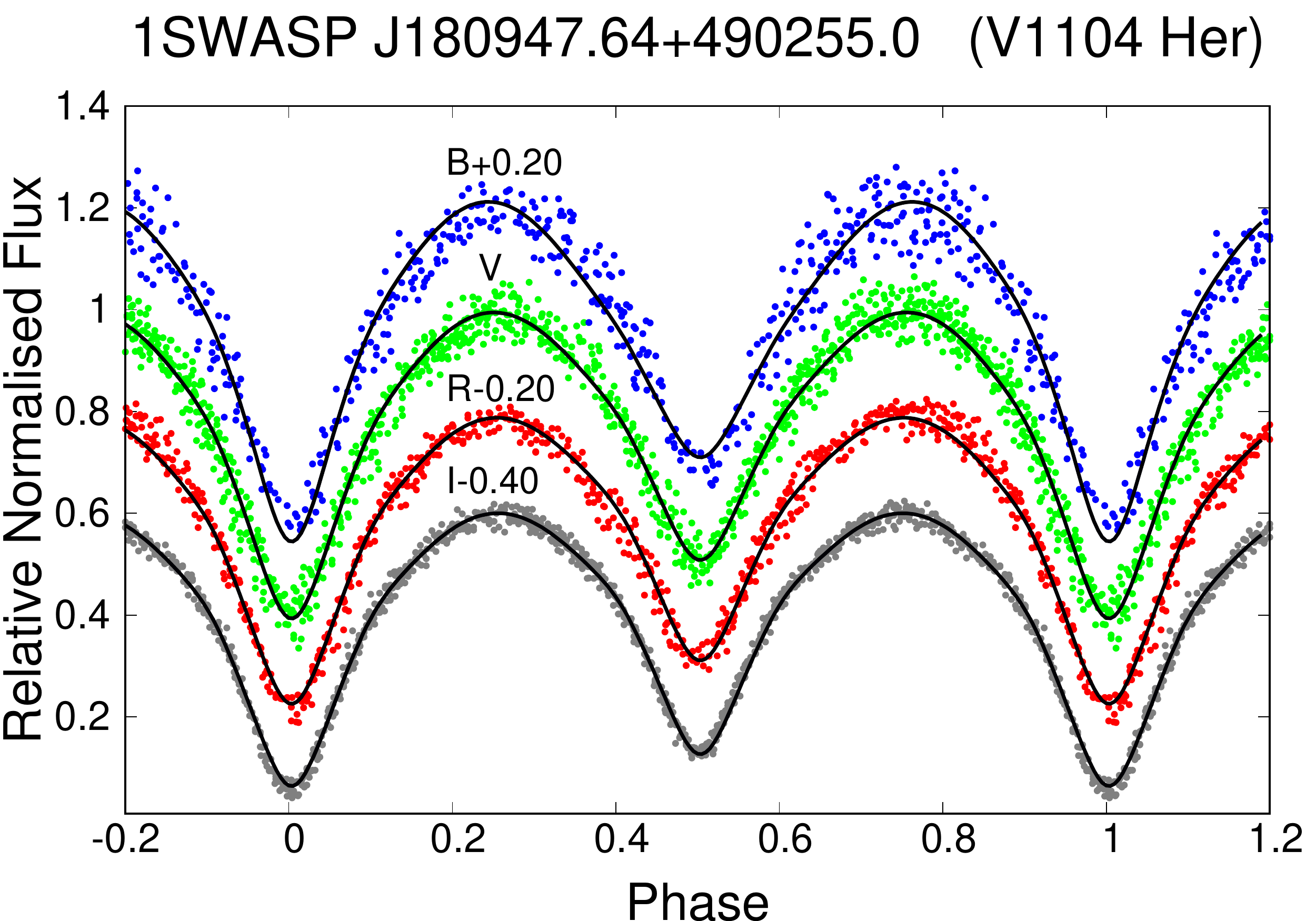}
\\
\vspace{10 pt}
\includegraphics[width=4.1cm,scale=1.0,angle=270]{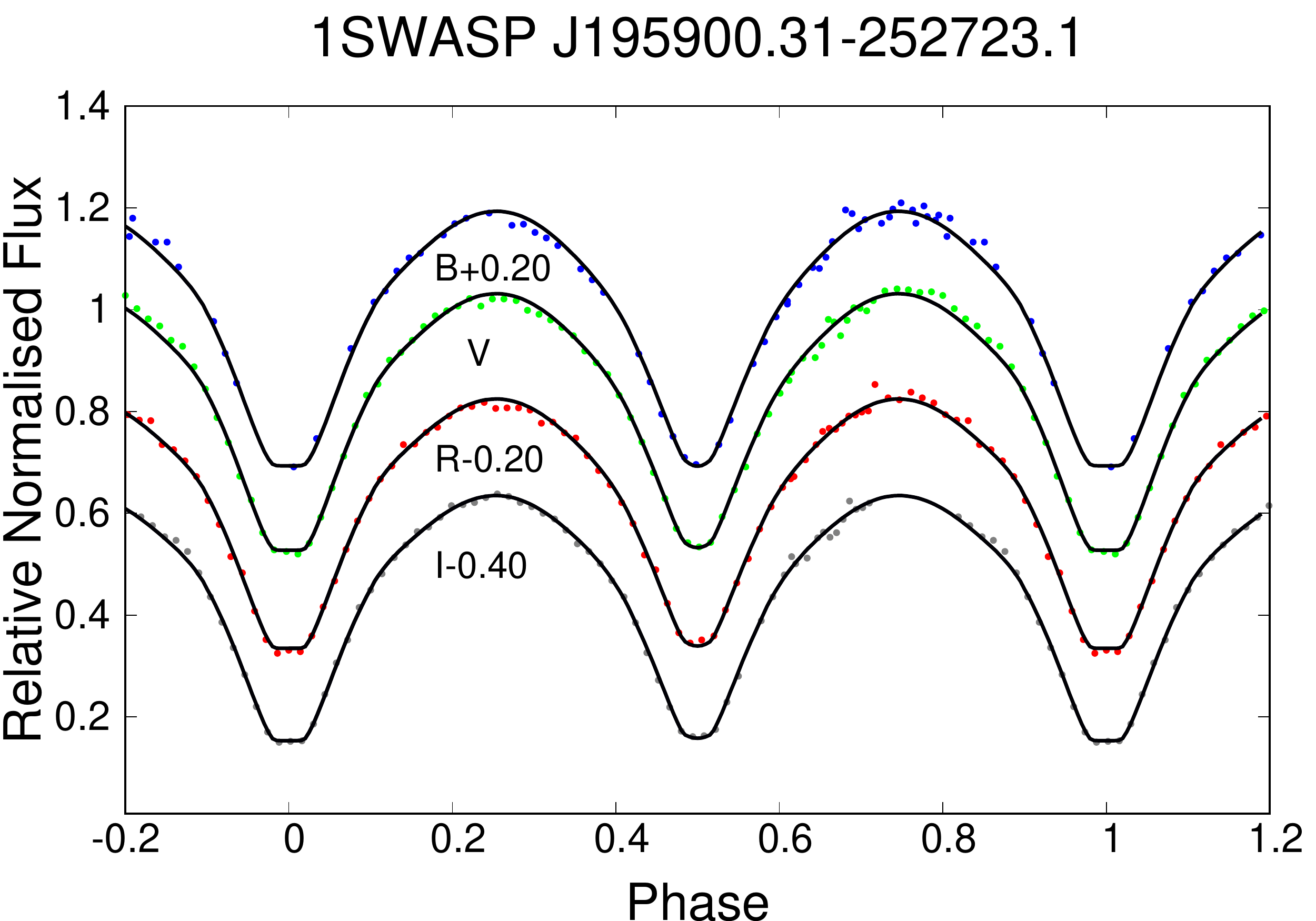}
\includegraphics[width=4.1cm,scale=1.0,angle=270]{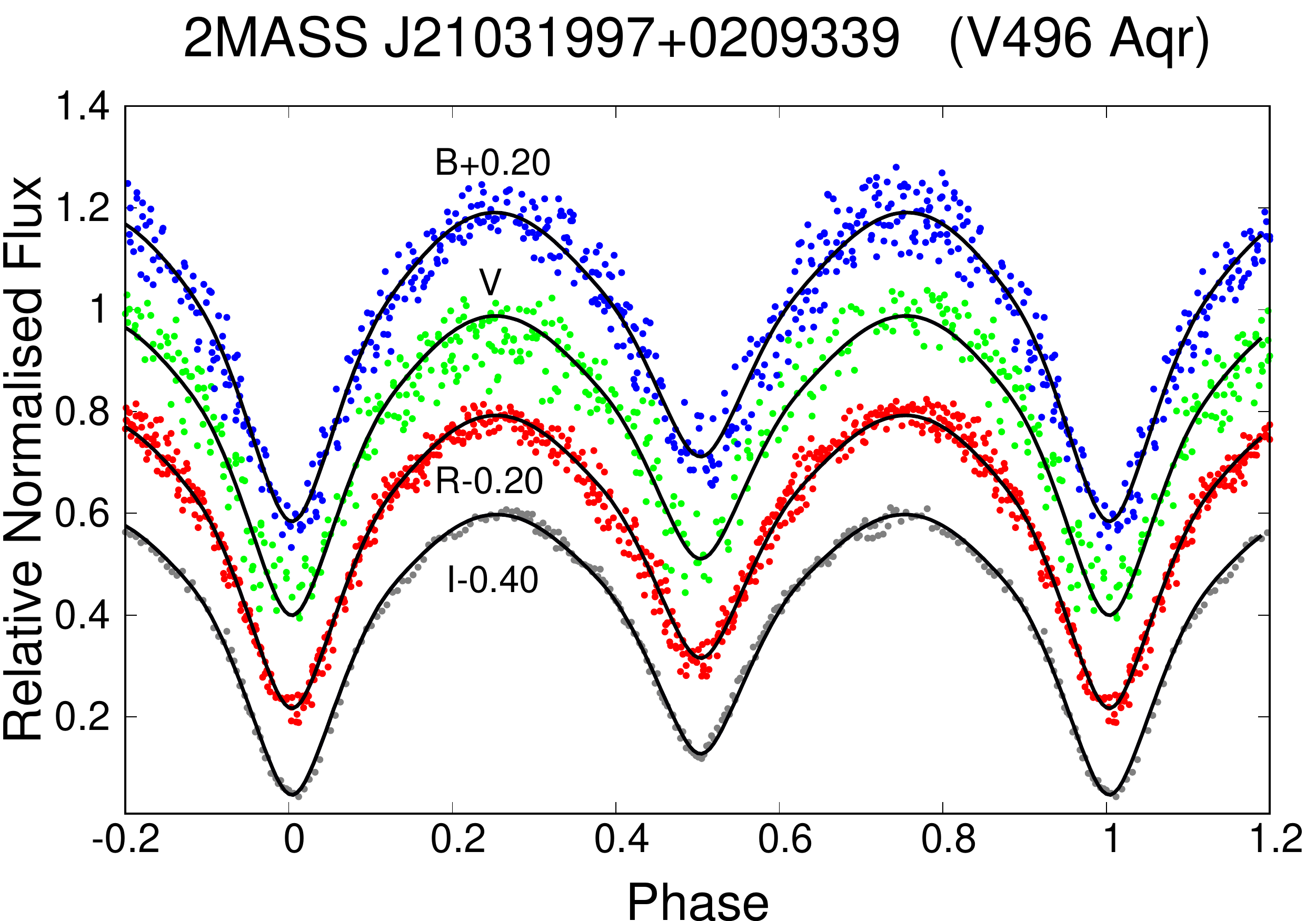}
\includegraphics[width=4.1cm,scale=1.0,angle=270]{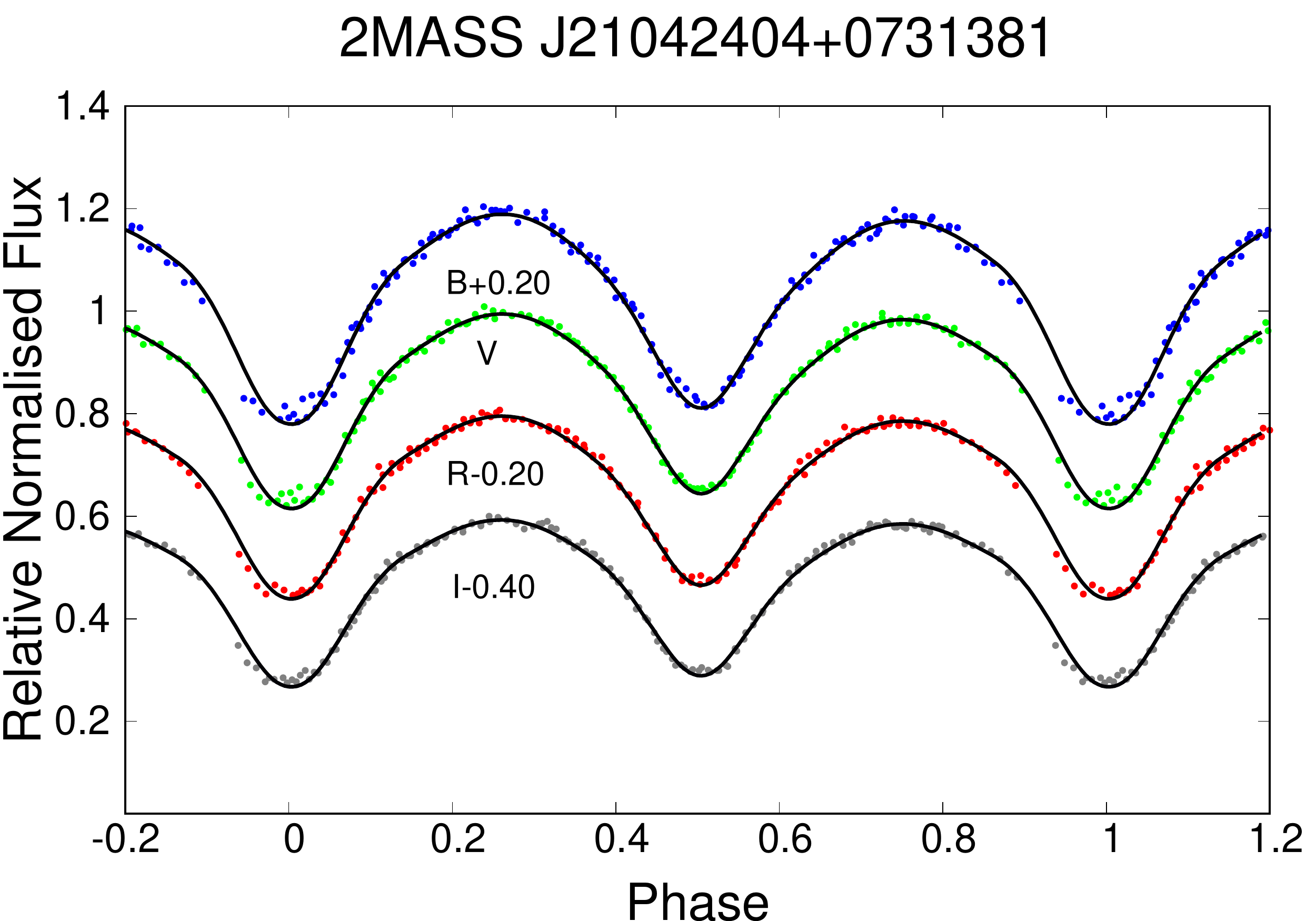}
\\
\vspace{10 pt}
\includegraphics[width=4.1cm,scale=1.0,angle=270]{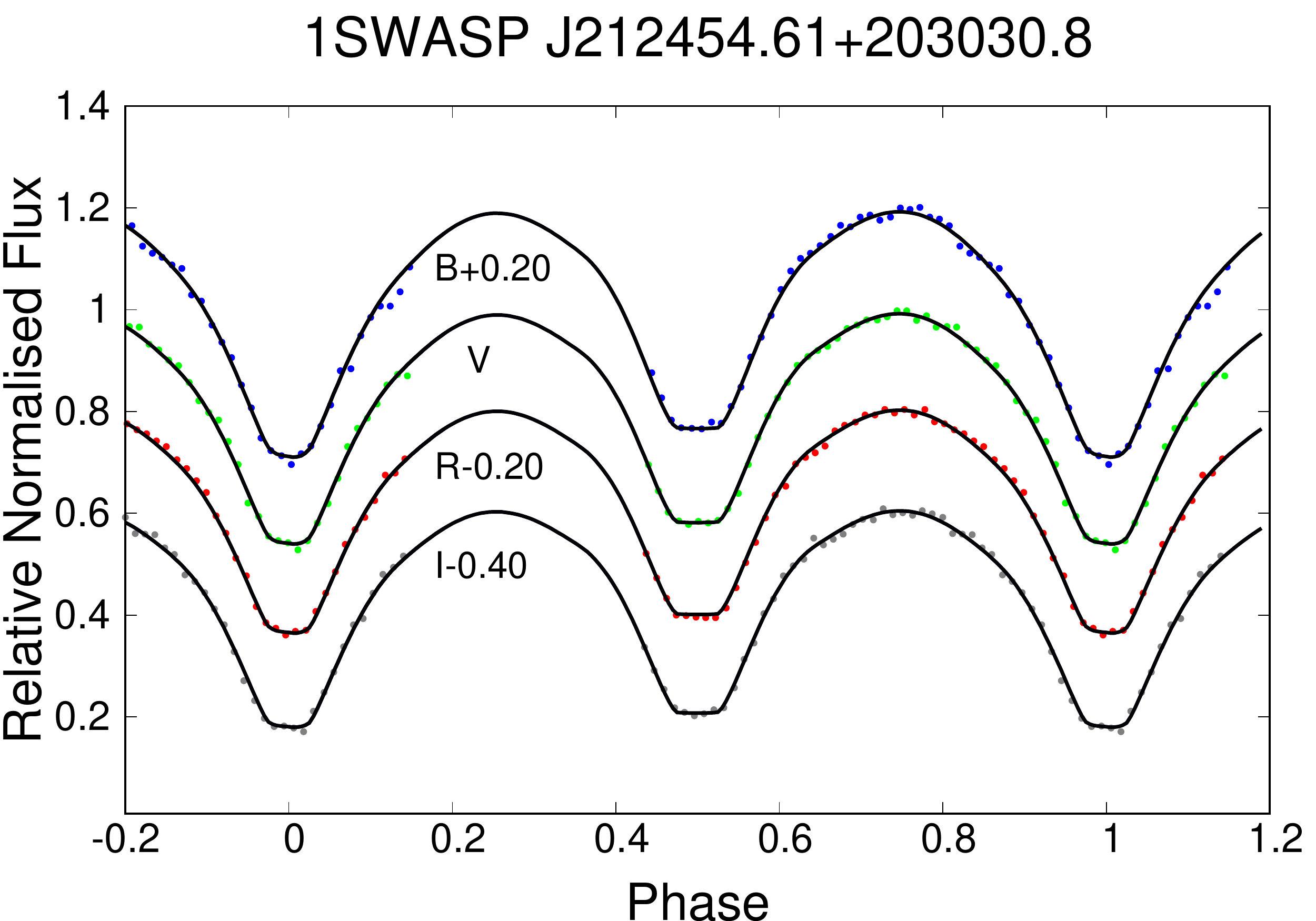}
\includegraphics[width=4.1cm,scale=1.0,angle=270]{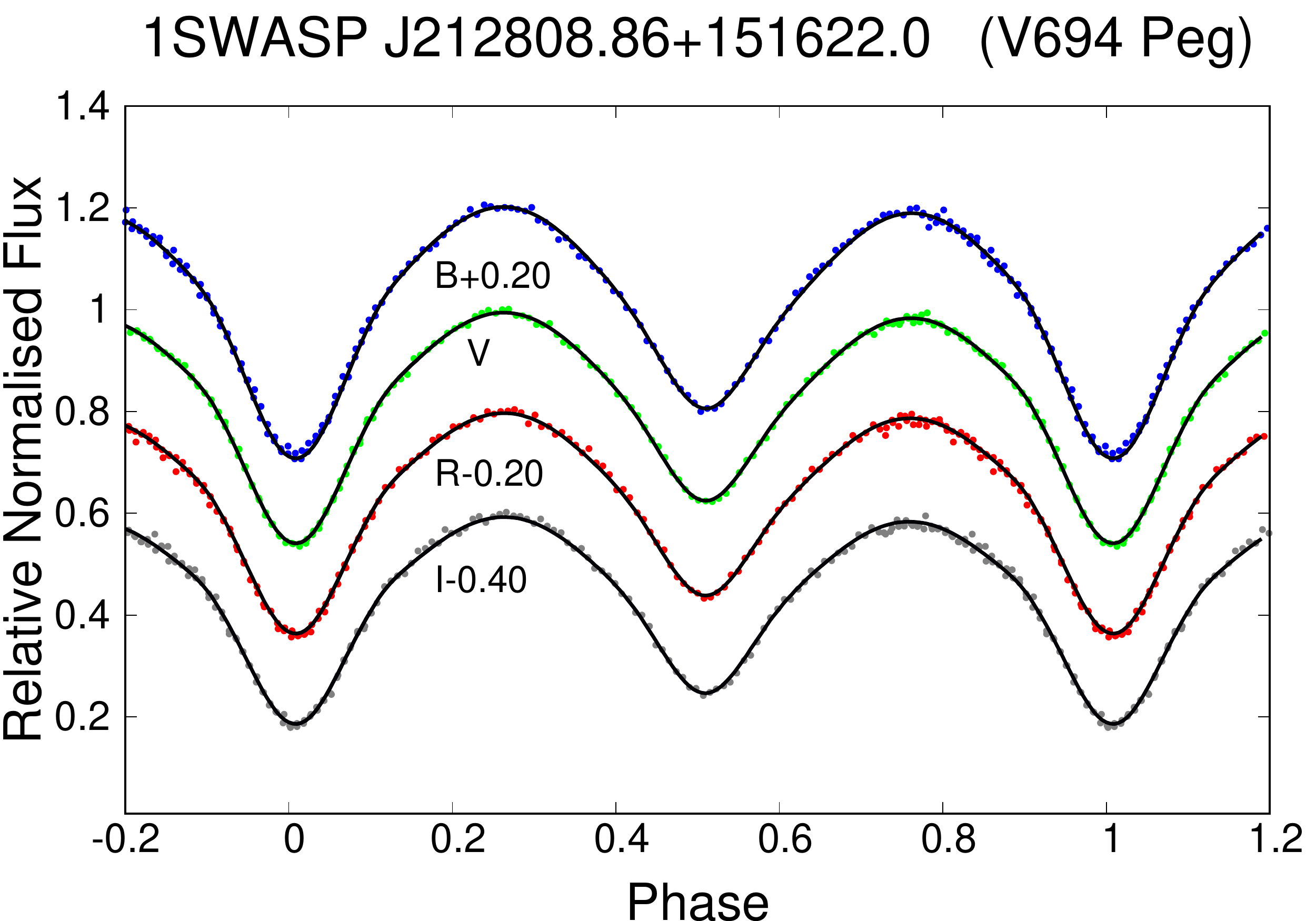}
\includegraphics[width=4.1cm,scale=1.0,angle=270]{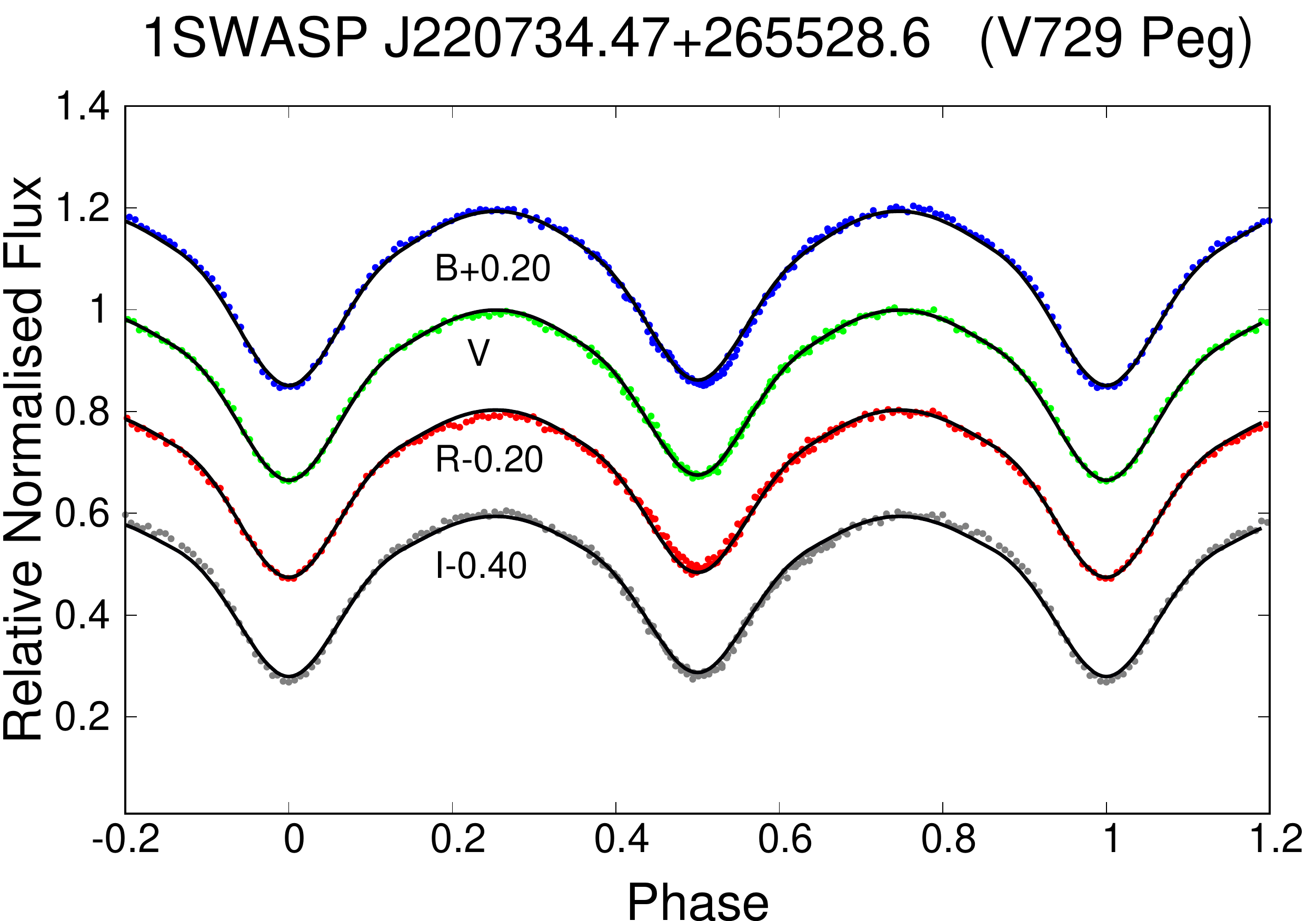}
\\
\vspace{10 pt}
\includegraphics[width=4.1cm,scale=1.0,angle=270]{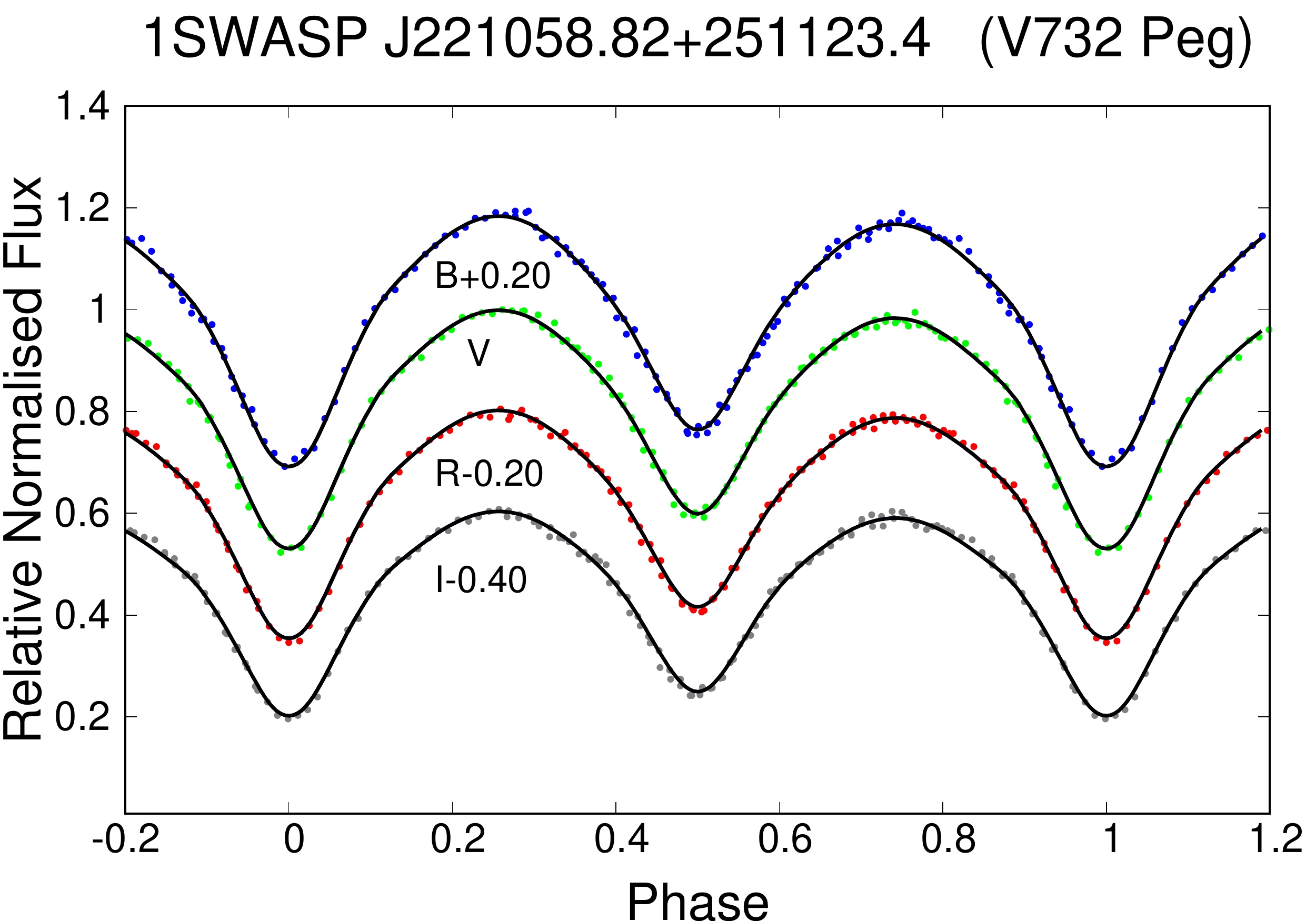}
\includegraphics[width=4.1cm,scale=1.0,angle=270]{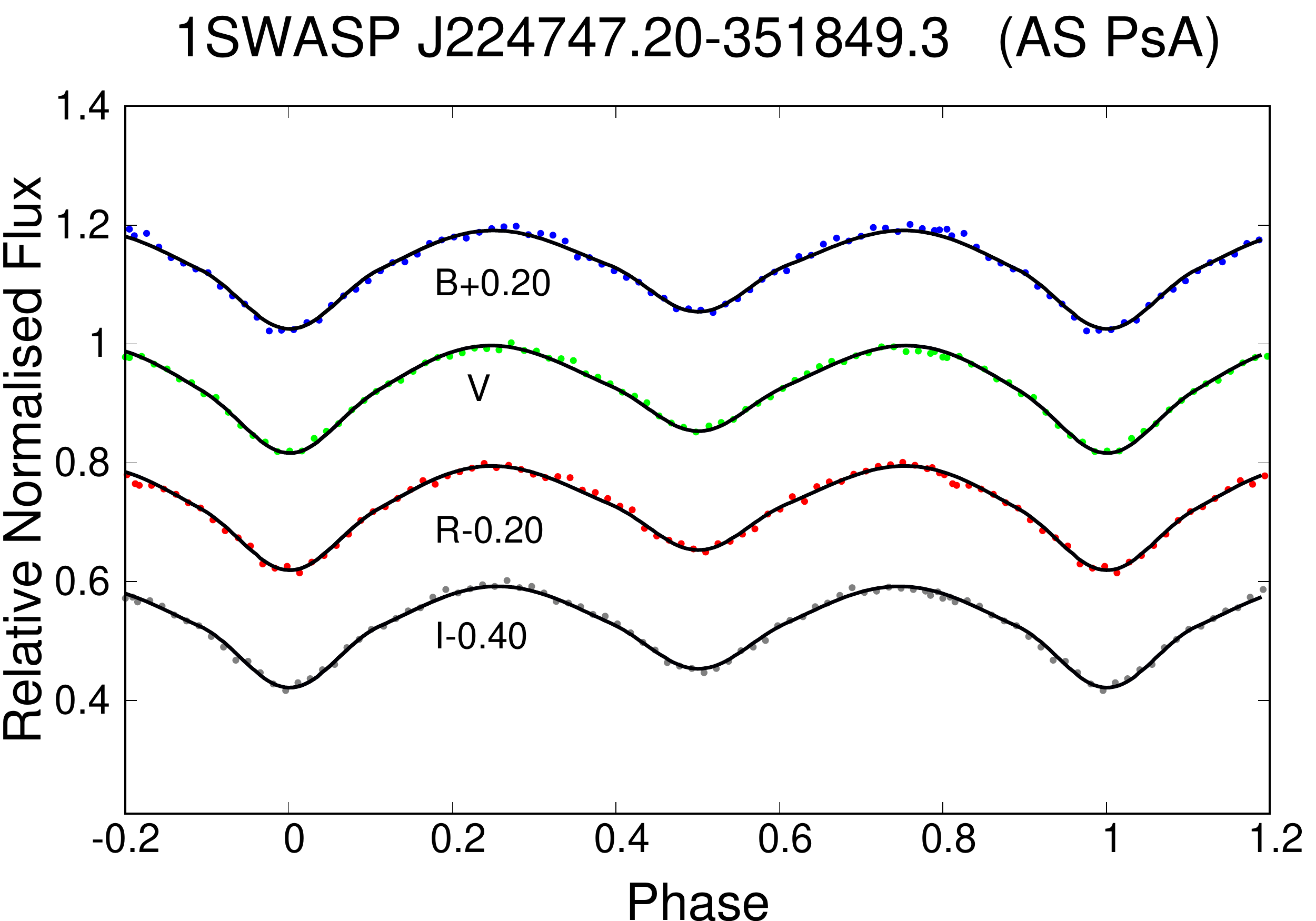}
\includegraphics[width=4.1cm,scale=1.0,angle=270]{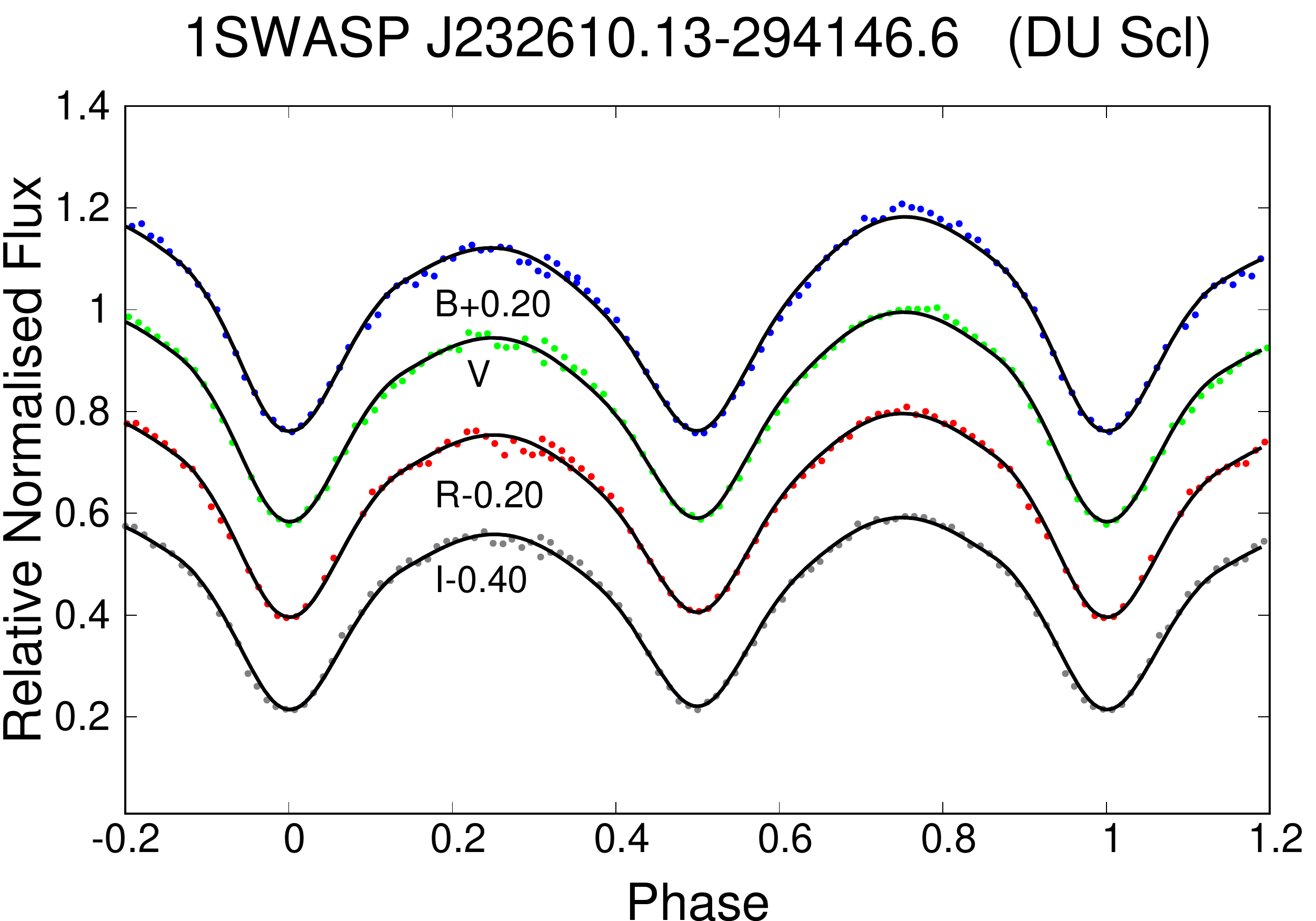}
\\
\vspace{10 pt}
\caption{The same as Fig. \ref{FigLC1} for the rest 15 systems in our sample.}
\label{FigLC2}
\end{figure*}

The resulting models provide the physical and geometrical parameters of the systems, as given in Tables \ref{Tab4.res1} - \ref{Tab9.res6}, together with their $2\sigma$ uncertainties. These models are shown together with the observed light curves in Figs. \ref{FigLC1} and \ref{FigLC2}.
In a few cases, the third light derived from the light curve modeling was rather low, close to $\sim$1\%, while no third body was found from the $O-C$ analysis (as described in detail in Section 5) and vice versa. This is acceptable within error range, especially in the case that we consider possible changes in the $O-C$ diagrams, when adding new data in the future.

\section{Absolute Physical Parameters}
Physical and geometrical parameters were estimated for both binary components using results from the best fits to the observed light curves, as described in the previous section. 
Physical parameters of contact binaries are highly constrained by their Roche geometry and therefore they follow certain empirical relations \citep{Maceroni1982, Hilditch1988,Gazeas2006,Gazeas2008,Michel2019}.
However, in the absence of radial velocity measurements for the majority of the targets, the mass of the primary components in each system had to be estimated empirically. For ultra-short contact binaries it is a serious obstacle for obtaining precise radial velocity measurements, due to the fast rotation of the components, which results in highly broadened and blended spectral lines.

Reasonable estimates for the mass and radius of the primary component can be derived using data for single MS stars, estimated from their temperature or color index \citep{Harmanec1988,Torres2010,Pecaut2013}. 
However, stellar mass is usually underestimated when models of single MS stars are used in calculations. 
When the binary configuration is imposed in a model, the mass is slightly different. 
By taking this argument into consideration, we used the empirical relations proposed by \cite{Gazeas2008,Gazeas2009} (Eq. \ref{eqgazeas1} and Eq. \ref{eqgazeas2} respectively), in order to determine the mass of the primary components and calculate an average value from these estimations. 
In these equations the mass is calculated in solar units, while orbital period in days, respectively.

\begin{equation}
\centering
\label{eqgazeas1}
    logM_{1} = 0.755(59)logP + 0.416(24)
\end{equation}
\begin{equation}
\centering
\label{eqgazeas2}
    logM_{1} = 0.725(59)logP - 0.076(32)logq + 0.365(32)
\end{equation}

\begin{table*}
\begin{center}
\caption[ ]{Absolute parameters (in solar units) and their errors of the systems studied in the present paper. Subscripts 1 refers to the larger and more massive component, while subscript 2 refers to the smaller and less massive one.}
\label{Tab10.Param}
\begin{tabular}{lcccccc}
\hline
System ID	&${\cal	M}_{\rm	1}$ $(M_{\odot})$		&	${\cal	M}_{\rm	2}$	$(M_{\odot})$&	$R_{\rm	1}$	$(R_{\odot})$	&	$R_{\rm	2}$	$(R_{\odot})$	&	$L_{\rm	1}$	$(L_{\odot})$	&	$L_{\rm 2}$ $(L_{\odot})$	\\
\hline																										
1SWASP J030749.87-365201.7	&$	0.838	\pm	0.007	$&$	0.465	\pm	0.055	$&$	0.782	\pm	0.029	$&$	0.570	\pm	0.021	$&$	0.267	\pm	0.050	$&$	0.128	\pm	0.024	$	\\
1SWASP J040615.79-425002.3	&$	0.821	\pm	0.009	$&$	0.536	\pm	0.034	$&$	0.735	\pm	0.015	$&$	0.576	\pm	0.012	$&$	0.279	\pm	0.047	$&$	0.128	\pm	0.023	$	\\
1SWASP J044132.96+440613.7	&$	0.837	\pm	0.010	$&$	0.545	\pm	0.022	$&$	0.739	\pm	0.009	$&$	0.667	\pm	0.009	$&$	0.401	\pm	0.061	$&$	0.260	\pm	0.042	$	\\
1SWASP J050904.45-074144.4	&$	0.863	\pm	0.003	$&$	0.281	\pm	0.006	$&$	0.813	\pm	0.005	$&$	0.461	\pm	0.004	$&$	0.371	\pm	0.059	$&$	0.118	\pm	0.019	$	\\
1SWASP J052926.88+461147.5	&$	0.850	\pm	0.002	$&$	0.327	\pm	0.003	$&$	0.797	\pm	0.003	$&$	0.490	\pm	0.004	$&$	0.484	\pm	0.072	$&$	0.174	\pm	0.026	$	\\
1SWASP J055416.98+442534.0	&$	0.844	\pm	0.010	$&$	0.193	\pm	0.009	$&$	0.812	\pm	0.007	$&$	0.393	\pm	0.005	$&$	0.449	\pm	0.069	$&$	0.107	\pm	0.016	$	\\
1SWASP J080150.03+471433.8	&$	0.819	\pm	0.003	$&$	0.369	\pm	0.010	$&$	0.751	\pm	0.006	$&$	0.496	\pm	0.005	$&$	0.237	\pm	0.041	$&$	0.109	\pm	0.019	$	\\
1SWASP J092328.76+435044.8	&$	0.870	\pm	0.003	$&$	0.356	\pm	0.005	$&$	0.824	\pm	0.004	$&$	0.523	\pm	0.004	$&$	0.688	\pm	0.095	$&$	0.275	\pm	0.038	$	\\
1SWASP J092754.99-391053.4	&$	0.837	\pm	0.005	$&$	0.428	\pm	0.003	$&$	0.769	\pm	0.002	$&$	0.539	\pm	0.003	$&$	0.450	\pm	0.067	$&$	0.186	\pm	0.029	$	\\
1SWASP J093010.78+533859.5	&$	0.852	\pm	0.005	$&$	0.338	\pm	0.006	$&$	0.803	\pm	0.054	$&$	0.502	\pm	0.034	$&$	0.282	\pm	0.061	$&$	0.12	\pm	0.026	$	\\
1SWASP J114929.22-423049.0	&$	0.855	\pm	0.002	$&$	0.293	\pm	0.042	$&$	0.810	\pm	0.029	$&$	0.473	\pm	0.017	$&$	0.174	\pm	0.036	$&$	0.055	\pm	0.012	$	\\
1SWASP J121906.35-240056.9	&$	0.847	\pm	0.001	$&$	0.344	\pm	0.024	$&$	0.787	\pm	0.044	$&$	0.538	\pm	0.030	$&$	0.260	\pm	0.053	$&$	0.092	\pm	0.020	$	\\
1SWASP J133105.91+121538.0	&$	0.816	\pm	0.005	$&$	0.424	\pm	0.043	$&$	0.754	\pm	0.023	$&$	0.533	\pm	0.016	$&$	0.358	\pm	0.060	$&$	0.140	\pm	0.025	$	\\
1SWASP J150822.80-054236.9	&$	0.931	\pm	0.007	$&$	0.475	\pm	0.030	$&$	0.874	\pm	0.052	$&$	0.611	\pm	0.036	$&$	0.481	\pm	0.094	$&$	0.232	\pm	0.045	$	\\
2MASS J15165453+0048263	    &$	0.814	\pm	0.005	$&$	0.237	\pm	0.016	$&$	0.742	\pm	0.021	$&$	0.555	\pm	0.016	$&$	0.706	\pm	0.100	$&$	0.381	\pm	0.055	$	\\
1SWASP J161335.80-284722.2	&$	0.840	\pm	0.011	$&$	0.572	\pm	0.070	$&$	0.766	\pm	0.031	$&$	0.612	\pm	0.025	$&$	0.225	\pm	0.044	$&$	0.120	\pm	0.024	$	\\
1SWASP J170240.07+151123.5	&$	0.948	\pm	0.001	$&$	0.328	\pm	0.002	$&$	0.915	\pm	0.003	$&$	0.536	\pm	0.004	$&$	0.468	\pm	0.075	$&$	0.204	\pm	0.031	$	\\
1SWASP J173003.21+344509.4	&$	0.825	\pm	0.010	$&$	0.543	\pm	0.028	$&$	0.749	\pm	0.013	$&$	0.590	\pm	0.010	$&$	0.245	\pm	0.043	$&$	0.125	\pm	0.023	$	\\
1SWASP J173828.46+111150.2	&$	0.922	\pm	0.005	$&$	0.264	\pm	0.001	$&$	0.879	\pm	0.004	$&$	0.463	\pm	0.005	$&$	0.526	\pm	0.080	$&$	0.126	\pm	0.020	$	\\
1SWASP J174310.98+432709.6	&$	0.921	\pm	0.010	$&$	0.538	\pm	0.034	$&$	0.859	\pm	0.017	$&$	0.640	\pm	0.013	$&$	0.522	\pm	0.081	$&$	0.229	\pm	0.038	$	\\
1SWASP J180947.64+490255.0	&$	0.824	\pm	0.017	$&$	0.797	\pm	0.019	$&$	0.764	\pm	0.005	$&$	0.701	\pm	0.005	$&$	0.141	\pm	0.028	$&$	0.095	\pm	0.020	$	\\
1SWASP J195900.31-252723.1	&$	0.870	\pm	0.007	$&$	0.474	\pm	0.023	$&$	0.811	\pm	0.011	$&$	0.586	\pm	0.008	$&$	0.501	\pm	0.074	$&$	0.232	\pm	0.037	$	\\
2MASS J21031997+0209339	    &$	0.825	\pm	0.017	$&$	0.819	\pm	0.019	$&$	0.743	\pm	0.004	$&$	0.706	\pm	0.004	$&$	0.186	\pm	0.034	$&$	0.136	\pm	0.026	$	\\
2MASS J21042404+0731381	    &$	0.807	\pm	0.005	$&$	0.247	\pm	0.033	$&$	0.751	\pm	0.023	$&$	0.412	\pm	0.013	$&$	0.269	\pm	0.048	$&$	0.079	\pm	0.014	$	\\
1SWASP J212454.61+203030.8	&$	0.849	\pm	0.002	$&$	0.364	\pm	0.006	$&$	0.782	\pm	0.004	$&$	0.503	\pm	0.004	$&$	0.416	\pm	0.064	$&$	0.174	\pm	0.027	$	\\
1SWASP J212808.86+151622.0	&$	0.838	\pm	0.004	$&$	0.398	\pm	0.040	$&$	0.779	\pm	0.023	$&$	0.527	\pm	0.016	$&$	0.265	\pm	0.048	$&$	0.089	\pm	0.017	$	\\
1SWASP J220734.47+265528.6	&$	0.845	\pm	0.010	$&$	0.560	\pm	0.016	$&$	0.745	\pm	0.005	$&$	0.588	\pm	0.004	$&$	0.287	\pm	0.047	$&$	0.173	\pm	0.029	$	\\
1SWASP J221058.82+251123.4	&$	0.809	\pm	0.002	$&$	0.356	\pm	0.037	$&$	0.750	\pm	0.022	$&$	0.491	\pm	0.015	$&$	0.302	\pm	0.052	$&$	0.102	\pm	0.019	$	\\
1SWASP J224747.20-351849.3	&$	0.806	\pm	0.012	$&$	0.607	\pm	0.014	$&$	0.680	\pm	0.006	$&$	0.548	\pm	0.006	$&$	0.141	\pm	0.026	$&$	0.061	\pm	0.013	$	\\
1SWASP J232610.13-294146.6	&$	0.864	\pm	0.002	$&$	0.291	\pm	0.033	$&$	0.813	\pm	0.022	$&$	0.469	\pm	0.013	$&$	0.327	\pm	0.057	$&$	0.080	\pm	0.015	$	\\

\hline
\end{tabular}
\end{center}
\end{table*}

The above empirical relations are based on combined spectroscopic and photometric models, as a result of the W~UMa Programme \citep{Kreiner2003} for contact binaries with orbital period ranging from 0.22 to 0.9 days. There is a paucity of ultra-short period systems under the orbital period cut-off limit in the sample used to derive the empirical relations. However, it is very encouraging that in the cases of two systems for which independent spectroscopic measurements exist, the agreement is very good. More specifically, the systems 1SWASP~J093010.78+533859.5 and 1SWASP~J150822.80-054236.9, were observed spectroscopically by \cite{Lohr2015a} and \cite{Lohr2014} and the primary mass values retrieved from these studies were $M_1 =0.86 \pm 0.02$ and $M_1 = 1.07^{+0.12}_{-0.09}$ respectively. As seen in Table \ref{Tab10.Param} our approach concluded in similar masses of $M_1 =0.852 \pm 0.005$ and $M_1 = 0.931 \pm 0.007$ respectively.

The physical parameters of all systems studied in this work are given in Table \ref{Tab10.Param}. In this table, we use the designation "1" for the more massive component, resulting always in a mass ratio less than unity. In cases where the mass ratio was found greater than 1 (see Tables in Appendix A), the mass ratio was inverted in order to produce a uniform sample for all systems. This inversion occurred in 10 out of the 30 systems in this sample. There is no physical difference between systems with q<1 and q>1. This is a result of the photometric definition of the primary minimum, which is always set as the deeper one. Radial velocity measurements can only resolve this issue and clarify which component is the more massive one.

\begin{table*}
	\centering
	\caption{Orbital period modulation parameters, as derived from the $O-C$ diagram analysis, accompanied with the corresponding mass transfer and mass loss rates.}
	\label{Tab11.OCs}
	\begin{tabular}{lcccccccc} 
	\hline
	
System ID	&		$P_{3}$ 		&		A 		&		$M_{3~min}$		&	$a_{12}sini_{3}$ 		& $\dot{P} $		&	$dM_{T}/dt$ 	&	$dM_{L}/dt$ 	\\
 &  (yr) & (days) & ($M_{\odot}$) & (AU) & $(10^{-7}$~d~yr $^{-1})$ & $(10^{-7} M_{\odot}$yr$^{-1})$ & $(10^{-7} M_{\odot}$yr$^{-1})$ \\
																											
\hline																																
1SWASP J030749.87-365201.7	&	$	-	$	&	$	-	$	&	$	-	$	& $	-	$	&	$	-	$	&	$	-			$	&	$	-			$	\\
1SWASP J040615.79-425002.3	&	$	-	$	&	$	-	$	&	$	-	$	& $	-	$	&	$	-1.526 \pm 0.221	$	&	$	-2.85	\pm	0.73	$	&	$	-			$	\\
1SWASP J044132.96+440613.7	&	$	15.86 \pm 5.05	$	&	$	0.0104 \pm 0.0042	$	&	$	0.53 \pm 0.14	$	& $	2.039 \pm 0.817	$	&	$	-	$	&	$	-			$	&	$	-			$	\\
1SWASP J050904.45-074144.4	&	$	-	$	&	$	-	$	&	$	-	$	& $	 -	$	&	$	 0.350 \pm 0.241	$	&	$	0.21	\pm	0.15	$	&	$	-0.86	\pm	0.59	$	\\
1SWASP J052926.88+461147.5	&	$	-	$	&	$	-	$	&	$	-	$	& $	-	$	&	$	-	$	&	$	-			$	&	$	-			$	\\
1SWASP J055416.98+442534.0	&	$	8.59 \pm  0.24	$	&	$	0.0023 \pm  0.0001	$	&	$	0.18 \pm 0.01	$	& $	0.669 \pm  0.039	$	&	$	-	$	&	$	-			$	&	$	-			$	\\
1SWASP J080150.03+471433.8	&	$	9.76\pm 1.52	$	&	$	0.0109 \pm 0.0031	$	&	$	0.66 \pm 0.01	$	& $	1.989 \pm 0.561	$	&	$	0.797 \pm 0.067	$	&	$	0.82	\pm	0.08	$	&	$	-2.17	\pm	0.18	$	\\
1SWASP J092328.76+435044.8	&	$	-	$	&	$	-	$	&	$	-	$	& $	-	$	&	$	-5.329 \pm 0.579	$	&	$	-4.51	\pm	0.29	$	&	$	-			$	\\
1SWASP J092754.99-391053.4	&	$	17.62 \pm 1.12	$	&	$	0.0042 \pm 0.0007	$	&	$	0.13 \pm 0.01	$	& $	0.736 \pm 0.125	$	&	$	-	$	&	$	-			$	&	$	-			$	\\
1SWASP J093010.78+533859.5	&	$	14.50 \pm  0.01	$	&	$	0.0023 \pm  0.0001	$	&	$	0.08 \pm 0.01	$	& $	0.402 \pm  0.015	$	&	$	-	$	&	$	-			$	&	$	-			$	\\
1SWASP J114929.22-423049.0	&	$	-	$	&	$	-	$	&	$	-	$	& $	-	$	&	$	-2.562 \pm 0.217	$	&	$	-1.66	\pm	0.39	$	&	$	-			$	\\
1SWASP J121906.35-240056.9	&	$	14.90 \pm  1.74	$	&	$	0.0029 \pm  0.0018	$	&	$	0.11 \pm 0.02	$	& $	0.596 \pm  0.372	$	&	$	-	$	&	$	-			$	&	$	-			$	\\
1SWASP J133105.91+121538.0	&	$	150.18 \pm 84.99	$	&	$	0.0018 \pm 0.0001	$	&	$	0.15 \pm 0.02	$	& $	3.049 \pm 0.018	$	&	$	-	$	&	$	-			$	&	$	-			$	\\
1SWASP J150822.80-054236.9	&	$	3.86 \pm  0.04	$	&	$	0.0010 \pm  0.0001	$	&	$	0.10 \pm 0.01	$	& $	0.182 \pm  0.007	$	&	$	-	$	&	$	-			$	&	$	-			$	\\
2MASS J15165453+0048263	&	$	-	$	&	$	-	$	&	$	-	$	& $	-	$	&	$	-	$	&	$	-			$	&	$	-			$	\\
1SWASP J161335.80-284722.2	&	$	8.89 \pm 0.45	$	&	$	0.0025 \pm 0.0009	$	&	$	0.15 \pm 0.06	$	& $	0.461 \pm 0.176	$	&	$	-5.525 \pm 0.035	$	&	$	-44.6	\pm	32.4	$	&	$	-			$	\\
1SWASP J170240.07+151123.5	&	$	-	$	&	$	-	$	&	$	-	$	& $	-	$	&	$	2.408 \pm 0.444	$	&	$	1.52	\pm	0.28	$	&	$	-5.78	\pm	1.07	$	\\
1SWASP J173003.21+344509.4	&	$	9.55 \pm 0.01	$	&	$	0.0043 \pm 0.0002	$	&	$	0.27 \pm 0.01	$	& $	0.866 \pm 0.031	$	&	$	-5.848 \pm 0.001	$	&	$	-13.9	\pm	2.12	$	&	$	-			$	\\
1SWASP J173828.46+111150.2	&	$	-	$	&	$	-	$	&	$	-	$	& $	-	$	&	$	12.25 \pm 1.74	$	&	$	5.97	\pm	0.86	$	&	$	-28.7	\pm	4.07	$	\\
1SWASP J174310.98+432709.6	&	$	49.02 \pm 3.20	$	&	$	0.0092 \pm 0.0002	$	&	$	0.22 \pm 0.01	$	& $	2.099 \pm 0.038	$	&	$	-	$	&	$	-			$	&	$	-			$	\\
1SWASP J180947.64+490255.0	&	$	8.35 \pm 0.19	$	&	$	0.0012 \pm  0.0001	$	&	$	0.11 \pm 0.01	$	& $	0.324 \pm 0.016	$	&	$	-	$	&	$	-			$	&	$	-			$	\\
1SWASP J195900.31-252723.1	&	$	28.73 \pm 15.18	$	&	$	0.0081 \pm 0.0022	$	&	$	0.29 \pm 0.05	$	& $	1.568 \pm 0.416	$	&	$	-	$	&	$	-			$	&	$	-			$	\\
2MASS J21031997+0209339	&	$	10.64 \pm 0.51	$	&	$	0.0045 \pm  0.0006	$	&	$	0.25 \pm 0.01	$	& $	0.793 \pm 0.102	$	&	$	-	$	&	$	-			$	&	$	-			$	\\
2MASS J21042404+0731381	&	$	2.78 \pm 0.05	$	&	$	0.0027 \pm 0.0001	$	&	$	0.28 \pm 0.01	$	& $	0.460 \pm 0.001	$	&	$	3.291 \pm 0.001	$	&	$	1.86	\pm	0.36	$	&	$	-8.27	\pm	0.09	$	\\
1SWASP J212454.61+203030.8	&	$	8.78 \pm 0.94	$	&	$	0.0038 \pm 0.0020	$	&	$	0.28 \pm 0.01	$	& $	0.683 \pm 0.367	$	&	$	-	$	&	$	-			$	&	$	-			$	\\
1SWASP J212808.86+151622.0	&	$	13.57 \pm  1.52	$	&	$	0.0017 \pm  0.0004	$	&	$	0.06 \pm 0.01	$	& $	0.286 \pm  0.075	$	&	$	-	$	&	$	-			$	&	$	-			$	\\
1SWASP J220734.47+265528.6	&	$	-	$	&	$	-	$	&	$	-	$	& $	-	$	&	$	-	$	&	$	-			$	&	$	-			$	\\
1SWASP J221058.82+251123.4	&	$	-	$	&	$	-	$	&	$	-	$	& $	-	$	&	$	-4.098 \pm 1.875	$	&	$	-21	\pm	9.8	$	&	$	-			$	\\
1SWASP J224747.20-351849.3	&	$	-	$	&	$	-	$	&	$	-	$	& $	-	$	&	$	5.445 \pm 0.729	$	&	$	20.7	\pm	3.51	$	&	$	-17.7	\pm	2.38	$	\\
1SWASP J232610.13-294146.6	&	$	-	$	&	$	-	$	&	$	-	$	& $	-	$	&	$	-	$	&	$	-			$	&	$	-			$	\\
\hline
	\end{tabular}
\end{table*}

\section{Orbital Period Modulation}

It is very common for contact binaries to exhibit orbital period modulations. Detailed study of these modulations is a useful tool to investigate the dynamical evolution of binary systems and search for companions. The most frequently observed period modulations are either parabolic, cyclic, or both. The secular period modulations are associated either with mass transfer between the components or mass loss through stellar winds. The cyclic modulation of the orbital period can be attributed to a third body orbiting the system or to magnetic braking in the components' envelope caused by the Applegate mechanism \citep{Applegate1992}. 

A thorough study of $O-C$ diagrams is the best way to detect possible period variations and study their orbital parameters. Nonetheless, the accuracy of the parameters that could be derived from $O-C$ diagrams, depends significantly on the time span of observations. Very short time scales (of the order of a few years) could easily lead to wrong conclusions. That is why, a significant effort was made to gather as many times of minimum light as possible (following the procedure mentioned in section 3.1) for each target leading to a more robust conclusion.

In order to construct the $O-C$ diagrams, the linear ephemeris is used (equation \ref{Eq1}) with the updated values of orbital period from Table \ref{Tab2.ObsLog}. When the linear ephemeris could not describe long-term trends in the $O-C$ diagrams, equations \ref{Eq2} and \ref{Eq3} were used to account for parabolic trends and cyclic variations, respectively. 
This procedure was carried out by using the \textit{LITE} software \citep{Zasche2009} that calculates period modulations, while taking into consideration the statistical weight of each time of minimum. The coefficients that appear in equations \ref{Eq2} and \ref{Eq3} are described as follows: $b$ is the quadratic coefficient linked to the orbital period change rate, $a_{12}$ is the projected semi-major axis, $i_3$ the orbital inclination of the tertiary component with respect to the system's orbital plane, c the speed of light, $e_3$ the eccentricity of the orbit of the tertiary component around the centre of mass, $\omega_3$ the longitude of periastron and $\nu$ is the true anomaly around the centre of mass of the triple system.

\begin{equation}
    \centering
        (O-C)_1 = T - (T_{0} + P \times E )
    \label{Eq1}
\end{equation}
\begin{equation}
    \centering
      (O-C)_2 = (O-C)_1 - b \times E^2
    \label{Eq2}
\end{equation}
\begin{equation}
    \centering
      (O-C)_3 = (O-C)_2 - \frac{a_{12}\sin i_3}{c} \left[\frac{1-e_{3}^2}{1+e_{3}\cos\nu}\sin(\nu+\omega_{3})+e_{3}\sin\omega_{3}\right]
    \label{Eq3}
\end{equation}

\begin{figure*}[H]
\includegraphics[width=7.1cm,scale=1.0,angle=0]{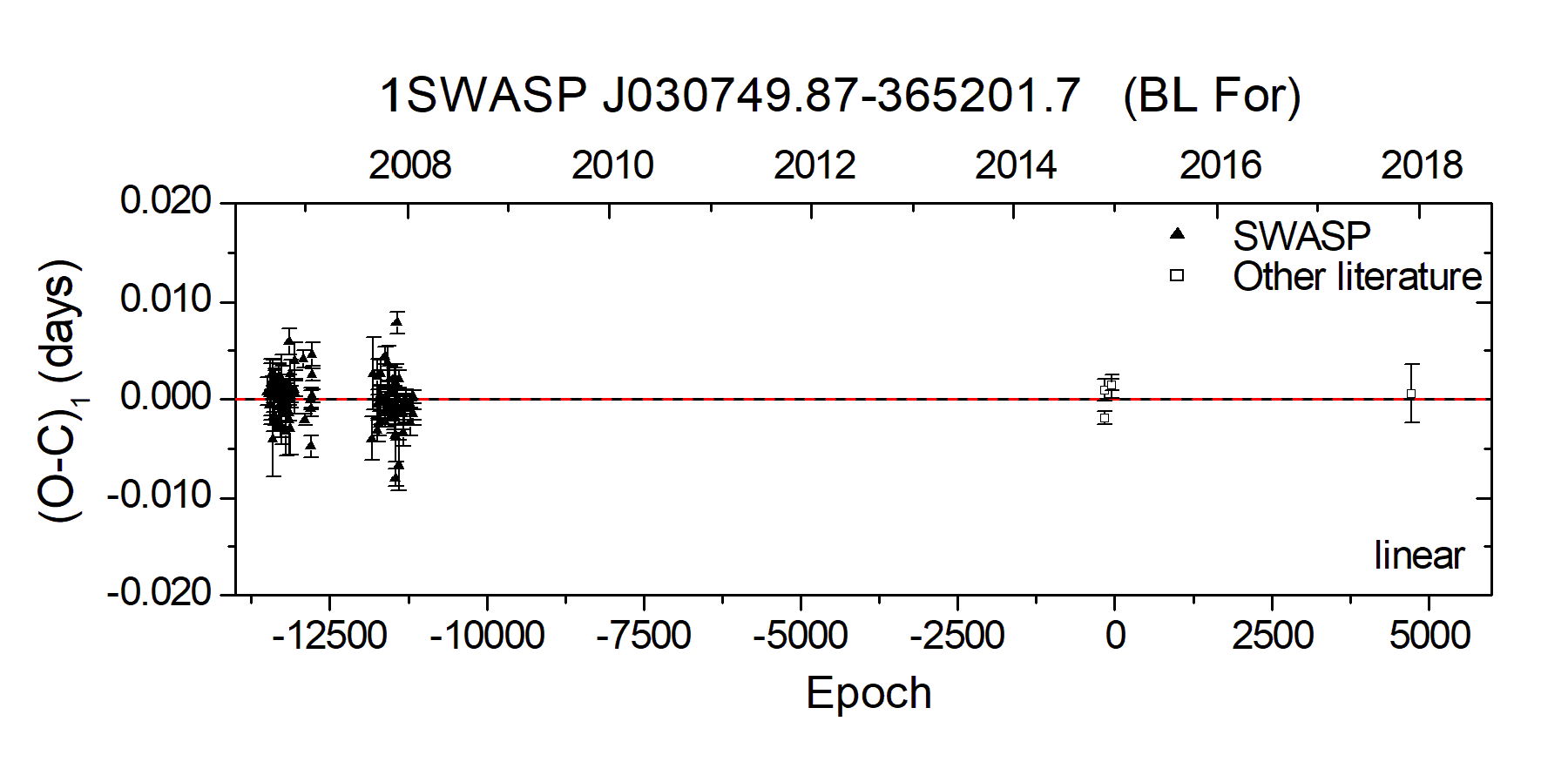}
\includegraphics[width=7.1cm,scale=1.0,angle=0]{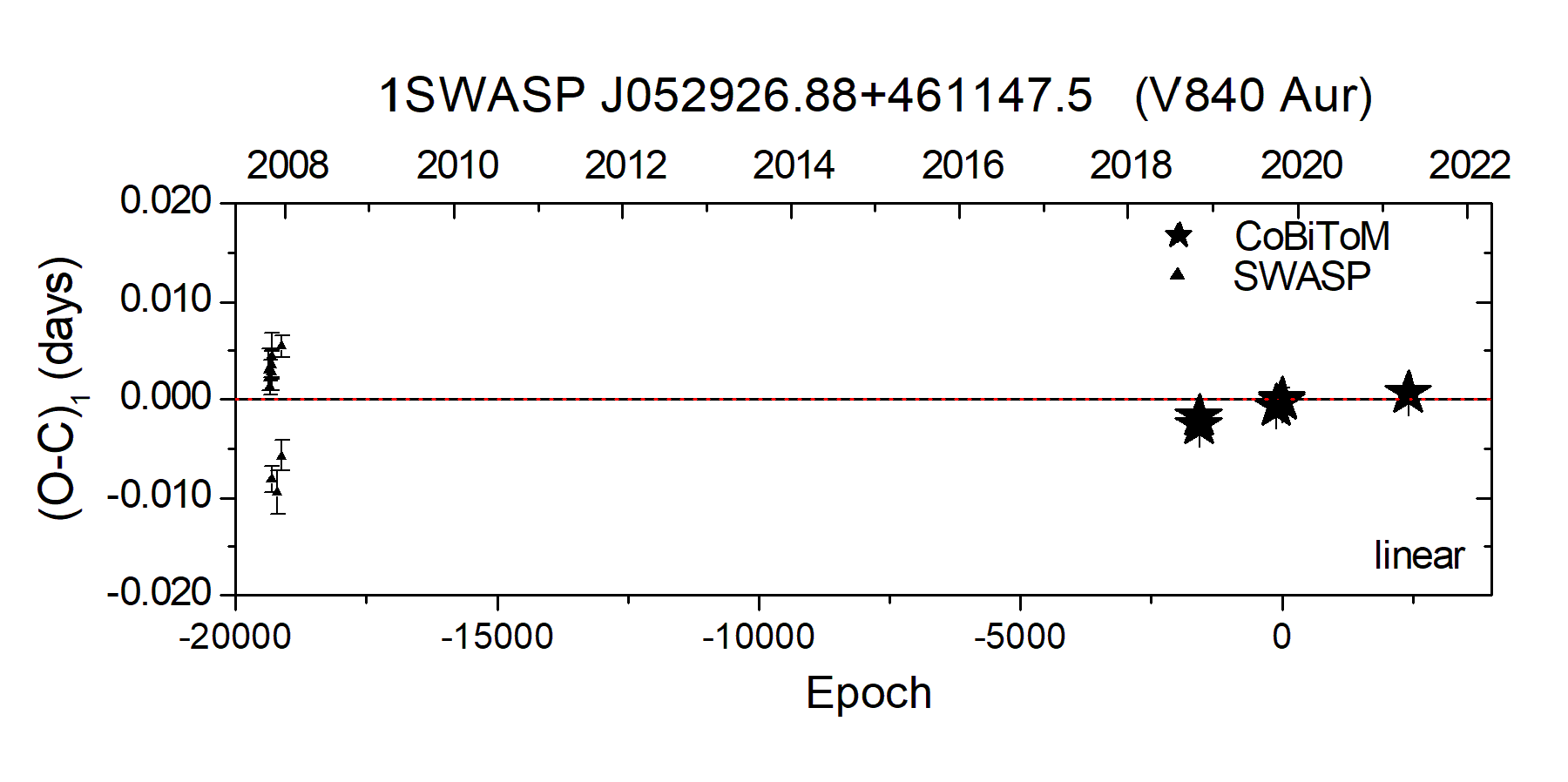}
\\
\includegraphics[width=7.1cm,scale=1.0,angle=0]{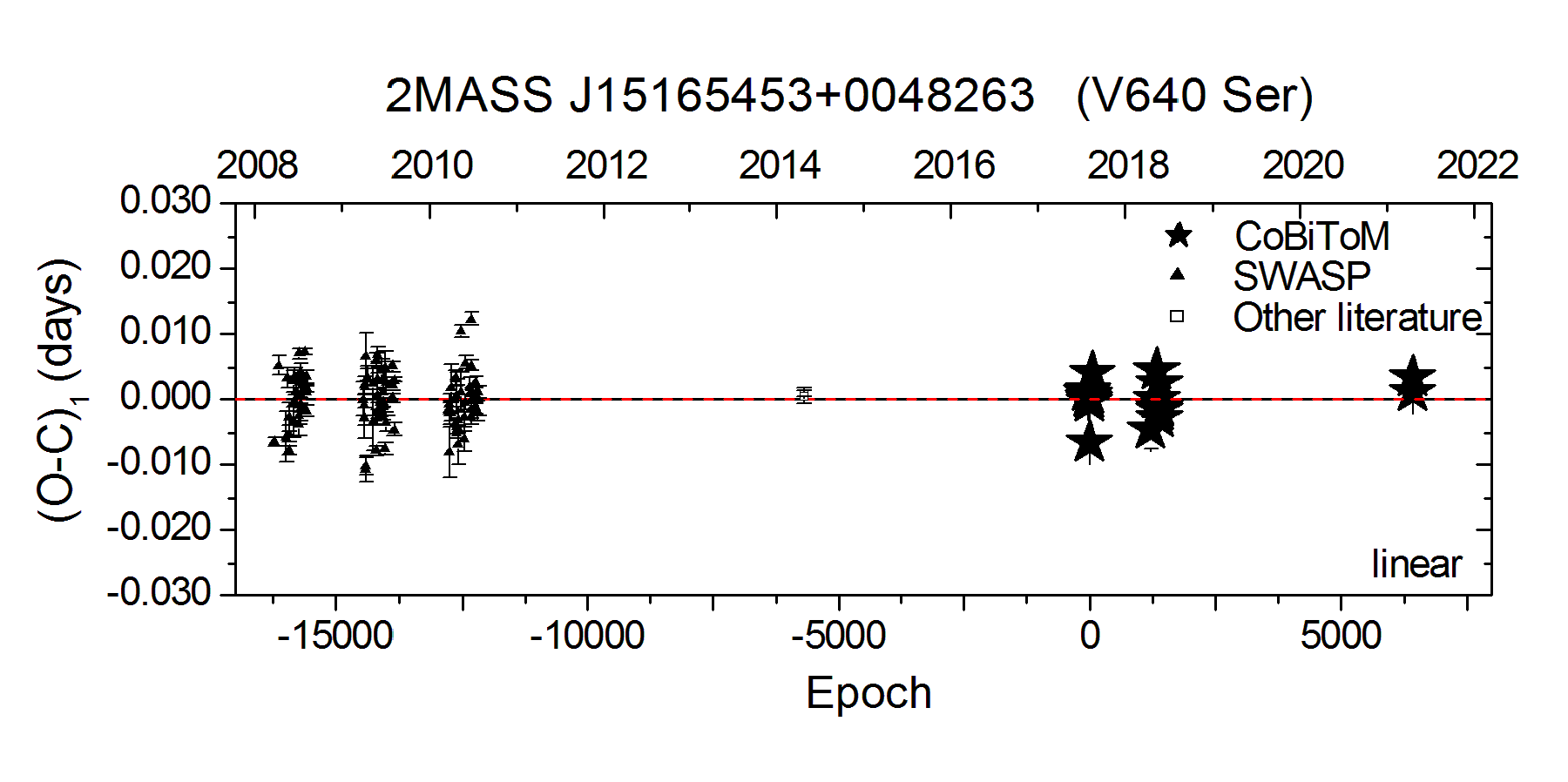}
\includegraphics[width=7.1cm,scale=1.0,angle=0]{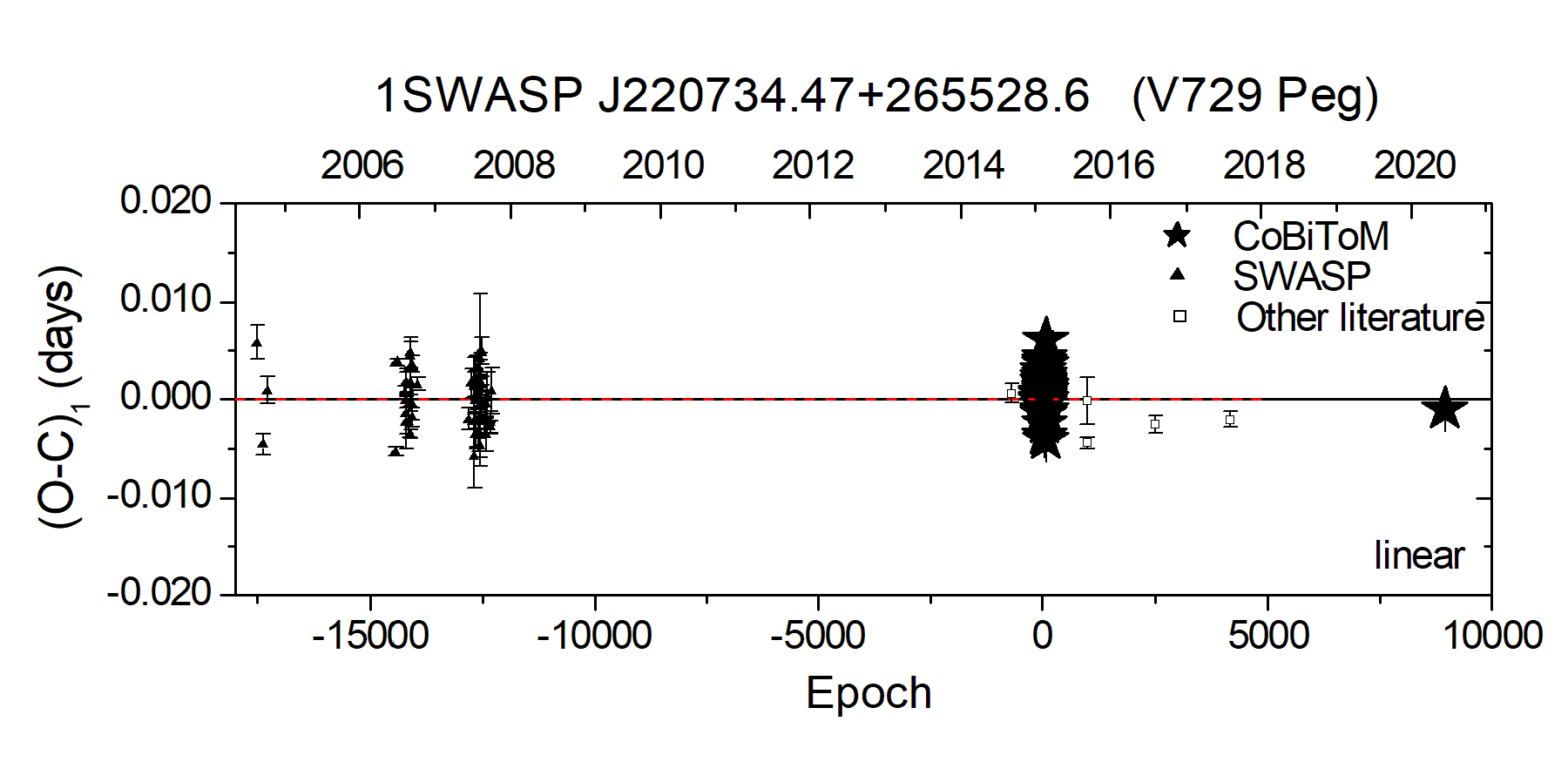}
\\
\includegraphics[width=7.1cm,scale=1.0,angle=0]{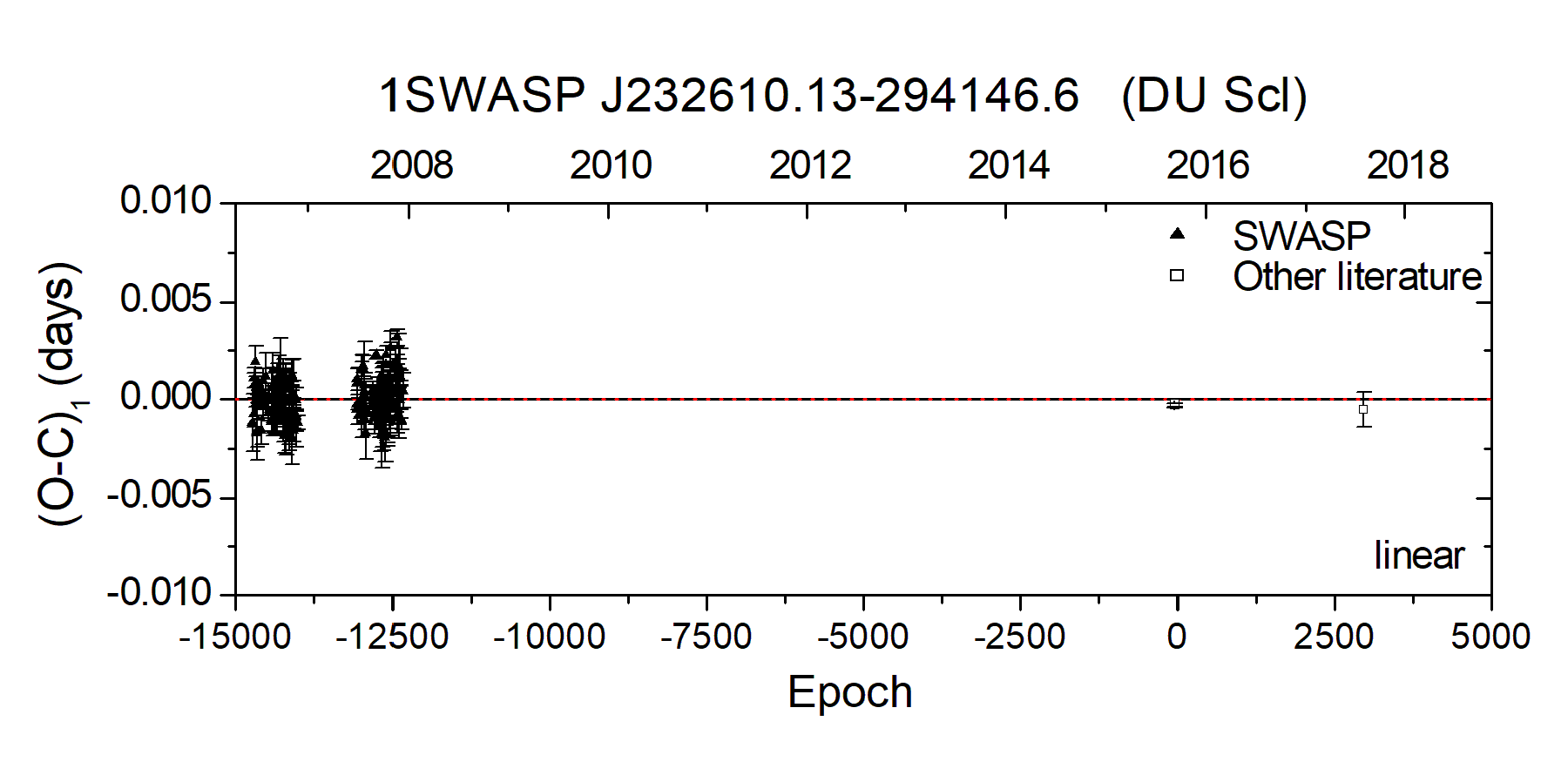}

\caption{The $O-C$ diagrams for the five systems of the sample that present circular orbits with no additional effects.}

\label{FigOC1}
\end{figure*}


\begin{figure*}[H]
\includegraphics[width=7.1cm,scale=1.0,angle=0]{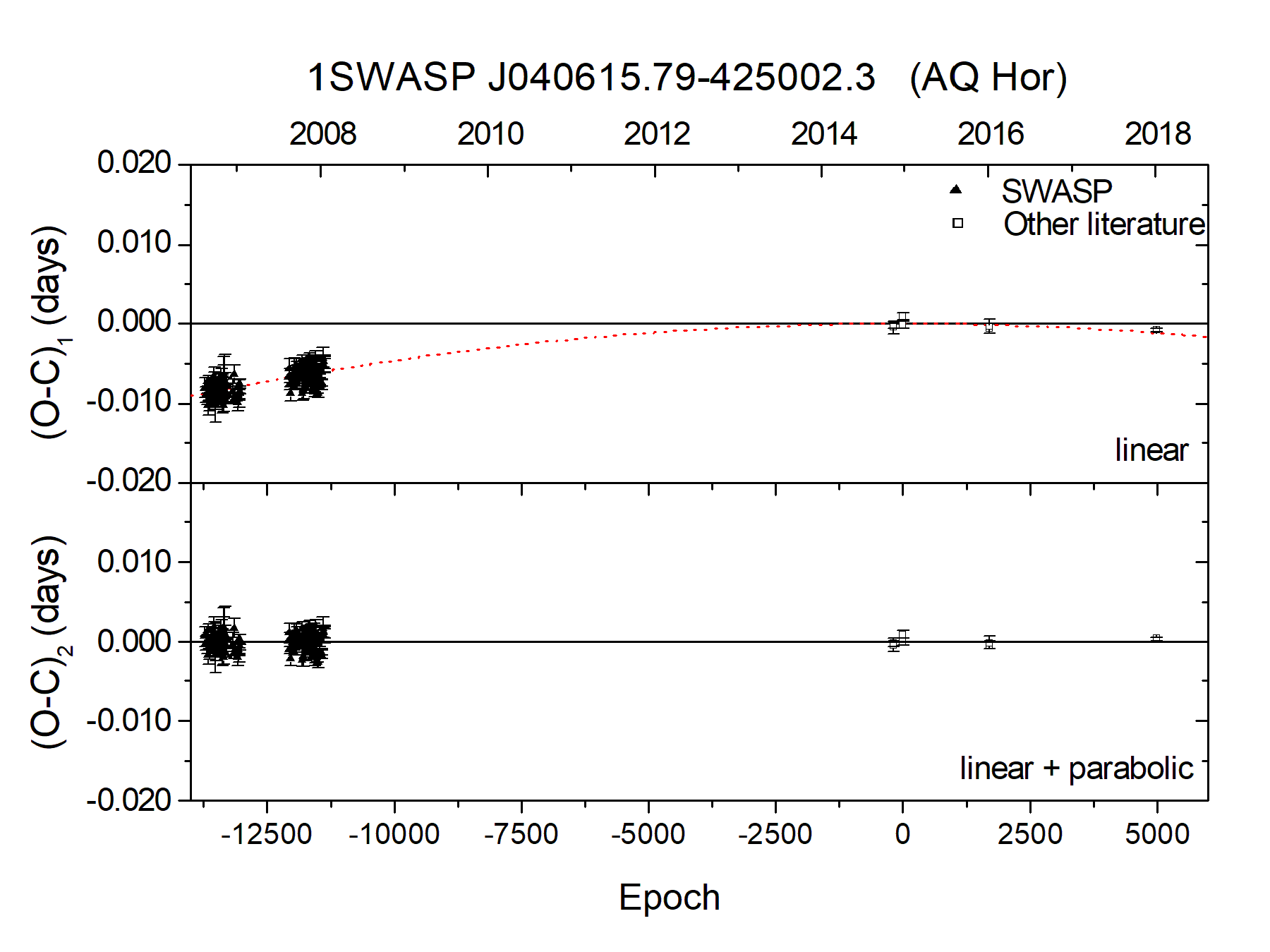}
\includegraphics[width=7.1cm,scale=1.0,angle=0]{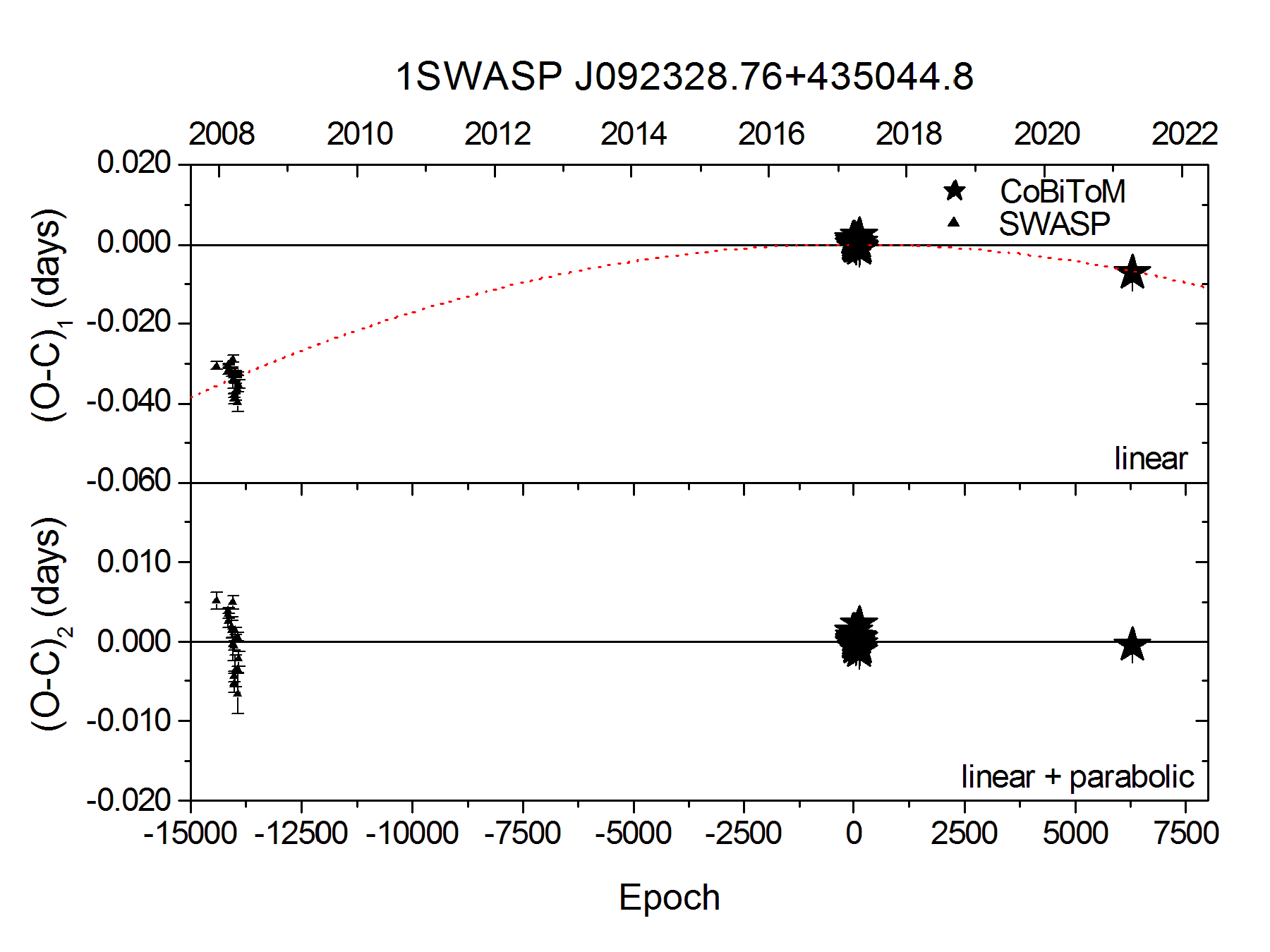}
\\
\includegraphics[width=7.1cm,scale=1.0,angle=0]{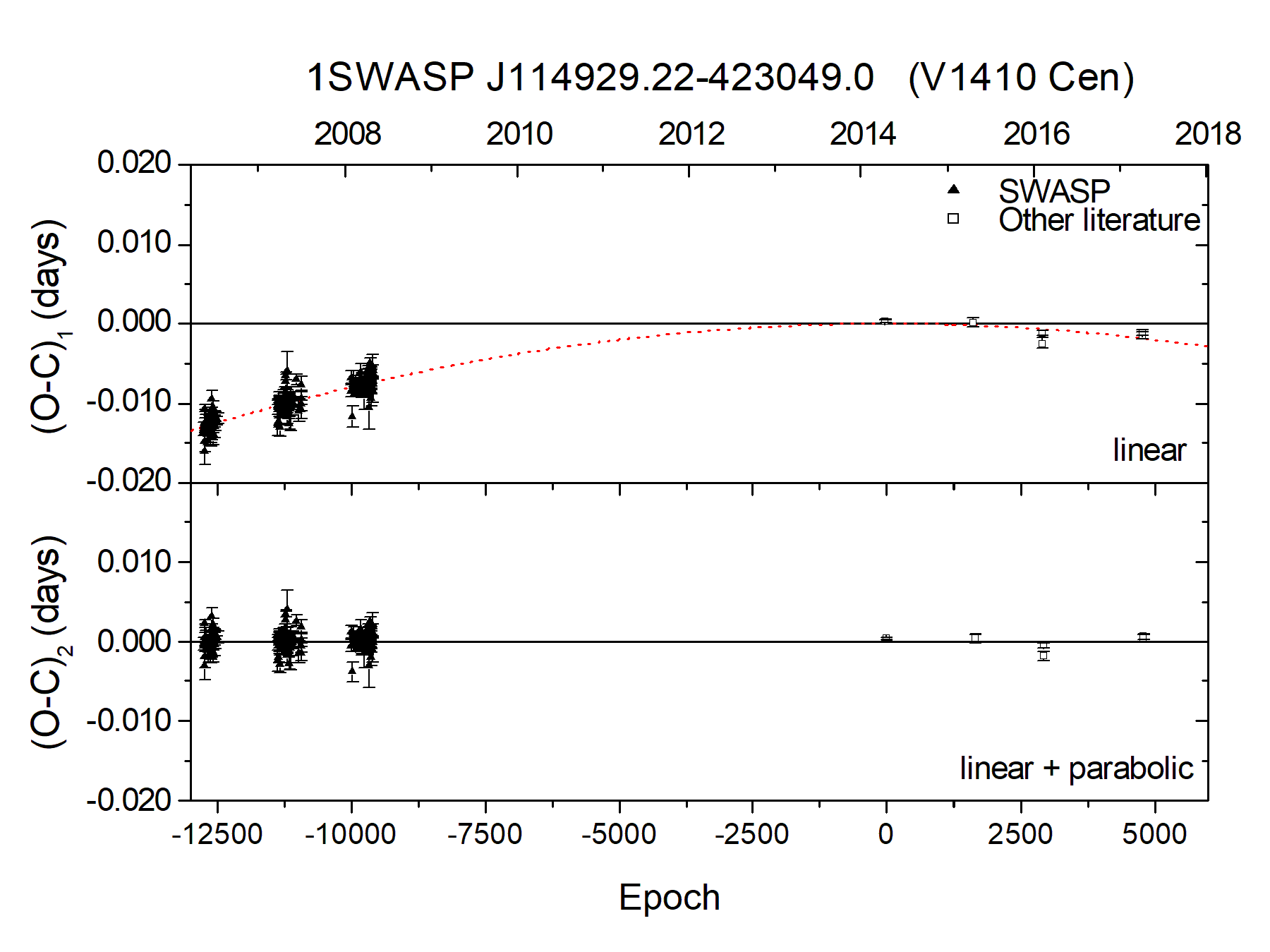}
\includegraphics[width=7.1cm,scale=1.0,angle=0]{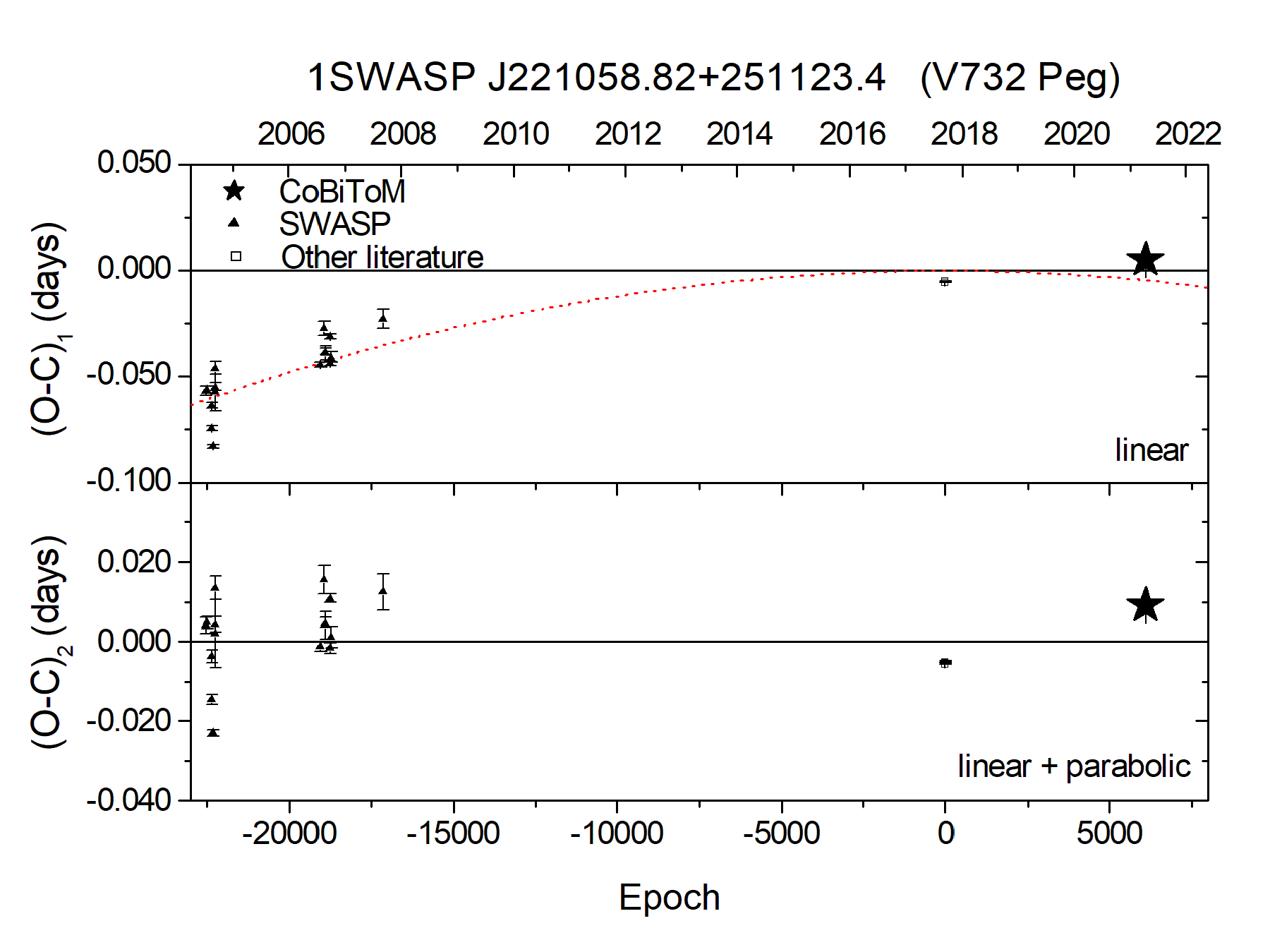}

\caption{The $O-C$ diagrams for the four systems that present circular orbits with prominent negative second order period modulation.}

\label{FigOC2a}
\end{figure*}

\begin{figure*}
\includegraphics[width=7.1cm,scale=1.0,angle=0]{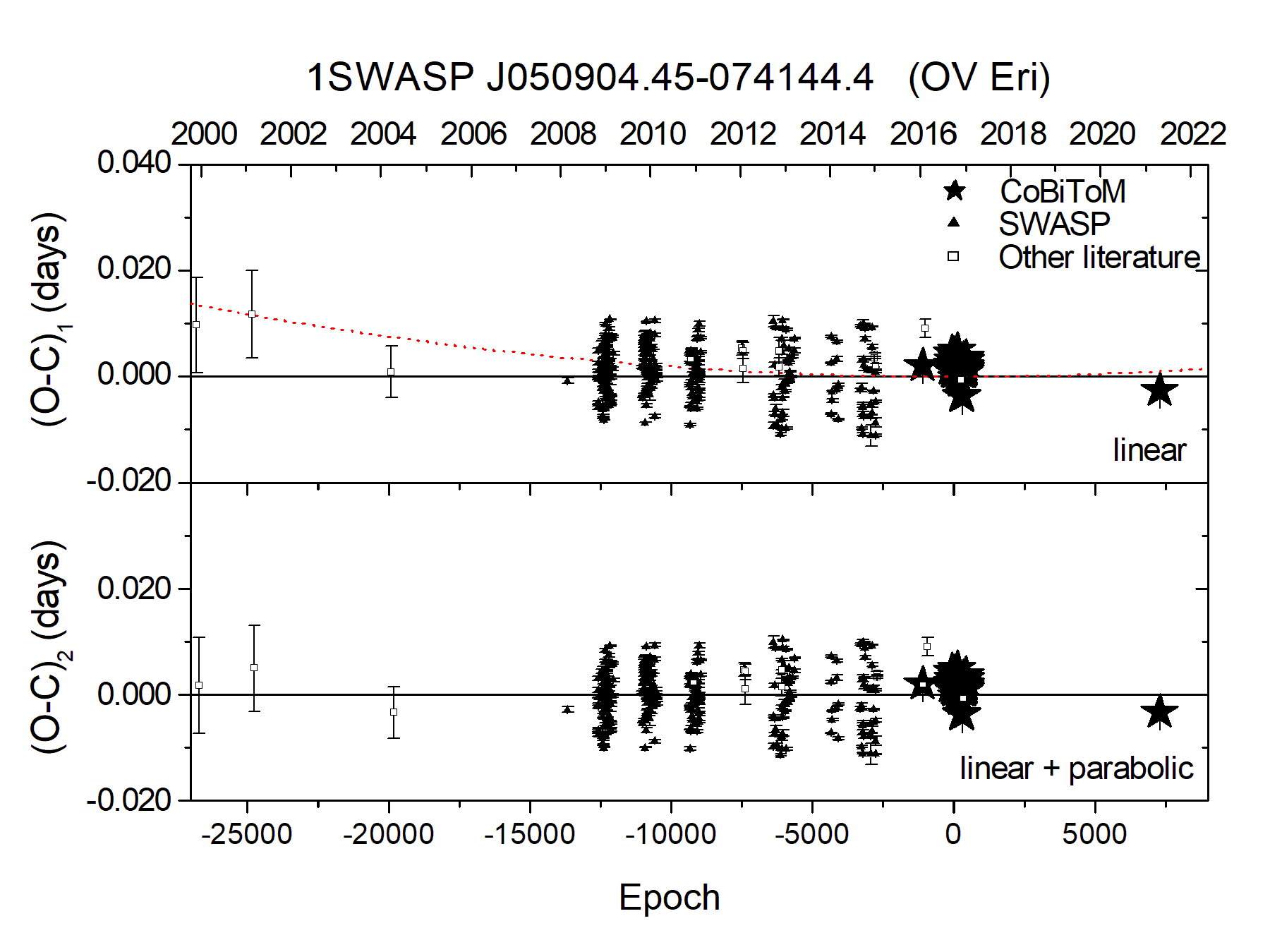}
\includegraphics[width=7.1cm,scale=1.0,angle=0]{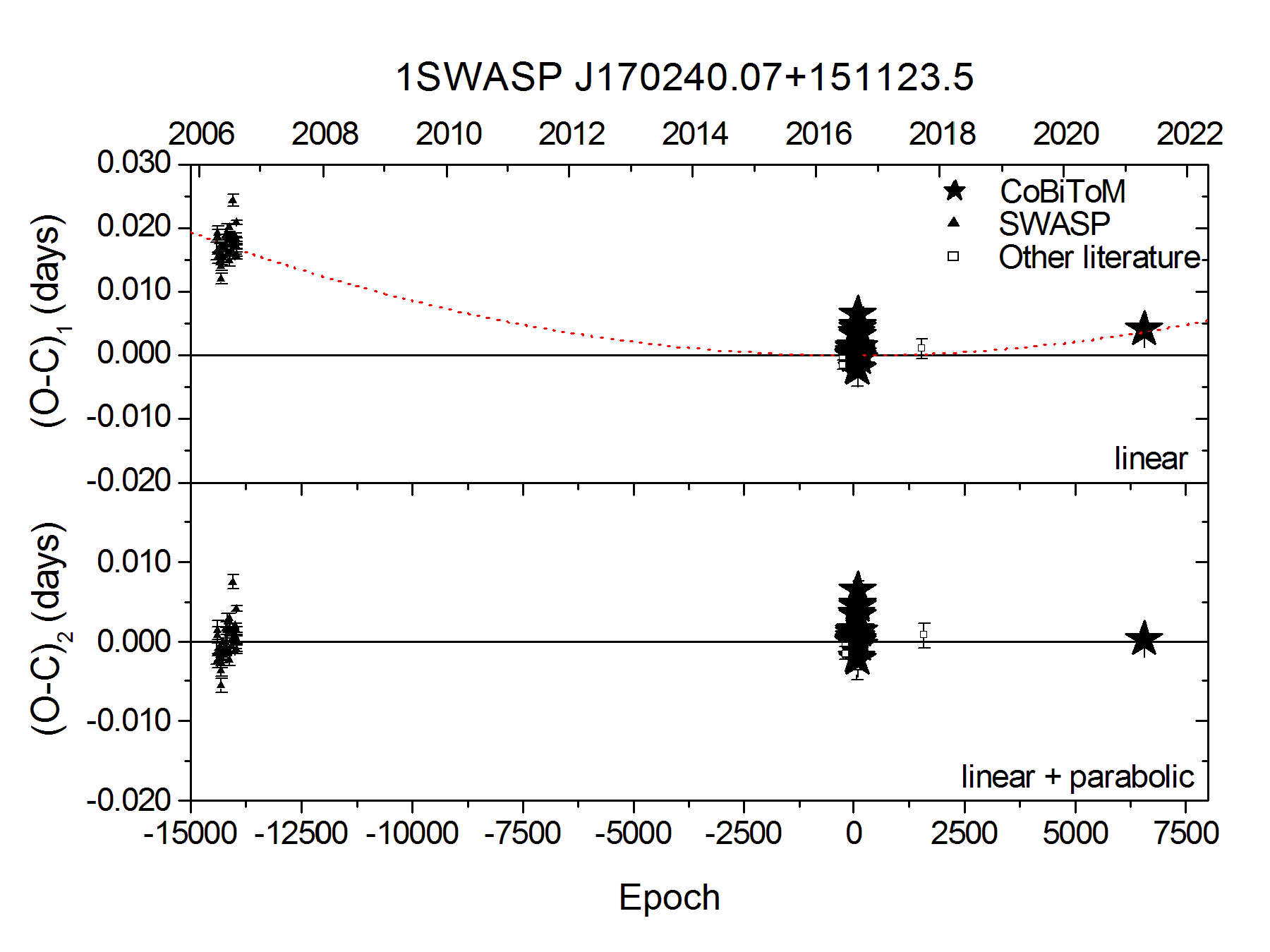}
\\
\includegraphics[width=7.1cm,scale=1.0,angle=0]{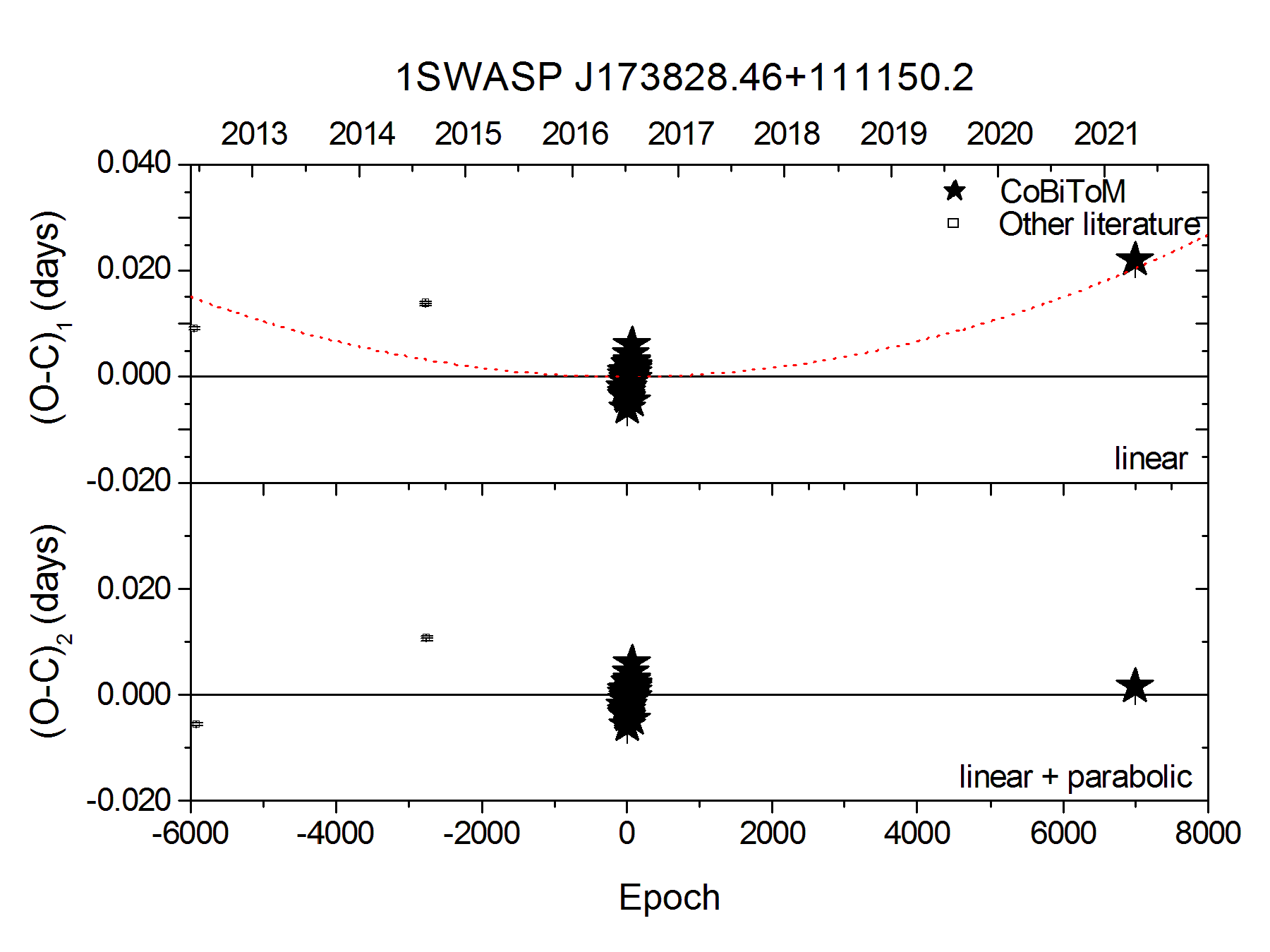}
\includegraphics[width=7.1cm,scale=1.0,angle=0]{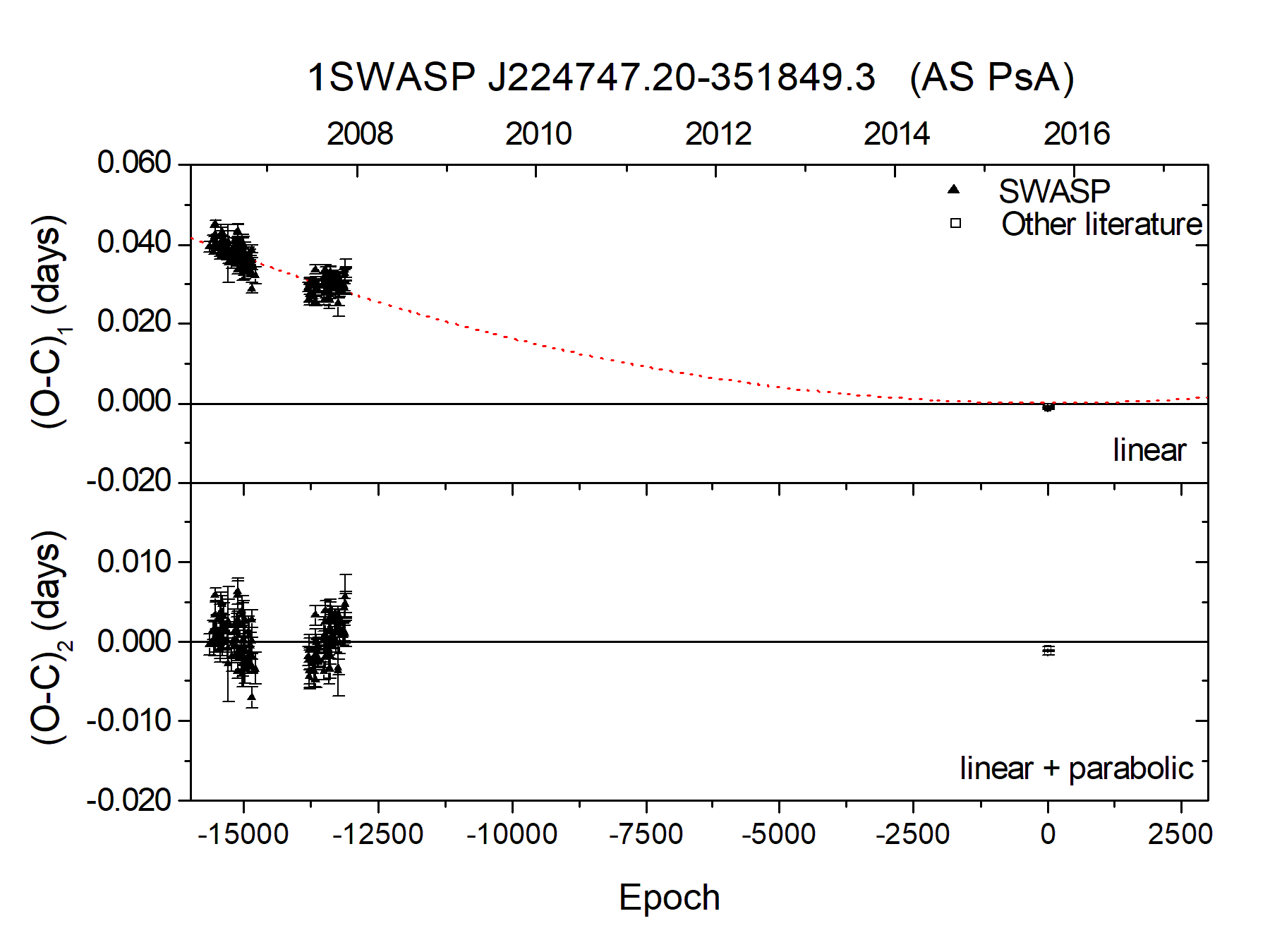}

\caption{The $O-C$ diagrams for the four systems that present circular orbits with prominent positive second order period modulation.}

\label{FigOC2b}
\end{figure*}

\begin{figure*}
\includegraphics[width=7.1cm,scale=1.0,angle=0]{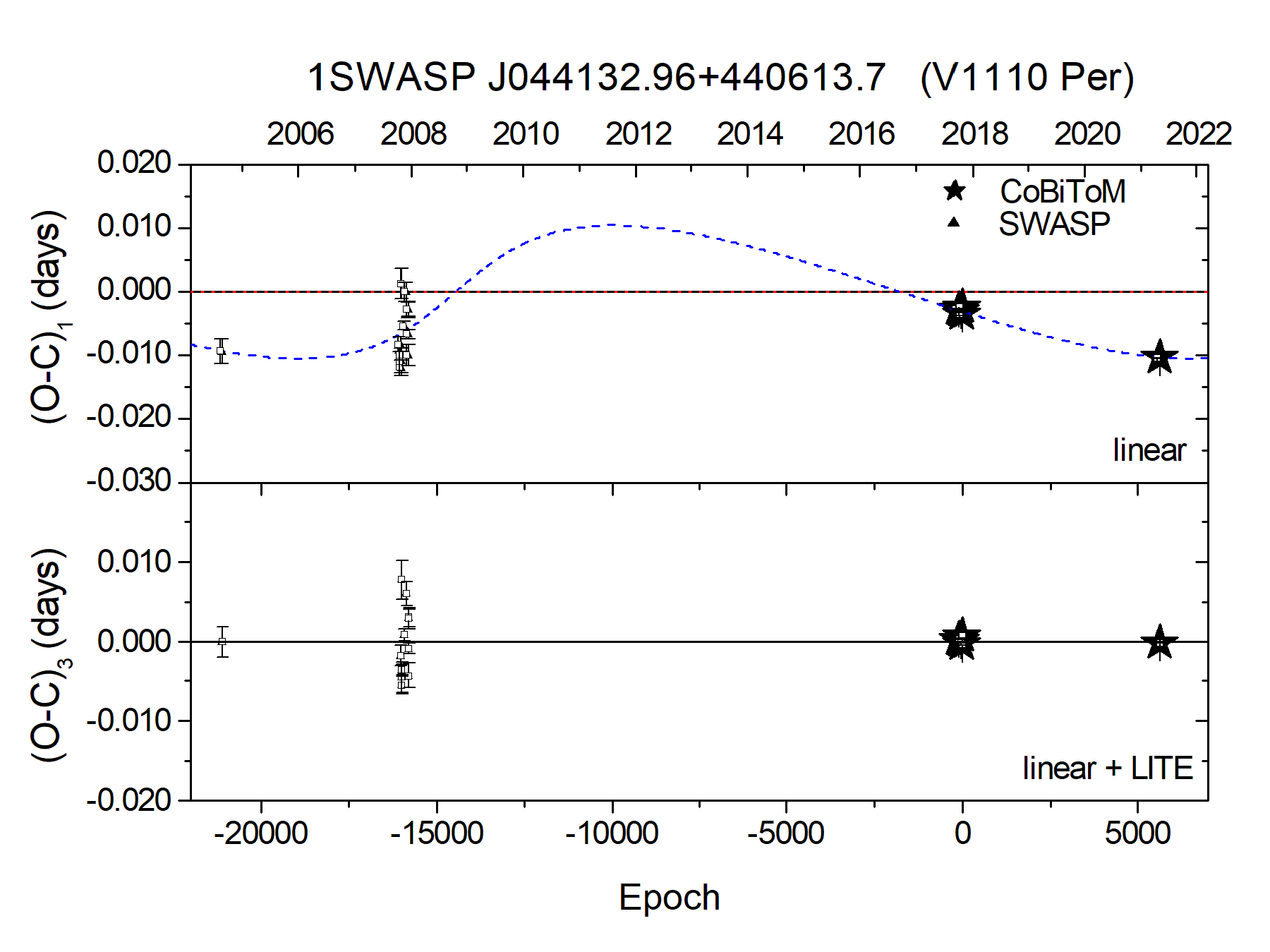}
\includegraphics[width=7.1cm,scale=1.0,angle=0]{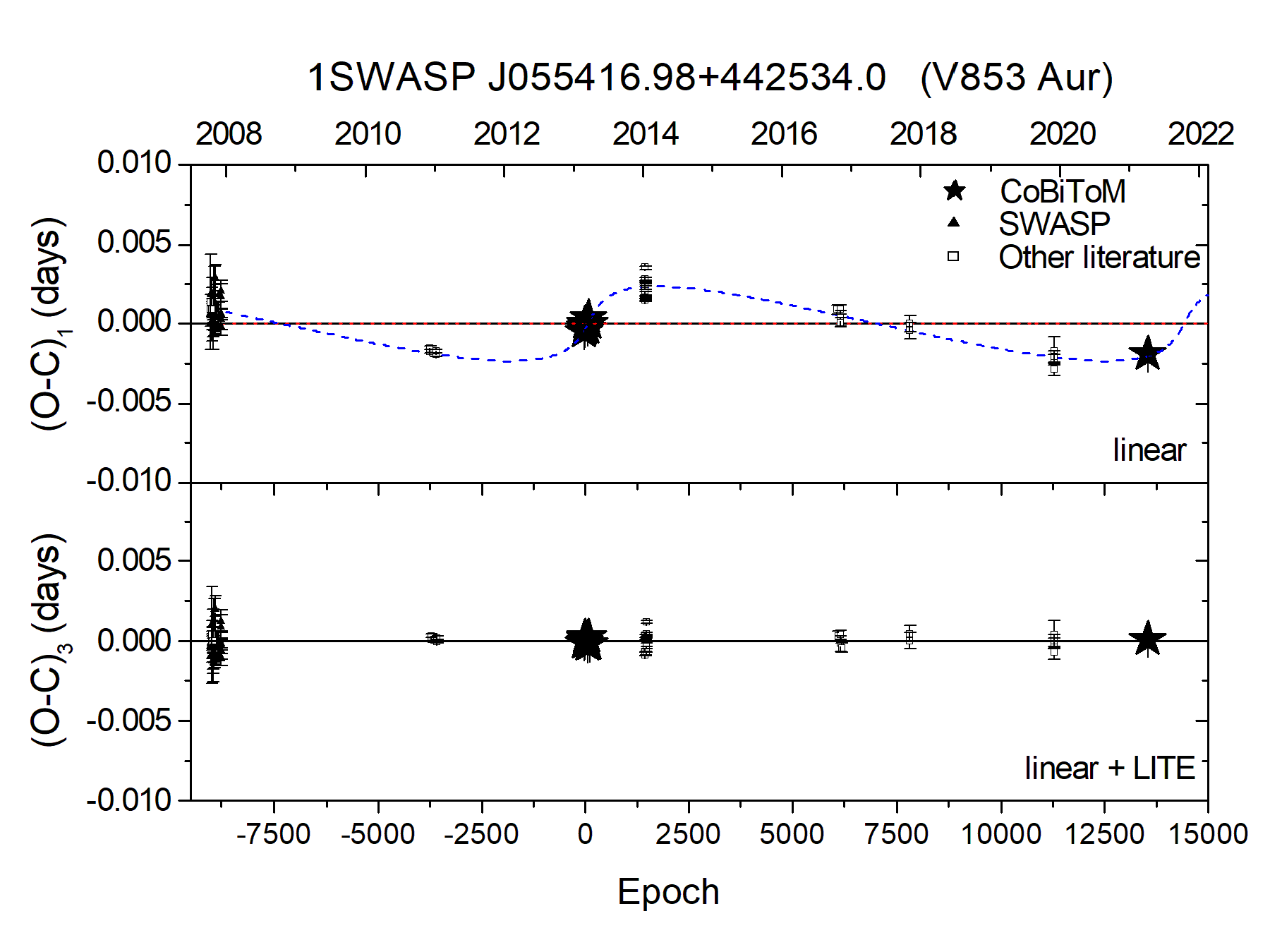}
\\
\includegraphics[width=7.1cm,scale=1.0,angle=0]{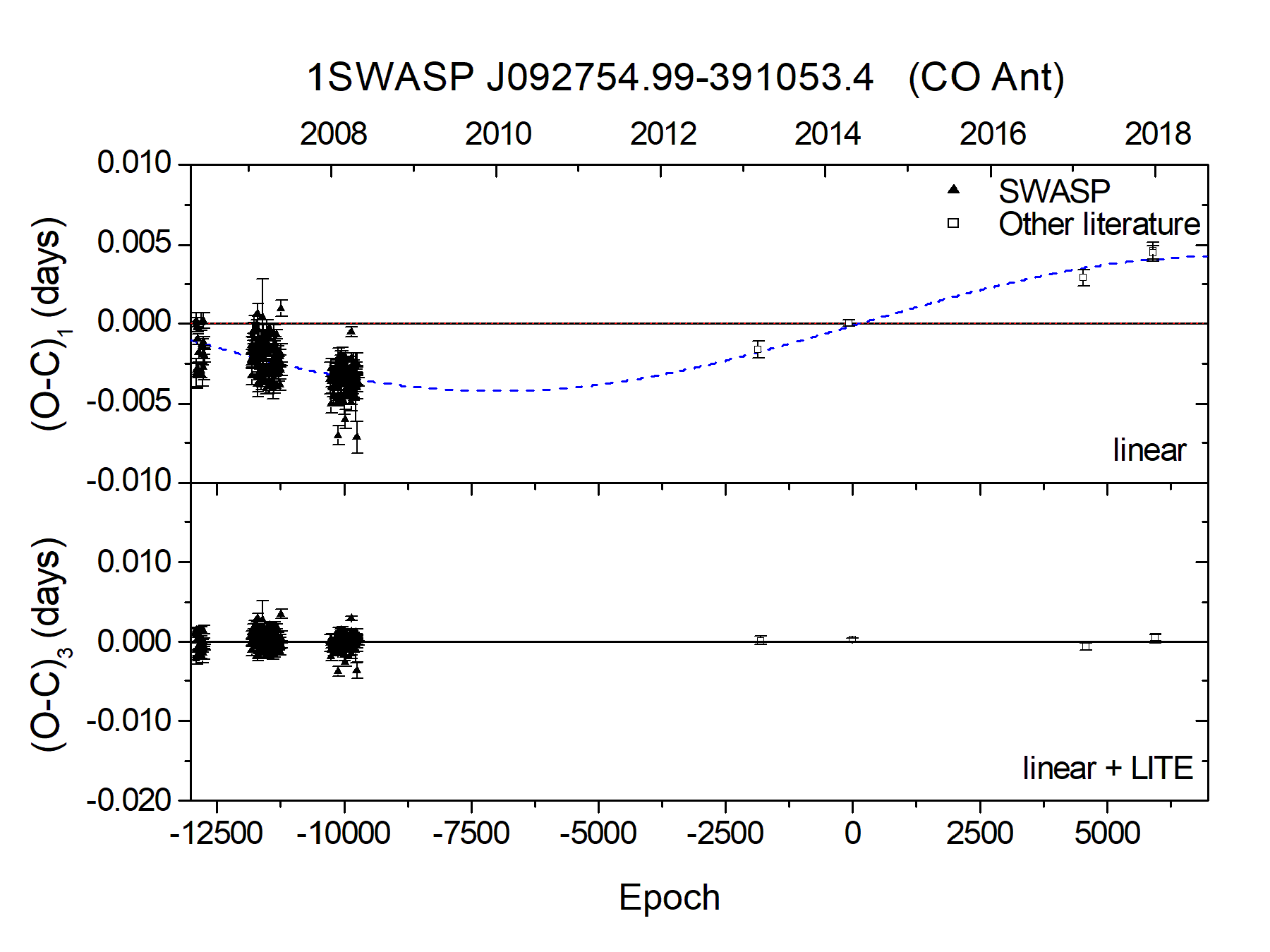}
\includegraphics[width=7.1cm,scale=1.0,angle=0]{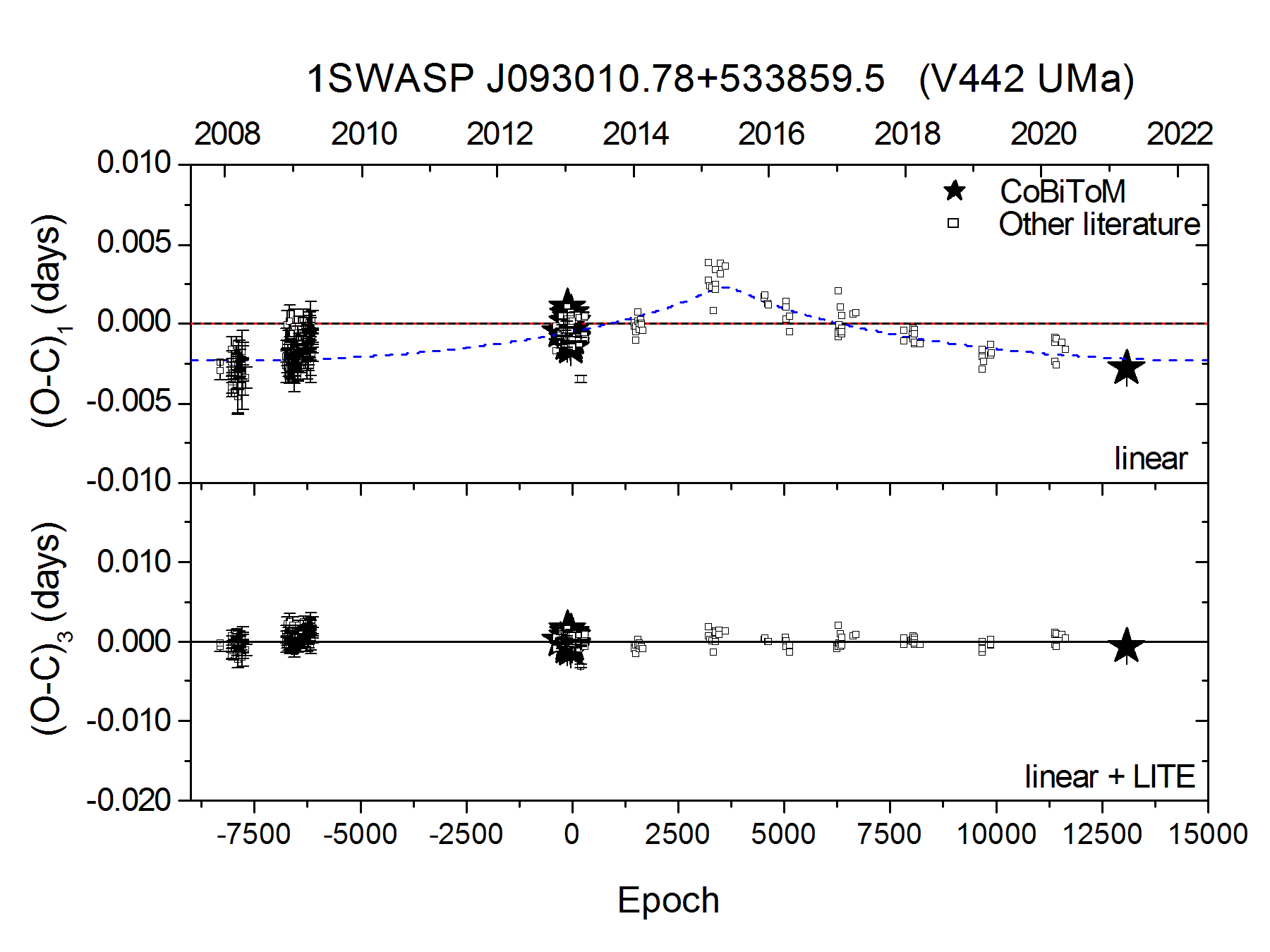}
\\
\includegraphics[width=7.1cm,scale=1.0,angle=0]{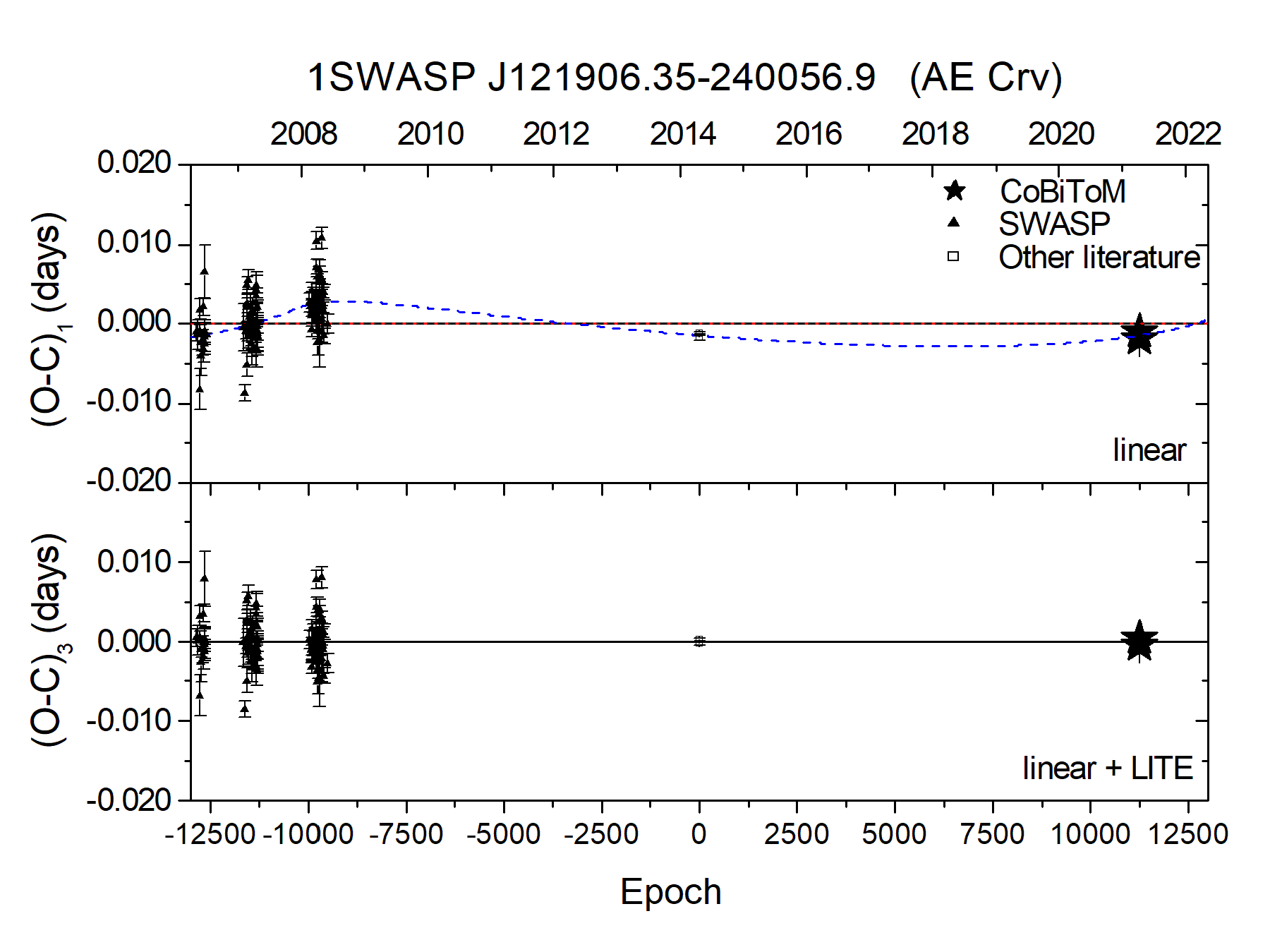}
\includegraphics[width=7.1cm,scale=1.0,angle=0]{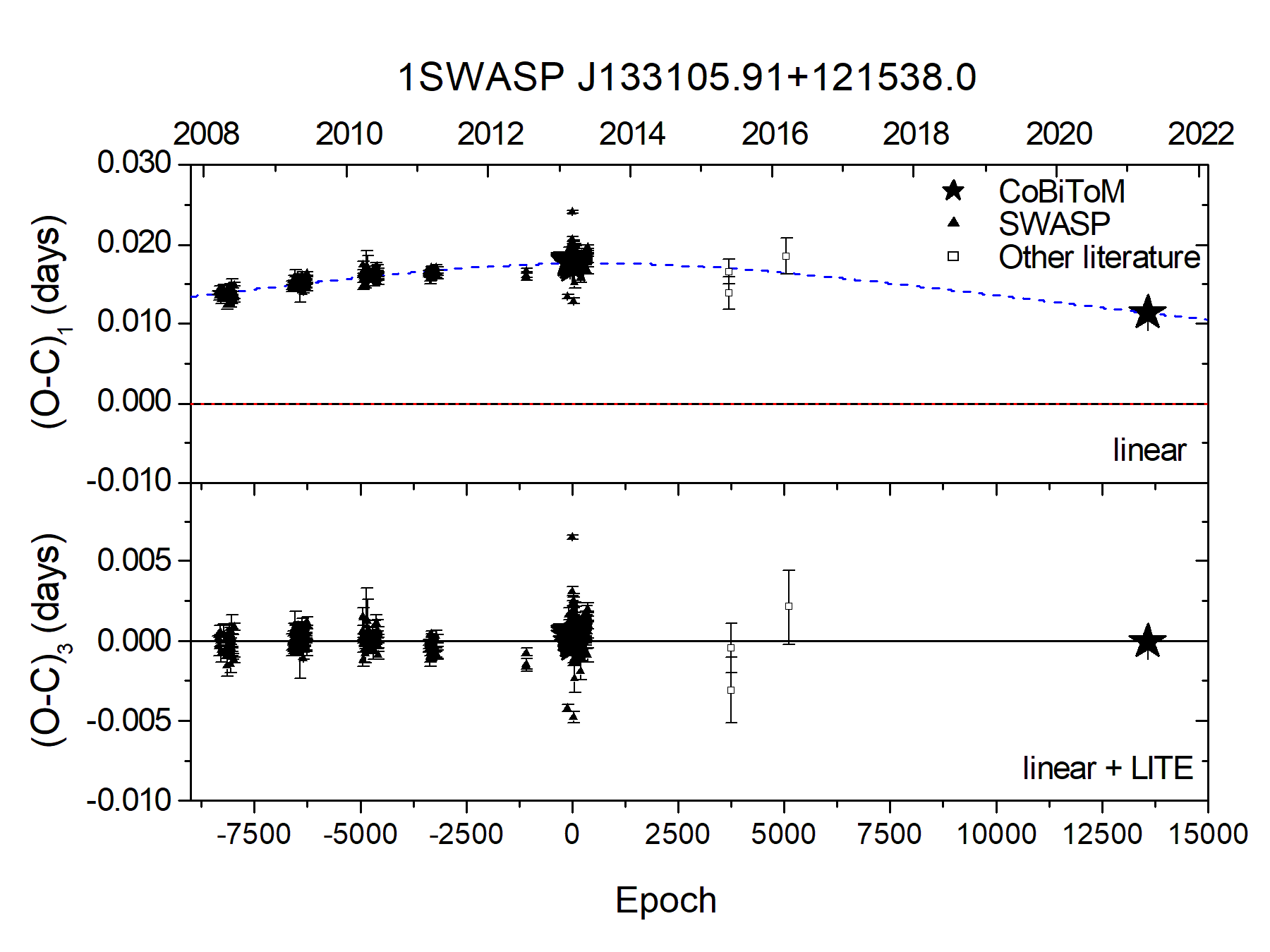}
\\
\includegraphics[width=7.1cm,scale=1.0,angle=0]{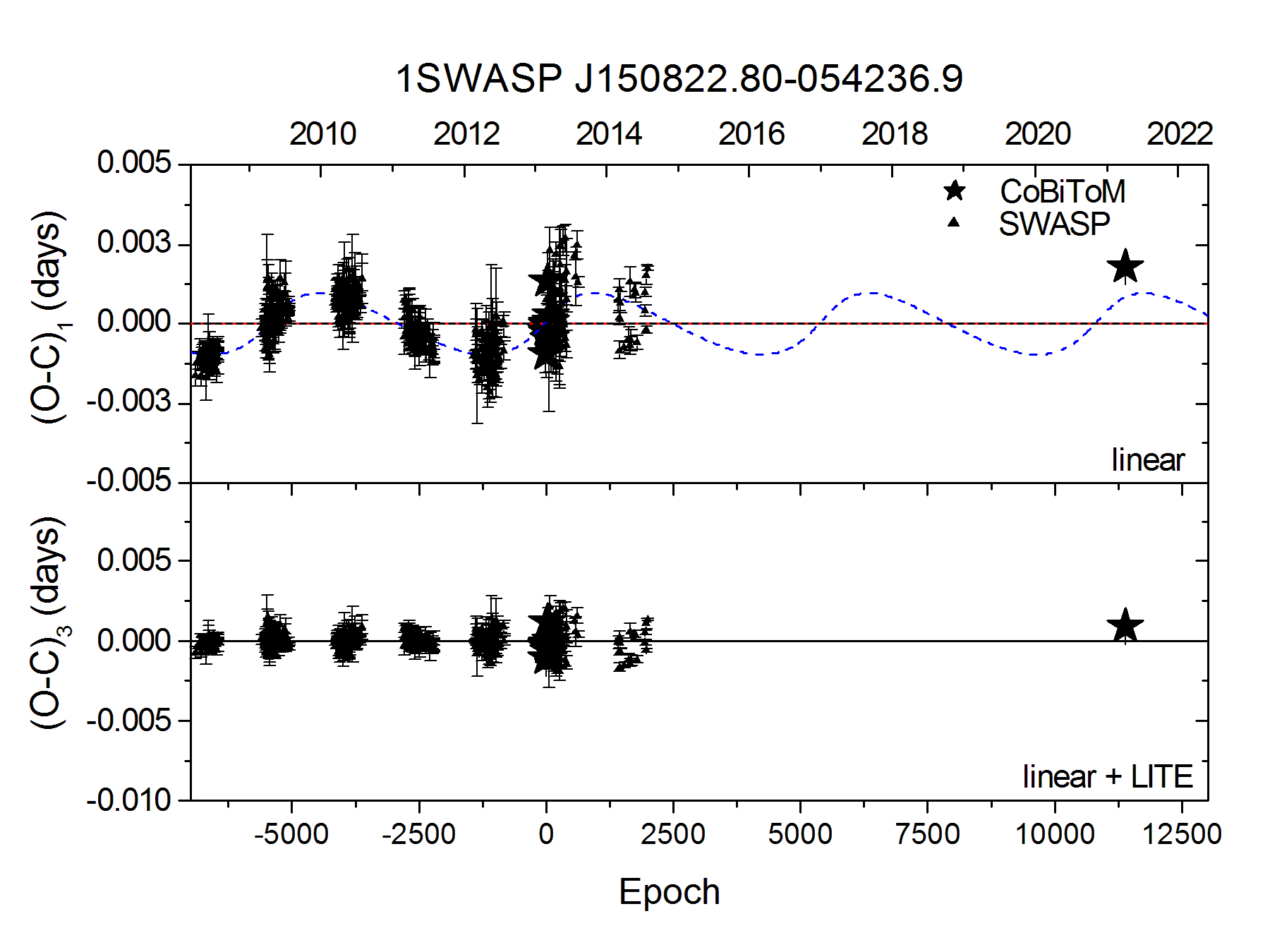}
\includegraphics[width=7.1cm,scale=1.0,angle=0]{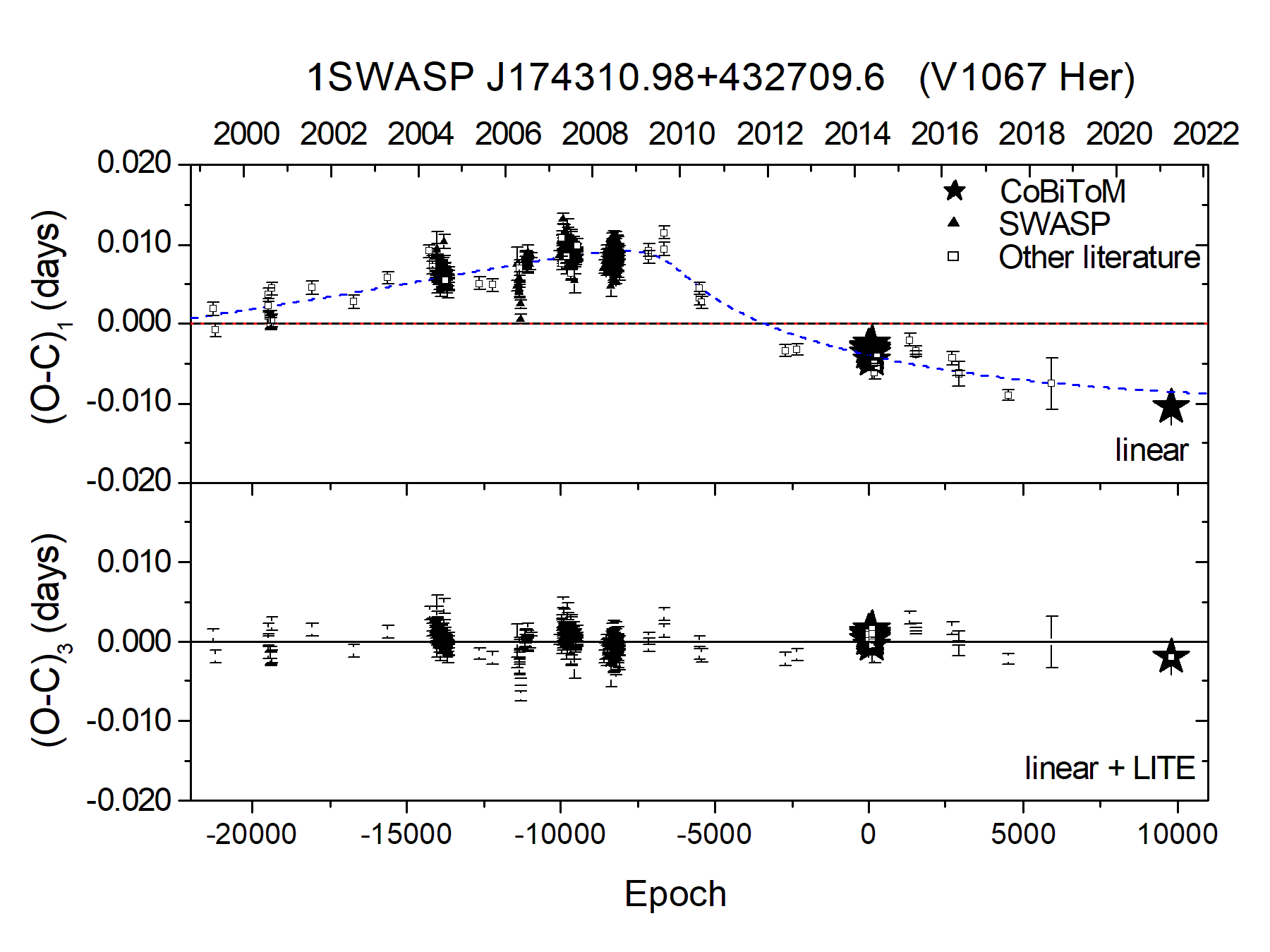}

\caption{The $O-C$ diagrams for the eight systems of the sample that present circular orbits with presence of an additional cyclic period modulation.}

\label{FigOC3a}
\end{figure*}

\begin{figure*}
\includegraphics[width=7.1cm,scale=1.0,angle=0]{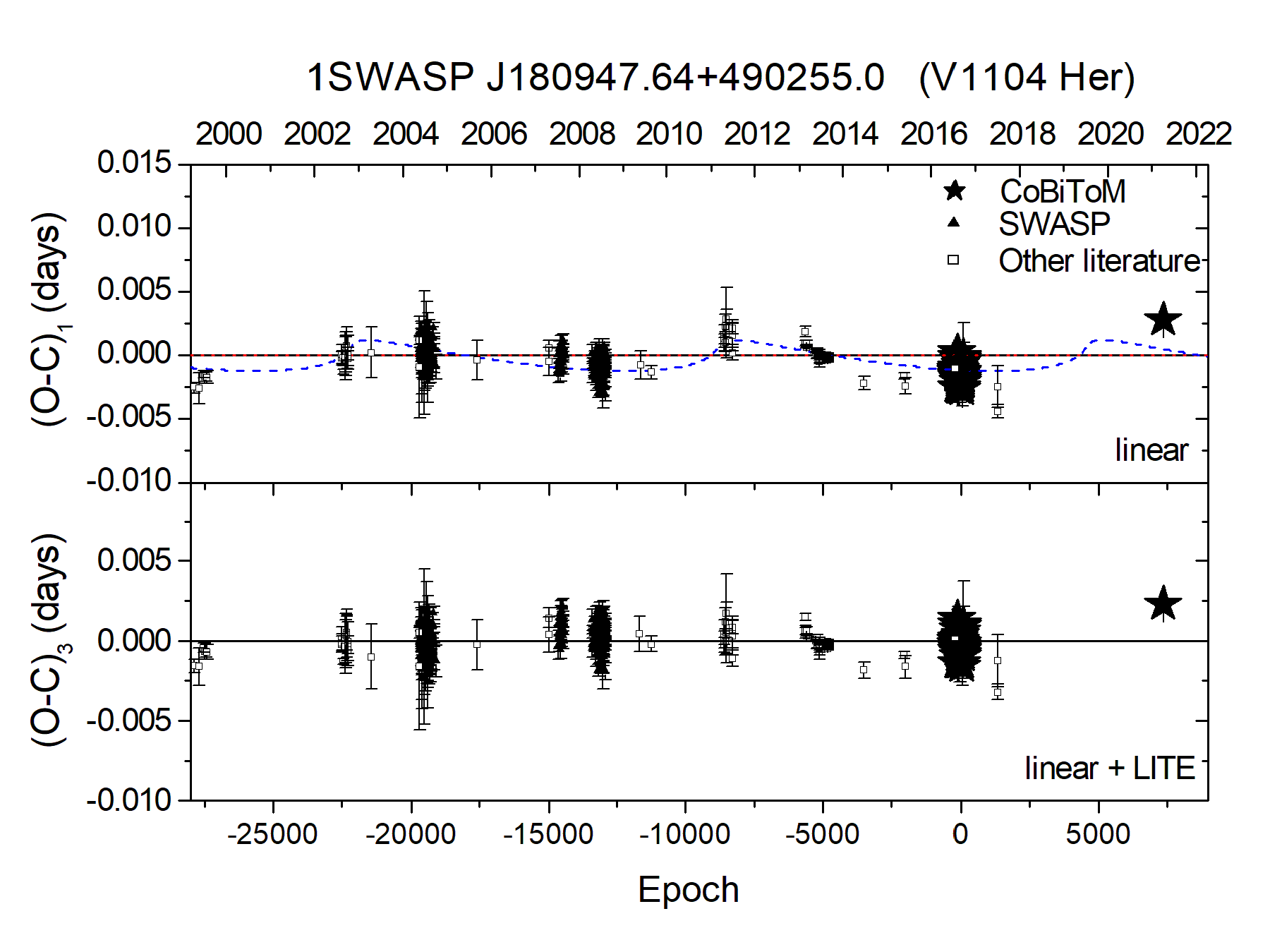}
\includegraphics[width=7.1cm,scale=1.0,angle=0]{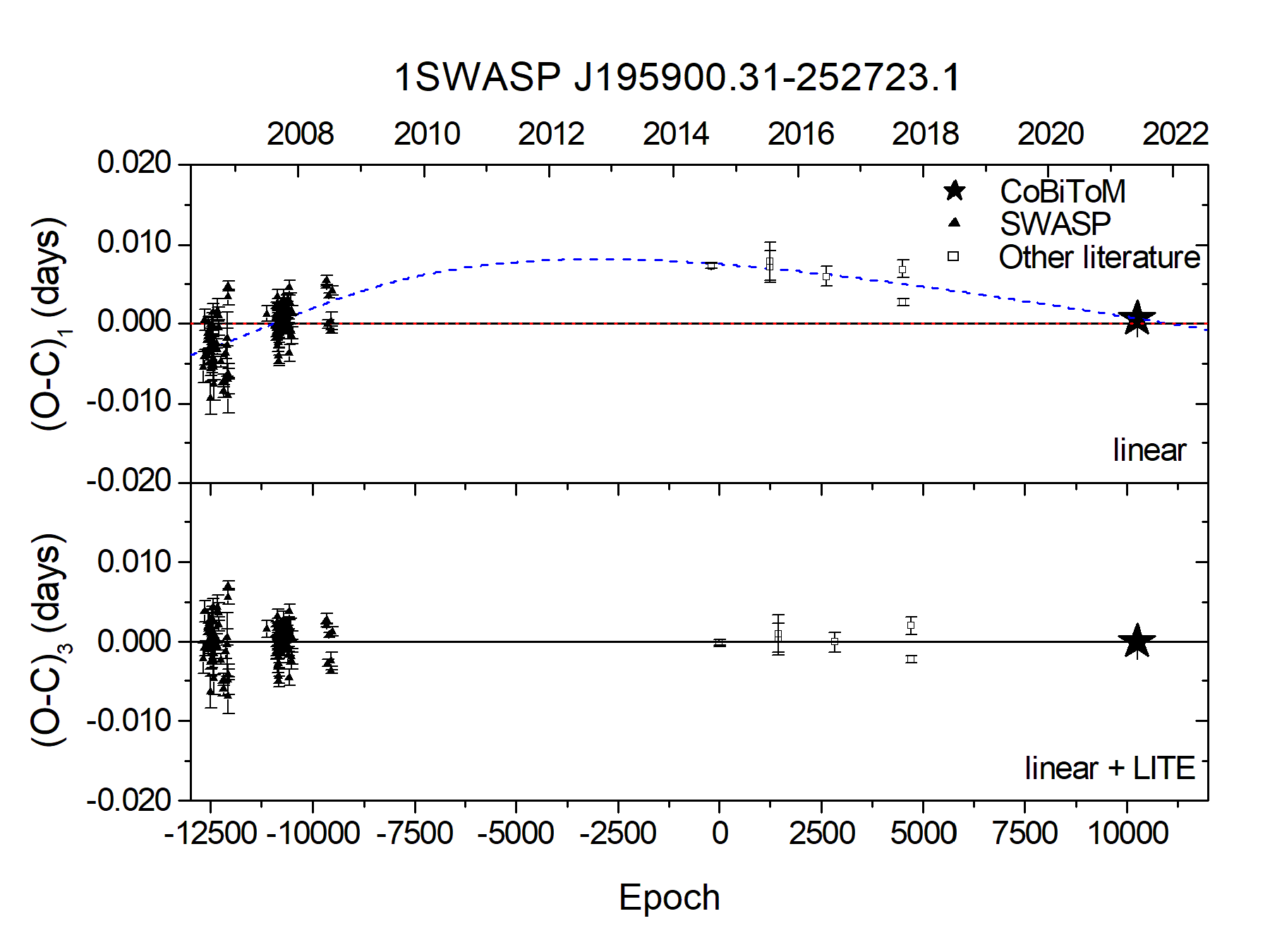}
\\
\includegraphics[width=7.1cm,scale=1.0,angle=0]{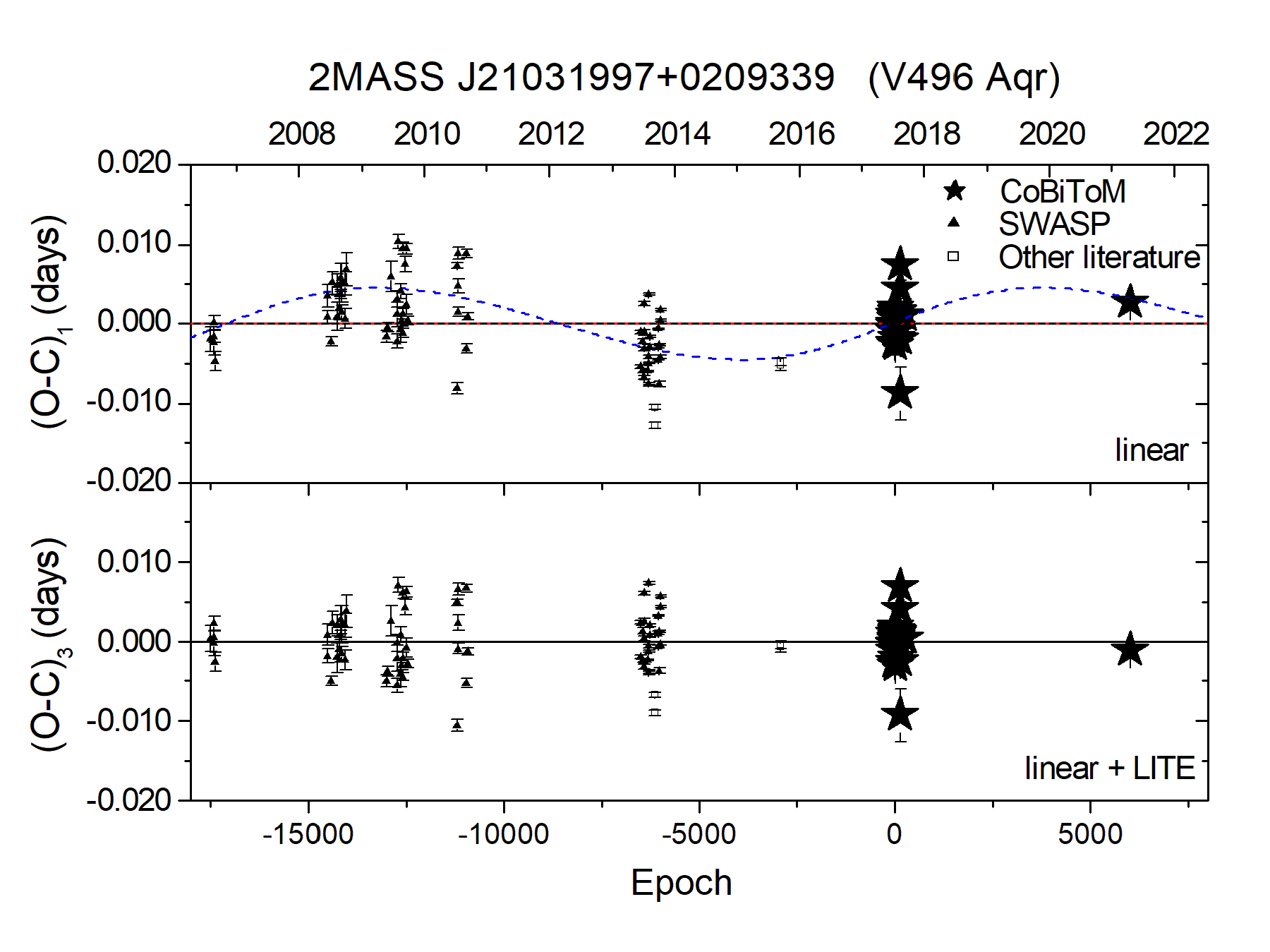}
\includegraphics[width=7.1cm,scale=1.0,angle=0]{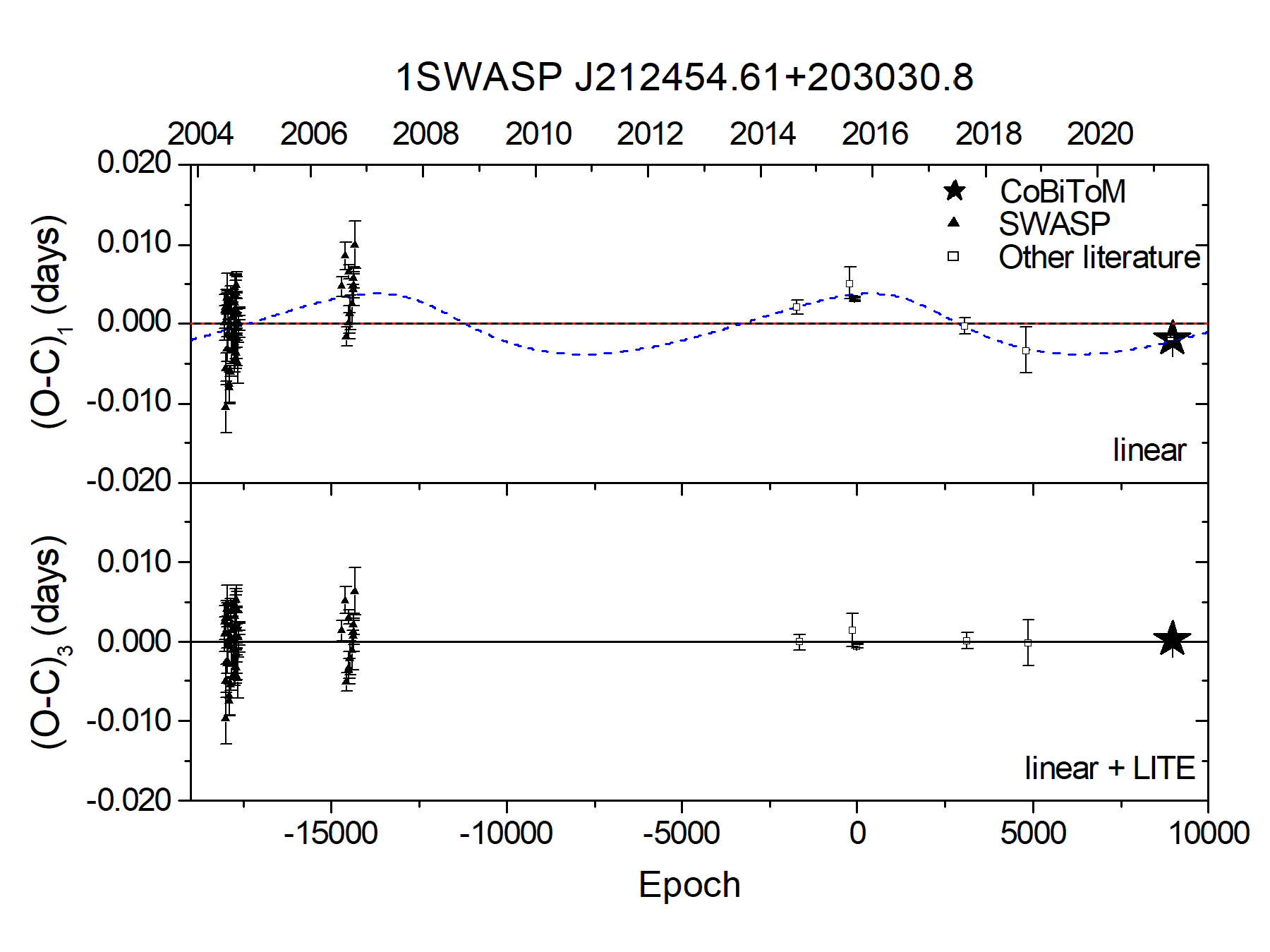}
\\
\includegraphics[width=7.1cm,scale=1.0,angle=0]{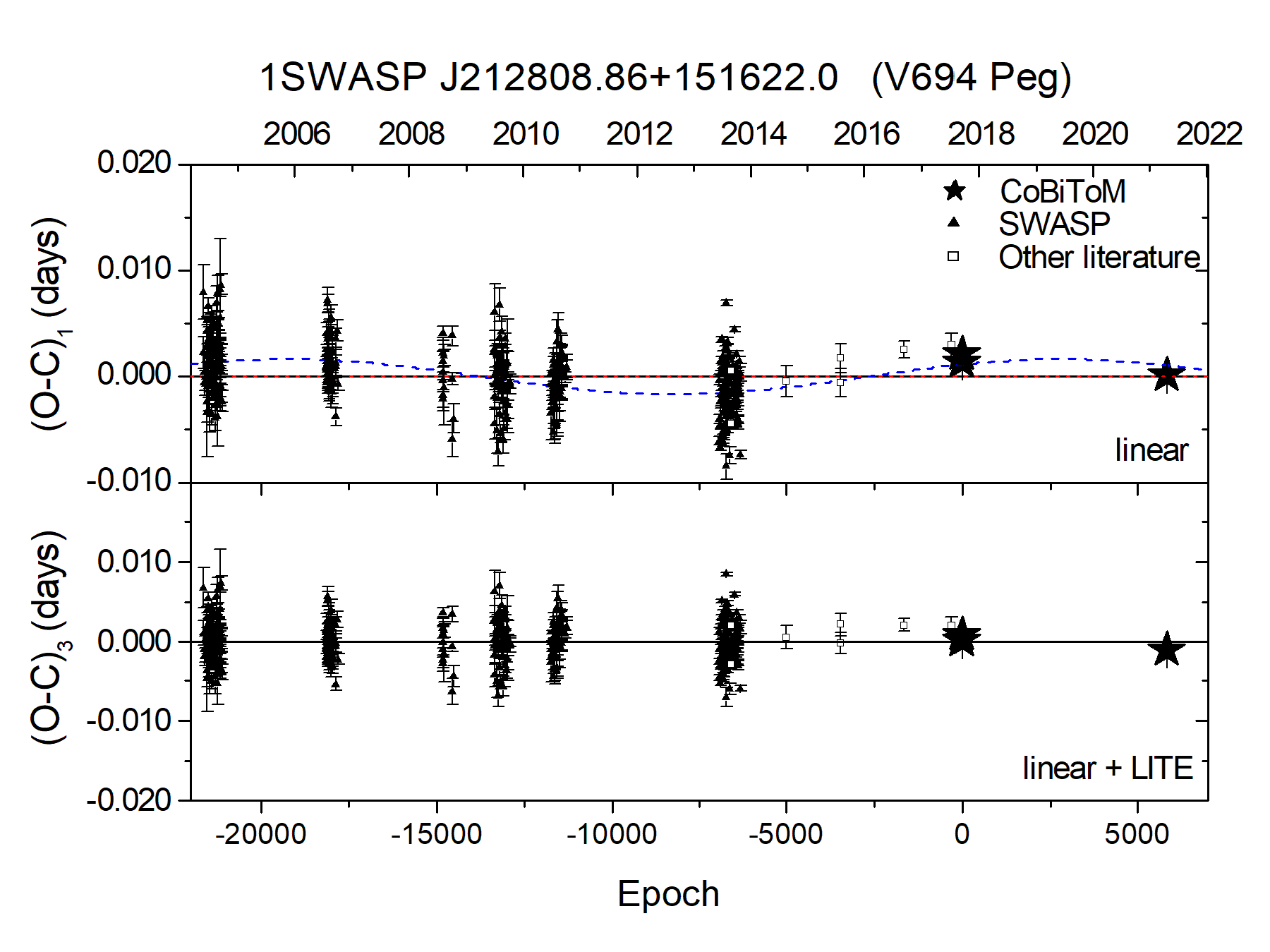}

\caption{The same as Fig. \ref{FigOC3a} for the rest five systems that present circular orbits with presence of an additional cyclic period modulation.}

\label{FigOC3b}
\end{figure*}

\begin{figure*}
\includegraphics[width=7.1cm,scale=1.0,angle=0]{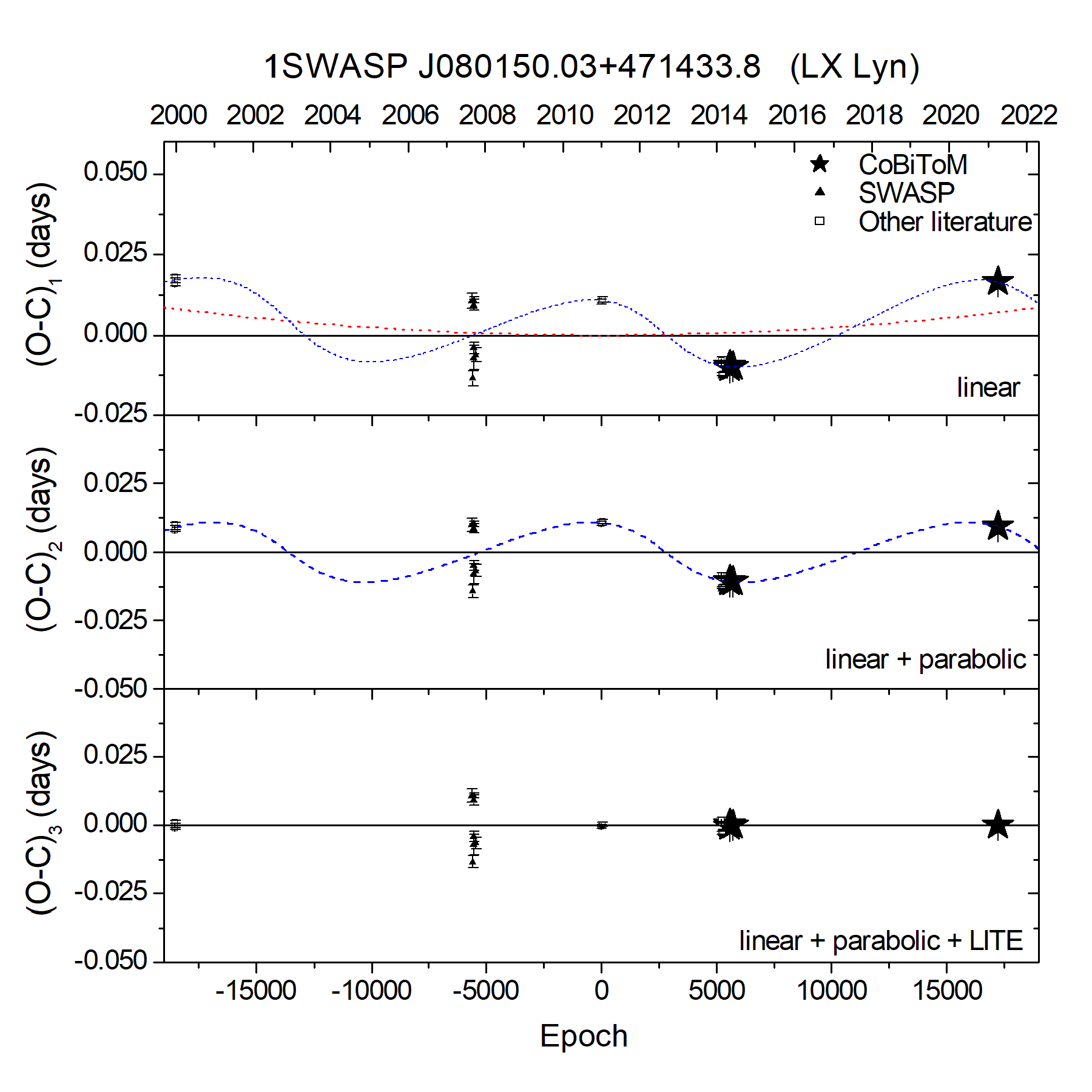}
\includegraphics[width=7.1cm,scale=1.0,angle=0]{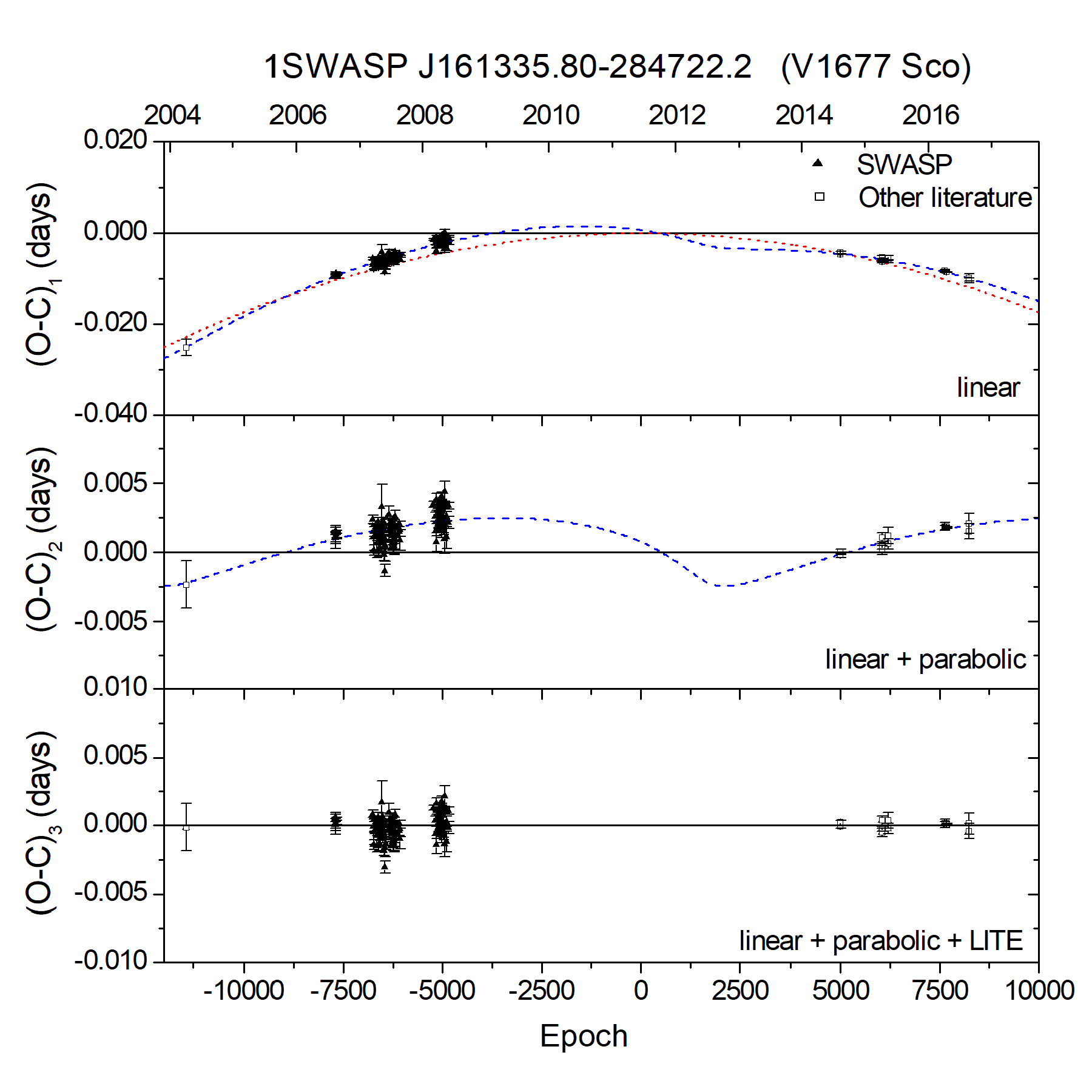}
\\
\includegraphics[width=7.1cm,scale=1.0,angle=0]{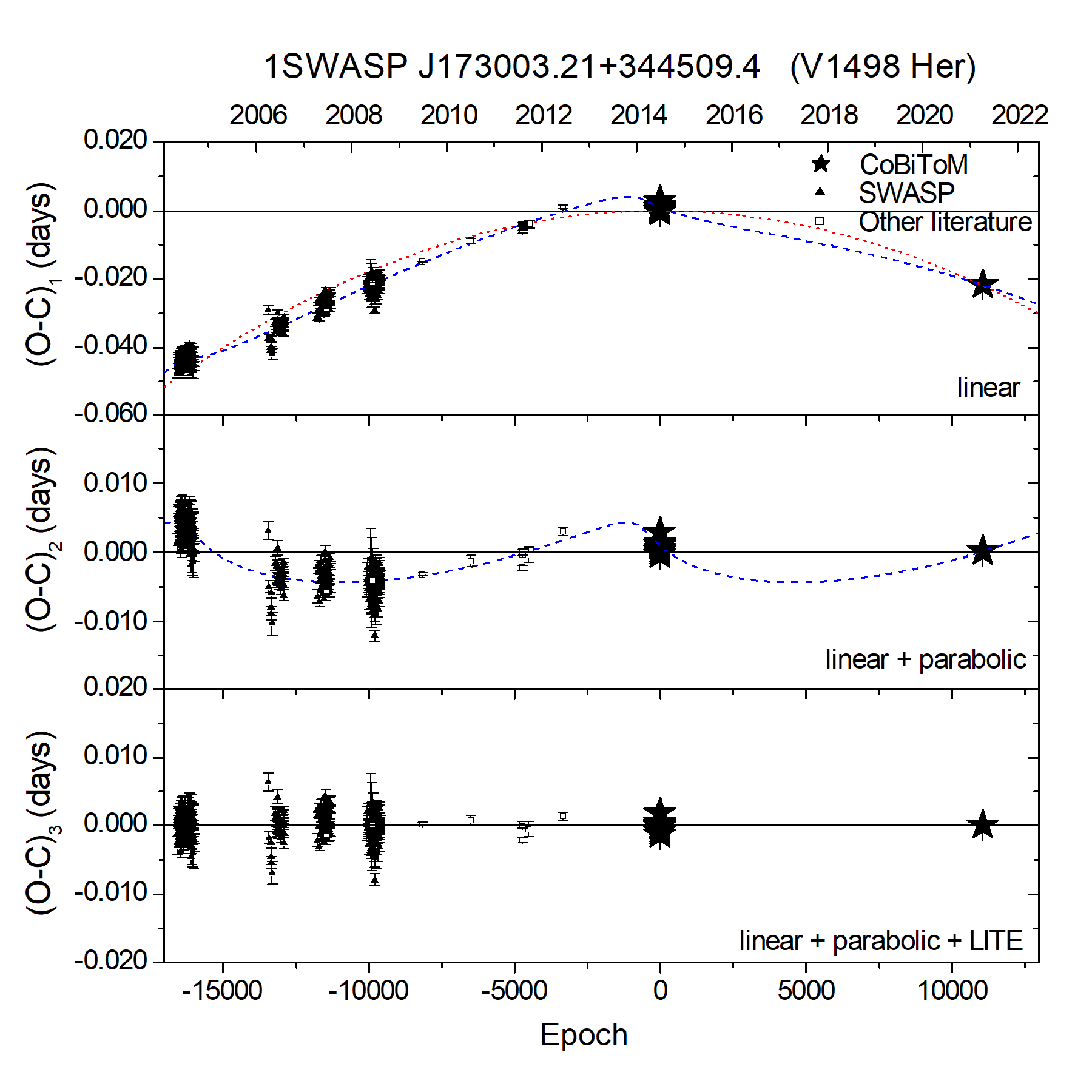}
\includegraphics[width=7.1cm,scale=1.0,angle=0]{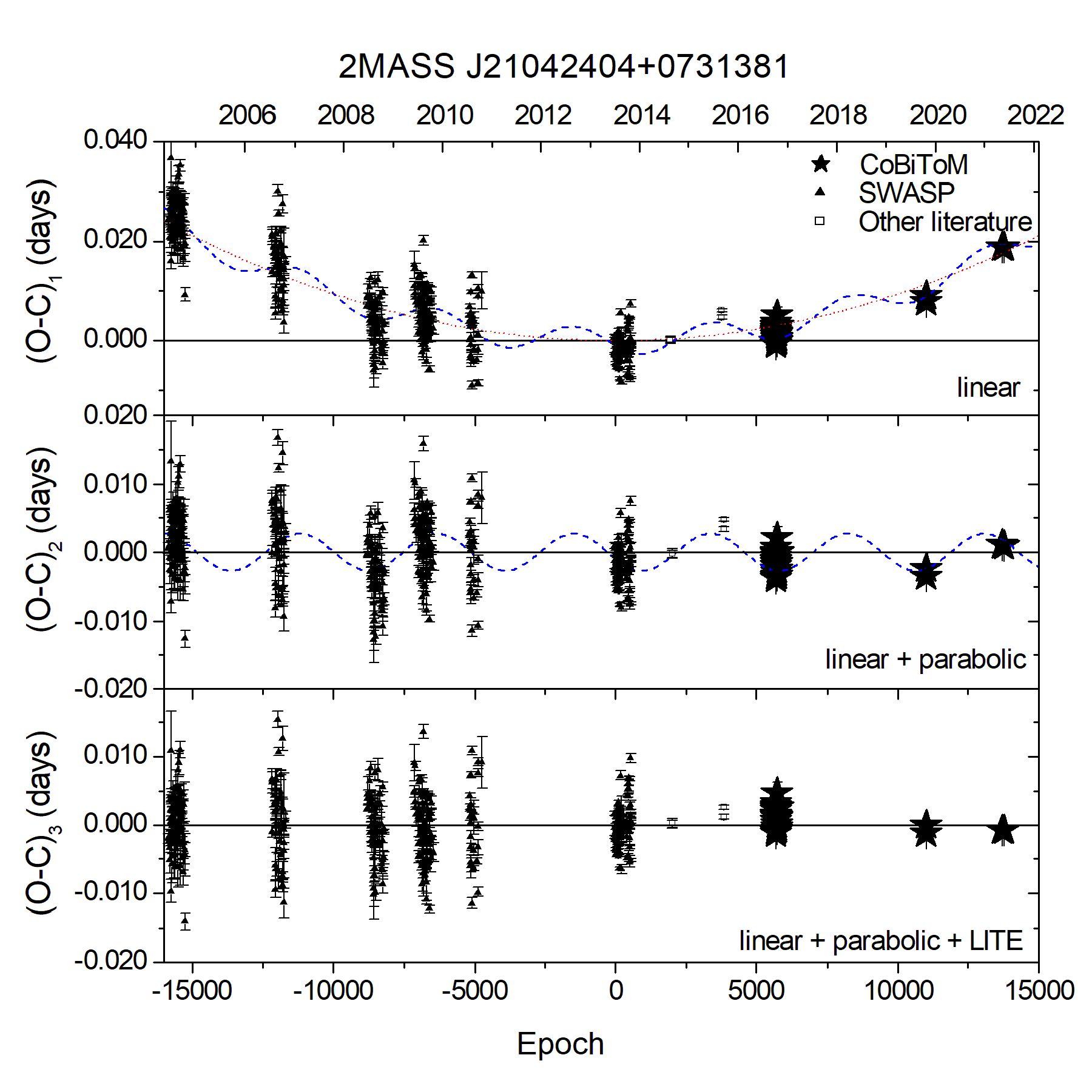}
\\

\caption{The $O-C$ diagrams for the four systems of the sample that present circular orbits with presence of a second order and a cyclic period modulation.}

\label{FigOC4}
\end{figure*}

In this study it is found that five systems out of 30 (i.e. 17 per cent) exhibit no change in their orbital period, as they show no modulation in their $O-C$ diagrams (Fig. \ref{FigOC1}). 

Orbital period modulations are shown in 12 out of 30 systems (40 per cent of the sample) out of which 6 present negative orbital period rates for downward parabola shape in the $O-C$ diagram and 6 positive period rates for upward parabola shape respectively. This finding, gives no conclusive evidence that the ultra-short orbital period systems tend to shrink their orbits, through the decrease of their orbital period. Eight of the above systems (27 per cent of the entire sample) present only secular orbital systematic period modulation 
(with negative or positive $dP/dt$ ) with no additional component, as shown in Figs. \ref{FigOC2a} and \ref{FigOC2b}.

Furthermore, from the $O-C$ analysis it was found that 13 systems (43 per cent) host a third component in a wider orbit around them, with no change in orbital period, while four systems (13 per cent) host a third component with change in orbital period (Figs. \ref{FigOC3a} with \ref{FigOC3b} and \ref{FigOC4}, respectively). 
The existence of the third component is also supported by the third light, found in the light curve models. 
The fact that the above 17 systems (57 per cent) host a third component, comes in agreement with \cite{D'Angelo2006} who found that more than 30 per cent of contact binary systems belong to triple systems and have a spectroscopic signature of a third component. 
It seems that more systems can be characterised as triples by means of the $O-C$ analysis, as it is more sensitive in detecting eclipse time variations. 
The results of the $O-C$ analysis, i.e. the period of the tertiary component ($P_{3}$), the amplitude of the cyclic variation ($A$), the value of $a_{12} sini_{3}$ and the possible minimum mass of the tertiary component in a co-planar orbit ($M_{3 min}$), are presented in Table \ref{Tab11.OCs}. 

For 12 systems that exhibit secular period changes, we have computed the mass transfer and mass loss rates, using the software provided by \cite{Liakos2015}. 
Secular orbital period increase (positive $dP/dt$) can be caused either by mass transfer from the less massive component to the more massive one or by mass loss from the system through stellar winds. 
On the contrary, orbital period decrease (negative $dP/dt$) is typically caused by mass transfer of the more massive component to the less massive one. 
For systems that present period increase in their $O-C$ diagrams, and given that we cannot be certain which is the driving mechanism, both mass transfer ($dM_{T}/dt$) and mass loss ($dM_{L}/dt$) mechanisms can be applied (Table \ref{Tab11.OCs}). It should be mentioned that these two mechanisms are not taking place in a binary at the same time. Typically, the values of mass transfer/loss rate found in contact binaries are of the order $10^{-7} M_{\odot}~yr^{-1}$  \citep{Kouzuma2018,Li2020}. 

According to previous studies \citep{Kubiak2006,Lohr2015,Pietrukowicz2017} contact binaries could exhibit both period decrease and period increase. 
\cite{Lohr2015} presented a statistical analysis on contact binaries from SWASP data, where the $\dot{P}$ is almost evenly distributed in both positive and negative values. Similar distribution of $\dot{P}$ is observed in the present study. 
However in the present study no extreme values of $\dot{P}$ were found compared to the study of \cite{Lohr2015}, which is probably due to the different time range.

$O-C$ diagrams could also be a very useful tool to determine whether the contact binary is a member of a triple or even a multiple system. Literature studies suggest that the number of the detected tertiary components in contact binary systems is increasing \citep[e.g][]{Pribulla2006,D'Angelo2006,Tokovinin2006,Rucinski2007a}. In our investigation, more than half of the sample (17 systems) have shown cyclic period variations, complying with the light curve model findings of third light. Interestingly, none of them satisfy the criterion of \cite{Lanza2002} regarding the quadrupole moment variation value of the components that is needed for the Applegate mechanism \citep{Applegate1992} to explain the observed cyclic orbital period changes. Therefore, the cyclic variations in our sample can be interpreted exclusively by the presence of tertiary components. 

In a few cases, the minimum mass of the possible third component is less than 0.1 $M_{\odot}$, which could indicate the presence of a brown dwarf or even a hot Jupiter. Therefore, long-term monitoring of these ultra-short contact binaries is essential, not only to detect any intrinsic period changes, but also to specify whether or not they belong to triple or even multiple stellar systems.


\section{Discussion}
In the present study we have provided results from 4-band photometry for 30 ultra-short orbital period contact binaries, which are very close to the period cut-off (less than 0.26~d). A strong asset of observing campaigns such as the \textit{CoBitoM Project}, is the long term monitoring of several contact binaries and the thorough investigation derived from the light curve and orbital period modulation analysis. Future spectroscopy for the determination of the radial velocities of both components is certainly desirable and will help to verify our results. In the following subsections, the topics that have been investigated in our study are summarised. 

\subsection{Darwin Instability}
Darwin instability is one of the physical mechanisms proposed to lead a binary to a merger  \citep[e.g. the case of the red nova progenitor V1309~Sco merger][\citeyear{Tylenda2013}]{Tylenda2011}. This instability appears when the binary exhibits orbital angular momentum loss and the total angular momenta exceed the value of one third of the spin of the primary component. Then, synchronous rotation ceases to exist. The binary enters a very unstable phase and the coalescence between the components is highly probable. Therefore, a check was made on the dynamical instability in each system of our sample.

The spin angular momentum can be calculated for both components, while the orbital one can be derived for the entire system, by using the Eq. \ref{eq:5} and Eq. \ref{eq:6}, respectively. 
In order to examine if this instability favors targets in our sample, we calculated the orbital and spin angular momenta. 
First, we consider that the systems in our sample are in synchronous rotation (as expected for contact binary systems) and we assume that the gyration radii are the same for the two components and equal to $k^2=0.06$ according to \citet{Rasio1995}. Then, the ratio of the spin angular momentum to the orbital angular momentum was computed using the Eq. \ref{eq:7}. 

\begin{equation}
\centering
\label{eq:5}
J_{spin} = (k_1^2~M_1~R_1^2 +k_2^2~M_2~ R_2^2)~{\omega}_s
\end{equation}
\begin{equation}
\centering
\label{eq:6}
J_{orb}= \frac{M_1~M_2}{M_1+M_2} a^2~{\omega}_o
\end{equation}

\begin{equation}
\centering
\label{eq:7}
\frac{J_{spin}}{J_{orb}}=k^2 \left( \frac{1+q}{q} \right) \left(\frac{R_1}{a} \right)^2 \left( 1 +q\left(\frac{R_2}{R_1} \right)^2 \right)
\end{equation}

Approximations on the effective radius of each Roche lobe ($r_{L}$) were used, using Eq. \ref{eqeggleton1} for filling the inner Roche lobe and Eq. \ref{eqeggleton2} for filling the outer Roche lobe, according to the expressions mentioned in \cite{Eggleton1983} and \cite{Yakut2005}. These boundaries were used to calculate respective limits of the spin angular momentum to the orbital angular momentum ratio, when the contact binary has filled the inner or the outer Roche lobes. 

\begin{equation}
\centering
\label{eqeggleton1}
    {r_L}_{~inner} = \frac{0.49 q^{2/3}}{0.6q^{2/3}+ln(1+q^{1/3})}
\end{equation}

\begin{equation}
\centering
\label{eqeggleton2}
    {r_L}_{~outer} = \frac{0.49 q^{2/3} + 0.27 q - 0.12 q^{4/3}}{0.6q^{2/3}+ln(1+q^{1/3})}
\end{equation}

\begin{figure}
\centering
\includegraphics[width=8.3cm,scale=1.0,angle=0]{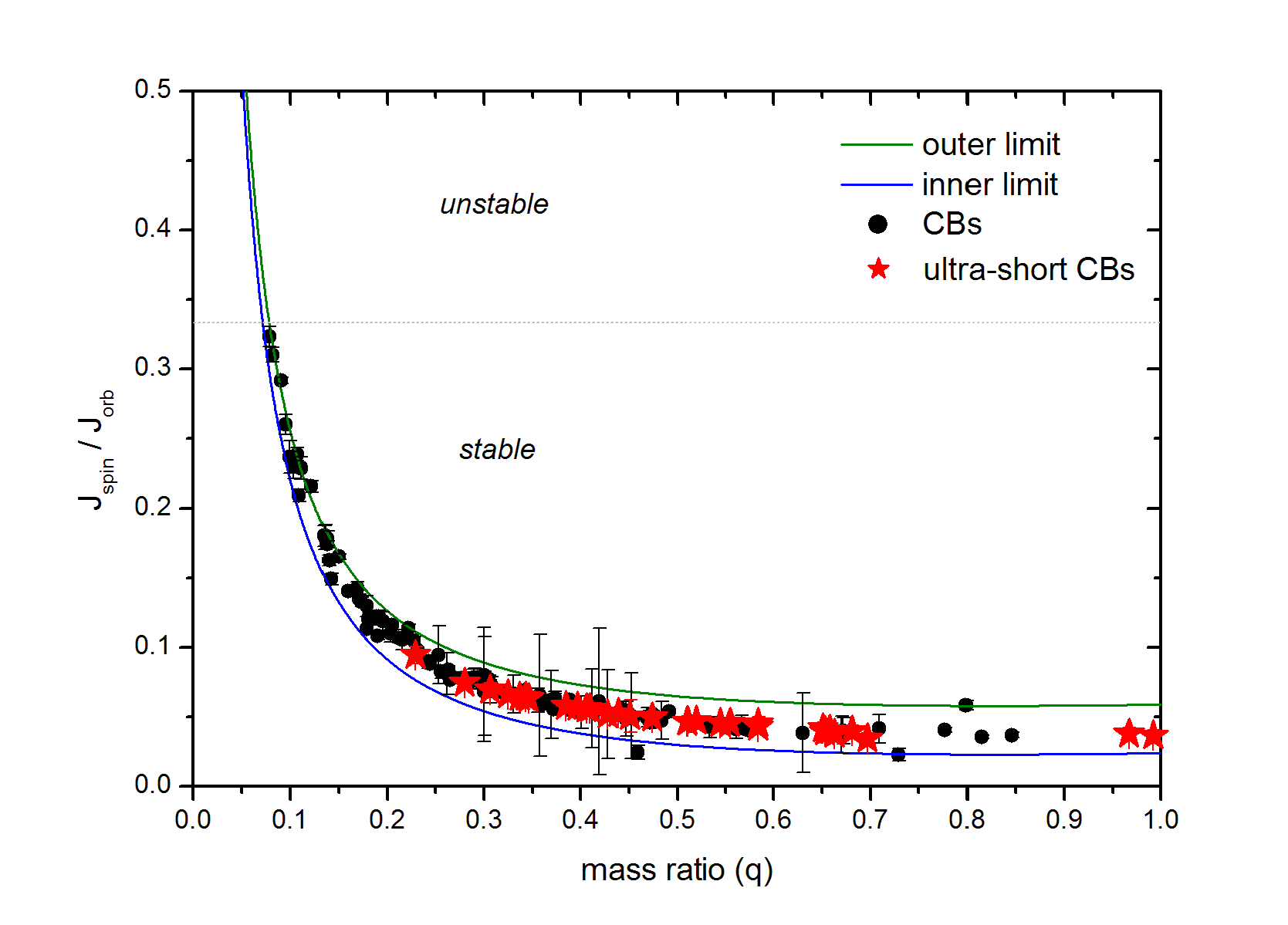}
  \caption{The spin to orbital angular momentum ratio is plotted against the mass ratio $q$. The solid lines indicate the inner and outer theoretical limits for the systems filling the inner and outer Roche lobe, respectively. The filled black points represent the sample of 138 W~UMa-type binaries from \citet{Gazeas2021a}, while the red star markers represent the ultra-short period binary systems of the present study. The uncertainty on the red star markers is smaller than the marker size.}
\label{fig:Darwin}
\end{figure}

In Fig. \ref{fig:Darwin} these theoretical boundaries are depicted for the uniform sample of contact binary systems from \citet{Gazeas2021a} and our sample. A similar investigation was made by \citet{Li2006}, but without accounting for ultra-short contact binaries. Darwin instability favours only systems with extreme mass ratio values close to 0.07 \citep{Li2006} and ultra-short contact binaries do not present extreme values. 
Fig. \ref{fig:Darwin} can be an observational confirmation of the theoretical 
prediction of \cite{Stepien2006,Stepien2012}, that the ultra-short binary systems are evolving with a very slow pace, without having enough time to reach the final coalescence, as this process takes several Gyrs.
A long evolutionary time is needed for ultra-short contact binaries to reach such an extremely small mass ratio and become dynamically unstable. Nevertheless, Darwin instability should take place during the very final stages of the possible merger process, as the binary could not maintain such an unstable orbit and asynchronous rotation.

\subsection{Orbital Period Modulations}
It is found that 40 per cent of the systems in the present study are accompanied by some amount of third light (either due to a close visual companion, or being members of triple or multiple systems). This result is also confirmed by the $O-C$ analysis, where 57 per cent of the entire sample is found to host a third component, causing the eclipse timings to vary through epochs (as seen through the LITE effect). 
The $O-C$ study shows that ultra-short period systems seem to have very stable orbits and they do not show evidence of coalescence. The existence of both negative and positive period modulation provides no conclusive evidence that the ultra-short orbital period systems shrink their orbits. 
The additional component does not seem to affect the orbit, at least in short time scales, which indicates a very slow process and orbital modulation.

The negative parabolic coefficient (i.e. downward parabolic trend in $O-C$ diagrams) is an indication of a shrinking orbit, but not a conclusive evidence. 
A parabolic trend could also be a part of a longer periodic modulation, while the $O-C$ data of the studied systems cover a period of 17-20 yrs in the best case.
This effect could also imply that an even larger fraction of systems containing tertiary components can exist, compared to the fraction that was found in the current or older studies.

This information can only be retrieved, when more data are collected in the forthcoming decades and the time span is increased significantly. Therefore, the results listed in Table \ref{Tab11.OCs} are based only on the currently available data.

\subsection{Magnetic Activity}
The systems in the current sample host cool stellar components of G and K spectral type, which frequently present spotted surfaces due to their magnetic activity. This is confirmed by the asymmetries and the temporal variability of the light curves in 19 out of 30 systems (63 per cent). 
As already mentioned, the Applegate mechanism does not seem to be applicable on the very low mass stars and it cannot explain the periodic modulation of $O-C$ diagrams. Even though most of the contact binaries of our sample include spots on their common envelopes, the Applegate mechanism is inadequate for explaining the orbital period modulation. Hence, the periodic behavior of $O-C$ diagrams are most likely a result of the LITE effect, caused by the existence of additional components orbiting around the systems.


\begin{figure}
\centering
\includegraphics[width=8.4cm,scale=1.0,angle=0]{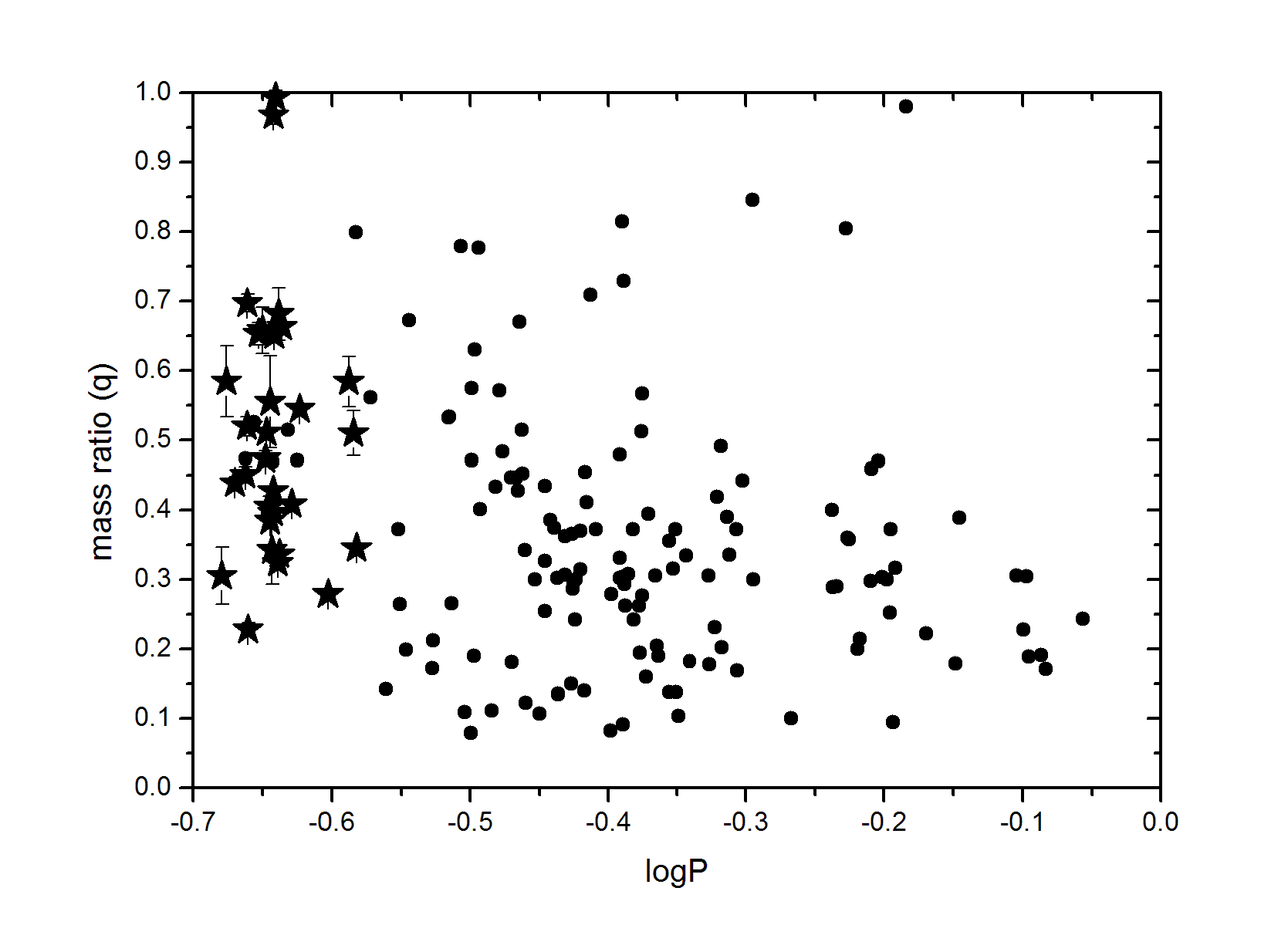}
\includegraphics[width=8.4cm,scale=1.0,angle=0]{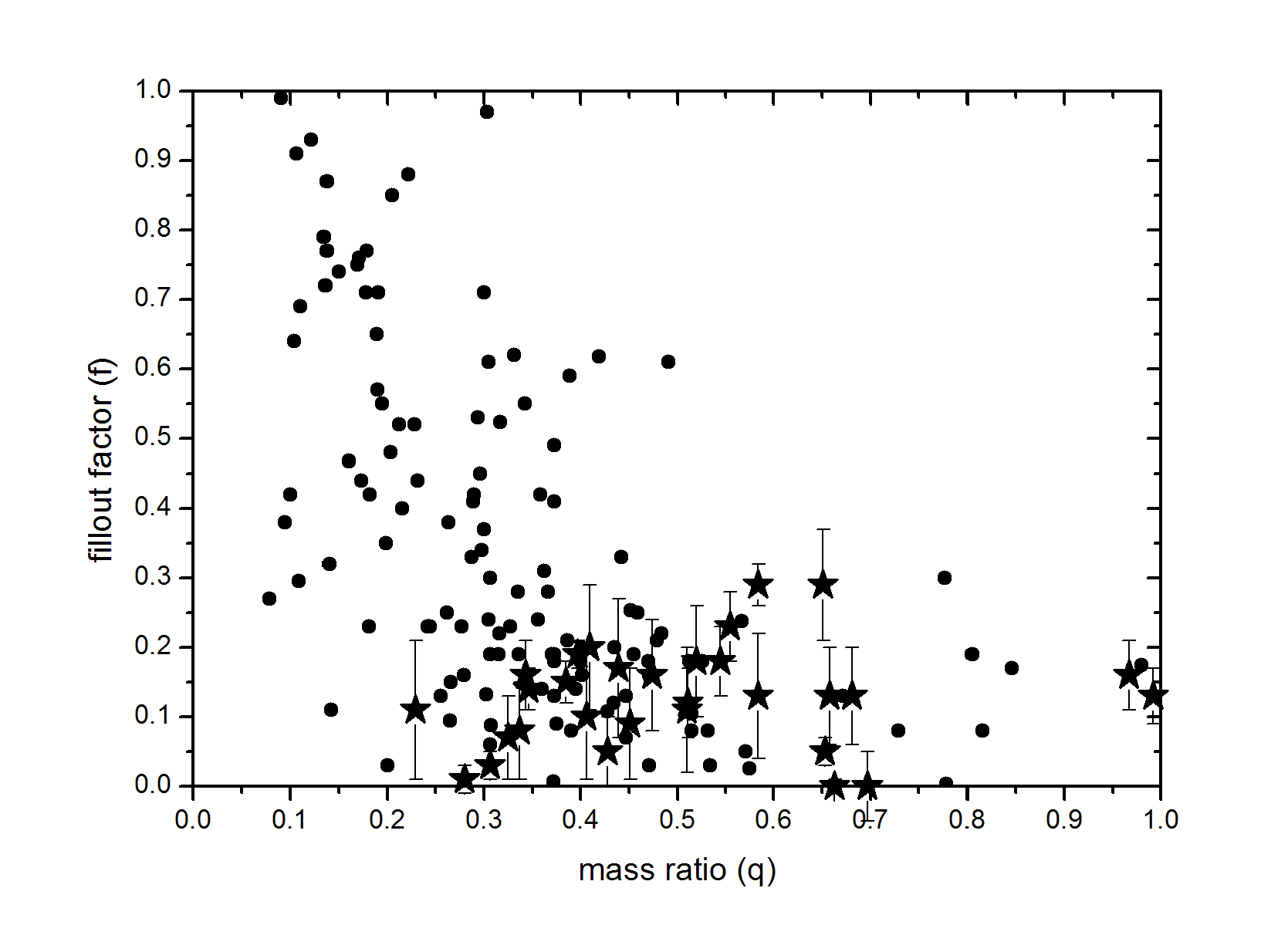}
\includegraphics[width=8.4cm,scale=1.0,angle=0]{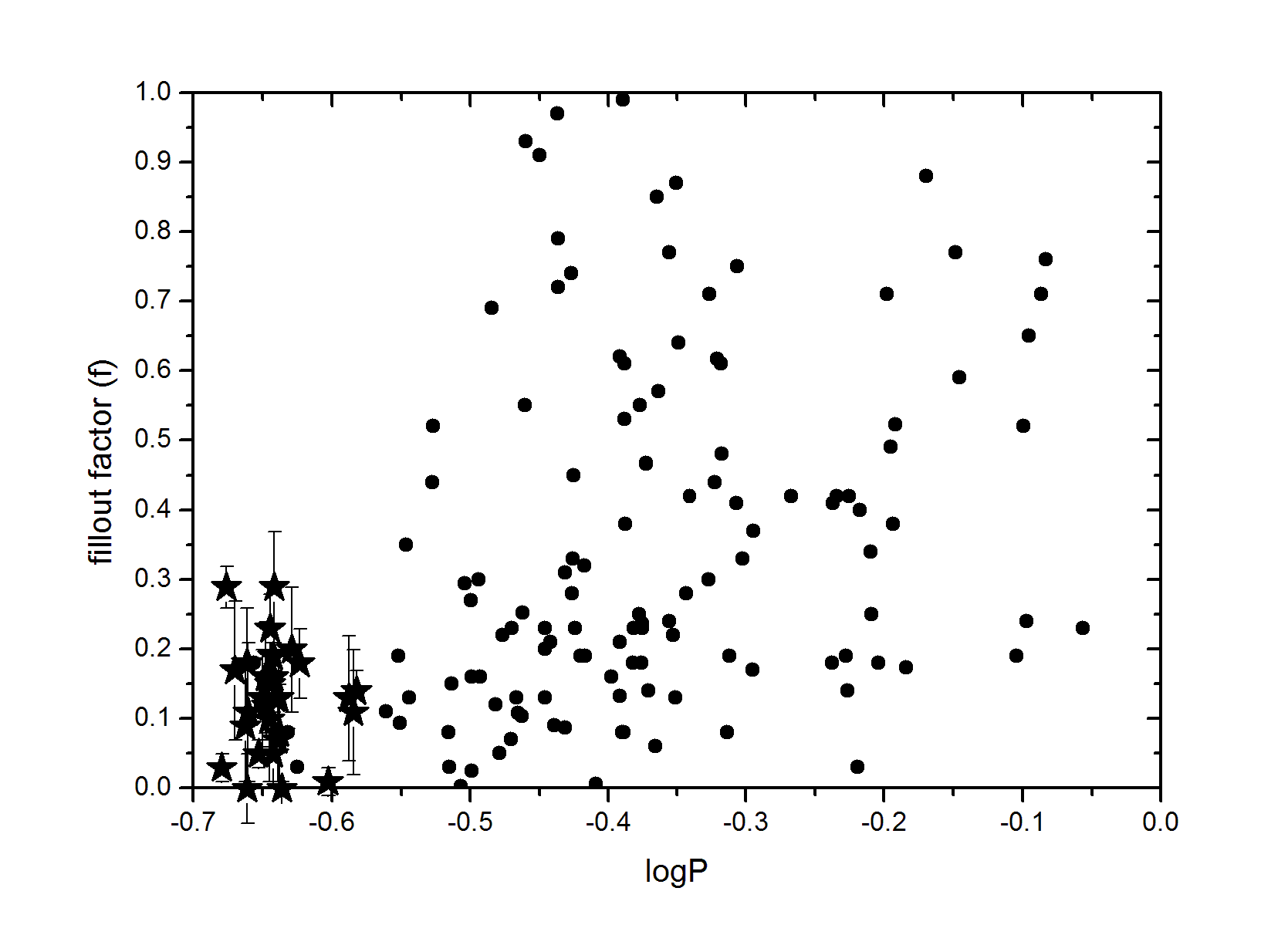}
  \caption{The upper panel shows the distribution of mass ratio as a function of the orbital period, the middle panel shows the distribution of fill-out factor across the mass ratio and the lower panel shows the distribution of fill-out factor across the orbital period. In all panels filled points represent the entire sample of 138 contact binaries, as derived from \textit{W~UMa Programme}, while star markers represent the 30 systems of our sample.}
\label{Fig.fqP}
\end{figure}

\subsection{Absolute Physical and Orbital Parameters}
Ultra-short orbital period contact binaries are among the systems with the smallest and faintest (in terms of size, mass and luminosity) low temperature components. \cite{Stepien2012} showed that these systems host evolved secondaries, which are most probably the product of a mass reversal episode some Gyrs ago.

In this study we found that ultra-short orbital period contact binaries do not reach extremely low mass ratio values and all of them range between the values 0.2 and 1.0 (Fig. \ref{Fig.fqP}). The primary mass ranges between 0.80 and 0.93~$M_{\odot}$ and the secondary mass ranges between 0.19 and 0.82~$M_{\odot}$. The primary's and secondary's radius is also less than 1~$R_{\odot}$, making the ultra-short orbital period contact binaries those with the smallest values of absolute physical parameters, amongst the entire sample of contact binaries known to date.
Our study also showed that the total mass in half of the systems in this sample is below the instability mass limit, as set by \cite{Jiang2012}. Moreover, six of these systems also have an orbital period shorter than 0.22~d.  
Although these systems were considered to have unstable orbits and are heading towards merging, we found that they are very stable, and in an quite early contact phase.

\cite{Gazeas2021a} showed an interesting trend in the fill-out factor parameter, as compared with the orbital period and mass ratio. It appears that systems with shorter orbital periods tend to have shallow contact configurations.
The few well-studied systems (up to that date) with orbital period less than 0.3~d are indeed in shallow contact with fill-out factors less than 25 per cent. 
The present study confirmed the above finding for all 30 systems. Fig. \ref{Fig.fqP} shows the correlation between the orbital and physical parameters, i.e. the mass ratio as a function of the orbital period, the fill-out factor as a function of the mass ratio and the mass ratio as a function of the orbital period.
The $f$ and $q$ parameters are derived from the light curve modeling (Tables in Appendix A). 
According to our results, none of the studied systems appears to be in deep contact configuration. 
Two systems (1SWASP~J220734.47+265528.6 \& 1SWASP~J224747.20-351849.3) appear to be in marginal contact, indicating that they have probably just entered the contact phase of their evolution. 
The low fill-out factor in these systems could plausibly be explained by assuming that the systems are evolutionary young, as suggested by \cite{Stepien2012} and \cite{Li2019}.

\section{Summary and Conclusions}
Summarising, contact binary systems with extremely short orbital periods are rare and only a handful of such systems are known and well studied up to date. Their components follow the MS trend and they are located within the ZAMS and TAMS limits.
The absence of deep contact configuration, the fact that there is no particular preference in orbital period modulation and the presence of very stable orbits in terms of angular momentum and Darwin criteria, lead to the conclusion that ultra-short period systems show no evidence of merging. Low mass-ratio systems seem more promising candidates for giving such an evidence, which could possibly lead towards the detection of red nova progenitors among them. Long-term monitoring of orbital parameters through $O-C$ diagrams could reveal the existence of a possible orbital substantial secular change, which could eventually lead to their merger into single fast-rotating stars. 

\section*{Acknowledgements}
This research is co-financed by Greece and the European Union (European Social Fund-ESF) through the Operational Programme `Human Resources Development, Education and Lifelong Learning' in the context of the project `Strengthening Human Resources Research Potential via Doctorate Research' (MIS-5000432), implemented by the State Scholarships Foundation (IKY).
This work utilizes data from the robotic and remotely controlled telescope of the University of Athens Observatory (UOAO), located at the National and Kapodistrian University of Athens, Greece. Part of this work is also based on observations obtained with the 1.2~m Kryoneri telescope, located at Corinthia, Greece and the 2.3~m Aristarchos telescope, located at Helmos Observatory, Achaia, Greece. Both telescopes are operated by the Institute for Astronomy, Astrophysics, Space Applications and Remote Sensing of the National Observatory of Athens.
The authors wish to thank Prof. A. Norton who reviewed the current manuscript and gave valuable comments that improved our work, as well as the collaborators at the observing facilities for their support and the telescope time allocation.

\section*{Data Availability}
The data underlying this article are available upon request to the corresponding author.

\bibliographystyle{mnras}
\bibliography{main}

\appendix

\section{Modeling Results}

\begin{table*}
\centering
\caption{Results derived from light curve modeling.}
\label{Tab4.res1}
\begin{flushleft}
\begin{small}
\begin{tabular}{lcccccc}
\hline
Parameters	&		1SWASP 		&		1SWASP 		&		1SWASP		&		1SWASP 		&		1SWASP 		\\
	&		J030749.87-365201.7		&		J040615.79-425002.3		&		J044132.96+440613.7		&		J050904.45-074144.4		&		J052926.88+461147.5		\\
\hline																					
Fill-out factor	&	$	23\pm 5 \%	$	&	$	5 \pm 2 \%	$	&	$	29 \pm 7 \%	$	&	$	7 \pm 6 \%	$	&	$	15 \pm 3 \%	$	\\
$i$ [deg]               	&	$	84.4\pm 0.9	$	&	$	72.6\pm 0.9	$	&	$	88.5\pm 0.7	$	&	$	82.9\pm 0.5	$	&	$	88.7\pm 0.8	$	\\
$T_{\rm 1}$ [K] 	&	$	4700^{*}	$	&	$	4900^{*}	$	&	$	5350^{*}	$	&	$	5000^{*}	$	&	$	5400^{*}	$	\\
$T_{\rm 2}$ [K] 	&	$	4578\pm 67	$	&	$	4559\pm 75	$	&	$	5056\pm 26	$	&	$	4985\pm 19	$	&	$	5328\pm 13	$	\\
$\Omega_{\rm 1}$=$\Omega_{\rm 2}$        	&	$	2.904\pm 0.017	$	&	$	4.546\pm 0.053	$	&	$	4.408\pm 0.039	$	&	$	2.507\pm 0.012	$	&	$	2.613\pm 0.007	$	\\
$q_{ph}$  	&	$	0.555\pm 0.066	$	&	$	1.532\pm 0.037	$	&	$	1.535\pm 0.022	$	&	$	0.325\pm 0.007	$	&	$	0.385\pm 0.003	$	\\
$L_{1}~(B)$             	&	$	7.873\pm 0.039	$	&	$	5.975\pm 0.063	$	&	$	5.557\pm 0.035	$	&	$	8.560\pm 0.048	$	&	$	8.411\pm 0.038	$	\\
$L_{1}~(V)$             	&	$	7.891\pm 0.038	$	&	$	5.884\pm 0.061	$	&	$	5.418\pm 0.048	$	&	$	8.653\pm 0.046	$	&	$	8.444\pm 0.034	$	\\
$L_{1}~(R)$             	&	$	8.002\pm 0.038	$	&	$	5.803\pm 0.060	$	&	$	5.292\pm 0.043	$	&	$	8.718\pm 0.045	$	&	$	8.517\pm 0.033	$	\\
$L_{1}~(I)$             	&	$	7.864\pm 0.037	$	&	$	5.641\pm 0.058	$	&	$	5.040\pm 0.028	$	&	$	8.831\pm 0.044	$	&	$	8.462\pm 0.029	$	\\
$L_{2}~(B)$         	&	$	3.937^{a}	$	&	$	5.599^{a}	$	&	$	5.887^{a}	$	&	$	3.031^{a}	$	&	$	3.309^{a}	$	\\
 $L_{2}~(V)$         	&	$	4.010^{a}	$	&	$	5.744^{a}	$	&	$	5.977^{a}	$	&	$	3.067^{a}	$	&	$	3.351^{a}	$	\\
 $L_{2}~(R)$         	&	$	4.128^{a}	$	&	$	5.892^{a}	$	&	$	6.029^{a}	$	&	$	3.093^{a}	$	&	$	3.405^{a}	$	\\
 $L_{2}~(I)$         	&	$	4.168^{a}	$	&	$	6.161^{a}	$	&	$	6.044^{a}	$	&	$	3.141^{a}	$	&	$	3.422^{a}	$	\\
$l_{3}~(B)$             	&	$	0	$	&	$	0	$	&	$	0.006\pm 0.004	$	&	$	0	$	&	$	0	$	\\
$l_{3}~(V)$             	&	$	0	$	&	$	0	$	&	$	0.028\pm 0.008	$	&	$	0	$	&	$	0	$	\\
$l_{3}~(R)$             	&	$	0	$	&	$	0	$	&	$	0.041\pm 0.008	$	&	$	0	$	&	$	0	$	\\
$l_{3}~(I)$             	&	$	0	$	&	$	0	$	&	$	0.069\pm 0.006	$	&	$	0	$	&	$	0	$	\\
\hline																					
r$_{1}~^{side}$         	&	$	0.4460	$	&	$	0.4177	$	&	$	0.4362	$	&	$	0.4859	$	&	$	0.4740	$	\\
r$_{2}~^{side}$         	&	$	0.3374	$	&	$	0.3391	$	&	$	0.3569	$	&	$	0.2818	$	&	$	0.2995	$	\\
\hline																					
co-latitude             	&	$	164.4\pm 1.4	$	&	$	91.6\pm 9.1	$	&	$	120.5\pm 10.4	$	&	$	162.4\pm 2.2	$	&	$	122.9\pm 9.7	$	\\
longitude               	&	$	47.4\pm 1.9	$	&	$	174.6\pm 3.7	$	&	$	131.2\pm 5.6	$	&	$	49.3\pm 3.7	$	&	$	121.7\pm 4.3	$	\\
radius                  	&	$	49.2\pm 0.9	$	&	$	21.1\pm 3.8	$	&	$	16.3\pm 2.6	$	&	$	42.2\pm 2.2	$	&	$	14.5\pm 3.0	$	\\
temp. factor            	&	$	0.163\pm 0.031	$	&	$	0.828\pm 0.071	$	&	$	0.757\pm 0.064	$	&	$	0.702\pm 0.064	$	&	$	0.763\pm 0.090	$	\\
\hline
\multicolumn{6}{l}{(*): Fixed parameter, (a): Calculated according to L$_{1}$} \\
\hline
\end{tabular}
\end{small}
\end{flushleft}

\end{table*}

\begin{table*}
\centering
\caption{Results derived from light curve modeling (continued).}
\label{Tab5.res2}
\begin{flushleft}
\begin{small}
\begin{tabular}{lcccccc}
\hline

Parameters	&		1SWASP 		&		1SWASP		&		1SWASP 		&		1SWASP		&		1SWASP 		\\
	&		J055416.98+442534.0		&		J080150.03+471433.8		&		J092328.76+435044.8		&		J092754.99-391053.4		&		J093010.78+533859.5		\\
\hline																					
Fill-out factor	&	$	11\pm 10\%	$	&	$	9 \pm 8\%	$	&	$	20\pm 9\%	$	&	$	12 \pm 5 \%	$	&	$	19 \pm 2 \%	$	\\
$i$ [deg]               	&	$	75.4\pm 0.6	$	&	$	86.3\pm 1.2	$	&	$	83.2\pm 0.7	$	&	$	72.1\pm 1.9	$	&	$	88.0\pm 1.3	$	\\
$T_{\rm 1}$ [K] 	&	$	5250^{*} 	$	&	$	4650^{*} 	$	&	$	5800^{*} 	$	&	$	5400^{*}	$	&	$	4700^{*}	$	\\
$T_{\rm 2}$ [K] 	&	$	5266\pm 37	$	&	$	4720\pm 12	$	&	$	5788\pm 32	$	&	$	5172\pm 11	$	&	$	4800\pm 36	$	\\
$\Omega_{\rm 1}$=$\Omega_{\rm 2}$        	&	$	2.288\pm 0.015	$	&	$	2.755\pm 0.023	$	&	$	2.646\pm 0.022	$	&	$	2.862\pm 0.016	$	&	$	2.627 \pm 0.004	$	\\
$q_{ph}$  	&	$	0.229\pm 0.010 	$	&	$	0.451\pm 0.012 	$	&	$	0.409\pm 0.011	$	&	$	0.511\pm 0.074	$	&	$	0.397^{**}	$	\\
$L_{1}~(U)$             	&	$	-	$	&	$	-	$	&	$	-	$	&	$	7.972\pm 0.053	$	&	$	-	$	\\
$L_{1}~(B)$             	&	$	9.318\pm 0.104	$	&	$	7.565\pm 0.141	$	&	$	8.245\pm 0.089	$	&	$	8.130\pm 0.056	$	&	$	2.384\pm 0.043	$	\\
$L_{1}~(V)$             	&	$	9.521\pm 0.099	$	&	$	7.644\pm 0.149	$	&	$	8.256\pm 0.083	$	&	$	8.186\pm 0.055	$	&	$	2.479\pm 0.043	$	\\
$L_{1}~(R)$             	&	$	9.524\pm 0.096	$	&	$	7.474\pm 0.157	$	&	$	8.207\pm 0.079	$	&	$	8.150\pm 0.052	$	&	$	2.556\pm 0.044	$	\\
$L_{1}~(I)$             	&	$	9.629\pm 0.093	$	&	$	7.546\pm 0.155	$	&	$	8.170\pm 0.071	$	&	$	-	$	&	$	2.661\pm 0.046	$	\\
$L_{2}~(U)$         	&	$	-	$	&	$	-	$	&	$	-	$	&	$	3.435^{a}	$	&	$	-	$	\\
$L_{2}~(B)$         	&	$	2.512^{a}	$	&	$	4.036^{a}	$	&	$	3.664^{a}	$	&	$	3.490^{a}	$	&	$	1.169^{a}	$	\\
 $L_{2}~(V)$         	&	$	2.560^{a}	$	&	$	4.043^{a}	$	&	$	3.674^{a}	$	&	$	3.620^{a}	$	&	$	1.203^{a}	$	\\
 $L_{2}~(R)$         	&	$	2.555^{a}	$	&	$	3.920^{a}	$	&	$	3.656^{a}	$	&	$	3.692^{a}	$	&	$	1.225^{a}	$	\\
 $L_{2}~(I)$         	&	$	2.575^{a}	$	&	$	3.897^{a}	$	&	$	3.644^{a}	$	&	$	-	$	&	$	1.249^{a}	$	\\
$l_{3}~(U)$             	&	$	-	$	&	$	-	$	&	$	-	$	&	$	0	$	&	$	-	$	\\
$l_{3}~(B)$             	&	$	0	$	&	$	0.005\pm 0.010	$	&	$	0	$	&	$	0	$	&	$	0.702\pm 0.045	$	\\
$l_{3}~(V)$             	&	$	0	$	&	$	0.005\pm 0.011	$	&	$	0	$	&	$	0	$	&	$	0.694\pm 0.046	$	\\
$l_{3}~(R)$             	&	$	0	$	&	$	0.040 \pm 0.012	$	&	$	0	$	&	$	0	$	&	$	0.685\pm 0.047	$	\\
$l_{3}~(I)$             	&	$	0	$	&	$	0.037 \pm 0.012	$	&	$	0	$	&	$	-	$	&	$	0.676\pm 0.049	$	\\
\hline																					
r$_{1}~^{side}$         	&	$	0.5215	$	&	$	0.4557	$	&	$	0.4717	$	&	$	0.4454	$	&	$	0.4736	$	\\
r$_{2}~^{side}$         	&	$	0.2566	$	&	$	0.3097	$	&	$	0.3077	$	&	$	0.3221	$	&	$	0.3044	$	\\
\hline																					
co-latitude             	&	$	157.0 \pm 10.5	$	&	$	-	$	&	$	150.7\pm 3.0	$	&	$	-	$	&	$	92.1\pm 1.3	$	\\
longitude               	&	$	265.2 \pm 5.5	$	&	$	-	$	&	$	115.7\pm 2.6	$	&	$	-	$	&	$	72.1\pm 0.5	$	\\
radius                  	&	$	24.0 \pm 5.1	$	&	$	-	$	&	$	20.2\pm 5.7	$	&	$	-	$	&	$ 	19.7\pm 0.2	$	\\
temp. factor            	&	$	0.208 \pm 0.051	$	&	$	-	$	&	$	0.360\pm 0.084	$	&	$	-	$	&	$	0.846\pm 0.001	$	\\

\hline
\multicolumn{6}{l}{(*): Fixed parameter, (**) Fixed mass ratio from spectroscopy \citep{Lohr2015a}, (a): Calculated according to L$_{1}$} \\
\hline
\end{tabular}
\end{small}
\end{flushleft}
\end{table*}

\begin{table*}
\centering
\caption{Results derived from light curve modeling (continued).}
\label{Tab6.res3}
\begin{flushleft}
\begin{small}
\begin{tabular}{lcccccc}
\hline

Parameters	&		1SWASP 		&		1SWASP 		&		1SWASP 		&		1SWASP 		&		2MASS 		\\
	&		J114929.22-423049.0		&		J121906.35-240056.9		&		J133105.91+121538.0		&		J150822.80-054236.9		&		J15165453+0048263		\\
\hline																					
Fill-out factor	&	$	16 \pm 5\%	$	&	$	10 \pm 9 \%	$	&	$	18 \pm 8 \%	$	&	$	11 \pm 9 \%	$	&	$	29 \pm 4 \%	$	\\
$i$ [deg]               	&	$	79.6\pm 0.2	$	&	$	73.0\pm 1.2	$	&	$	75.0\pm 0.1	$	&	$	86.8\pm 0.5	$	&	$	63.7\pm 0.9	$	\\
$T_{\rm 1}$ [K] 	&	$	4150^{*}	$	&	$	4650^{*}	$	&	$	5150^{*} 	$	&	$	5150^{*}	$	&	$	6150^{*}	$	\\
$T_{\rm 2}$ [K] 	&	$	4075\pm 34	$	&	$	4335\pm 99	$	&	$	4845\pm 15	$	&	$	5130\pm 16	$	&	$	6095\pm 71	$	\\
$\Omega_{\rm 1}$=$\Omega_{\rm 2}$        	&	$	2.525\pm 0.010	$	&	$	5.835\pm 0.255	$	&	$	5.032\pm 0.069	$	&	$	2.860\pm 0.029	$	&	$	2.933\pm 0.010	$	\\
$q_{ph}$  	&	$	0.343\pm 0.049	$	&	$	2.465\pm 0.283	$	&	$	1.923\pm 0.051	$	&	$	0.510^{**}	$	&	$	0.584\pm 0.051	$	\\
$L_{1}~(U)$             	&	$	-	$	&	$	-	$	&	$	-	$	&	$	-	$	&	$	6.902\pm 0.422	$	\\
$L_{1}~(B)$             	&	$	8.861\pm 0.035	$	&	$	5.039\pm 0.065	$	&	$	5.136\pm 0.064	$	&	$	7.807\pm 0.089	$	&	$	6.998 \pm 0.305	$	\\
$L_{1}~(V)$             	&	$	8.853\pm 0.033	$	&	$	4.894\pm 0.065	$	&	$	5.071\pm 0.064	$	&	$	7.664\pm 0.095	$	&	$	7.096\pm 0.300	$	\\
$L_{1}~(R)$             	&	$	8.871\pm 0.032	$	&	$	4.711\pm 0.067	$	&	$	4.975\pm 0.064	$	&	$	7.686\pm 0.100	$	&	$	7.072\pm 0.293	$	\\
$L_{1}~(I)$             	&	$	8.782\pm 0.032	$	&	$	4.533\pm 0.063	$	&	$	4.850\pm 0.064	$	&	$	7.766\pm 0.098	$	&	$	-	$	\\
$L_{2}~(U)$         	&	$	-	$	&	$	-	$	&	$	-	$	&	$	-	$	&	$	4.085^{a}	$	\\
$L_{2}~(B)$         	&	$	2.821^{a}	$	&	$	6.858^{a}	$	&	$	6.432^{a}	$	&	$	4.152^{a}	$	&	$	4.177^{a}	$	\\
 $L_{2}~(V)$         	&	$	2.970^{a}	$	&	$	7.089^{a}	$	&	$	6.602^{a}	$	&	$	4.085^{a}	$	&	$	4.258^{a}	$	\\
 $L_{2}~(R)$         	&	$	3.075^{a}	$	&	$	7.179^{a}	$	&	$	6.700^{a}	$	&	$	4.106^{a}	$	&	$	4.265^{a}	$	\\
 $L_{2}~(I)$         	&	$	3.117^{a}	$	&	$	7.476^{a}	$	&	$	6.919^{a}	$	&	$	4.163^{a}	$	&	$	-	$	\\
$l_{3}~(U)$             	&	$	-	$	&	$	-	$	&	$	-	$	&	$	-	$	&	$	0	$	\\
$l_{3}~(B)$             	&	$	0	$	&	$	0	$	&	$	0	$	&	$	0.005\pm 0.001	$	&	$	0	$	\\
$l_{3}~(V)$             	&	$	0	$	&	$	0	$	&	$	0	$	&	$	0.011\pm 0.004	$	&	$	0	$	\\
$l_{3}~(R)$             	&	$	0	$	&	$	0	$	&	$	0	$	&	$	0.016\pm 0.007	$	&	$	0	$	\\
$l_{3}~(I)$             	&	$	0	$	&	$	0	$	&	$	0	$	&	$	0.010\pm 0.007	$	&	$	-	$	\\
\hline																					
r$_{1}~^{side}$         	&	$	0.4860	$	&	$	0.4664	$	&	$	0.4485	$	&	$	0.4455	$	&	$	0.4458	$	\\
r$_{2}~^{side}$         	&	$	0.2909	$	&	$	0.3014	$	&	$	0.3281	$	&	$	0.3217	$	&	$	0.3468	$	\\
\hline																					
co-latitude             	&	$	-	$	&	$	164.2\pm 9.5	$	&	$	-	$	&	$	171.5\pm 1.2	$	&	$	136.0\pm 2.4	$	\\
longitude               	&	$	-	$	&	$	206.2\pm 3.6	$	&	$	-	$	&	$	39.1\pm 4.6	$	&	$	230.7\pm 1.7	$	\\
radius                  	&	$	-	$	&	$	72.5\pm 3.7	$	&	$ 	-	$	&	$	40.4\pm 1.5	$	&	$	13.2\pm 1.3	$	\\
temp. factor            	&	$	-	$	&	$	0.861\pm 0.017	$	&	$	-	$	&	$	0.572\pm 0.044	$	&	$	0.653\pm 0.087	$	\\
\hline
\multicolumn{6}{l}{(*): Fixed parameter, (**): Fixed mass ratio from spectroscopy \citep{Lohr2014}, (a): Calculated according to L$_{1}$} \\
\hline
\end{tabular}
\end{small}
\end{flushleft}
\end{table*}

\begin{table*}
\centering
\caption{Results derived from light curve modeling (continued).}
\label{Tab7.res4}
\begin{flushleft}
\begin{small}
\begin{tabular}{lcccccc}
\hline

Parameters	&		1SWASP 		&		1SWASP 		&		1SWASP 		&		1SWASP 		&		1SWASP		\\
	&		J161335.80-284722.2		&		J170240.07+151123.5 		&		J173003.21+344509.4		&		J173828.46+111150.2		&		J174310.98+432709.6		\\
\hline																					
Fill-out factor	&	$	13 \pm 7\%	$	&	$	14 \pm 3 \%	$	&	$	13 \pm 7 \%	$	&	$	1 \pm 2\%	$	&	$	13 \pm 9\%	$	\\
$i$ [deg]               	&	$	78.3\pm 1.5	$	&	$	89.4 \pm 0.4	$	&	$	67.7\pm 0.7	$	&	$	67.9\pm 0.7	$	&	$	77.2 \pm 0.6	$	\\
$T_{\rm 1}$ [K] 	&	$	4550^{*}	$	&	$	5000^{*} 	$	&	$	4700^{*} 	$	&	$	5250^{*}	$	&	$	5300^{*} 	$	\\
$T_{\rm 2}$ [K] 	&	$	4347\pm 40	$	&	$	5307\pm 17	$	&	$	4475\pm 18	$	&	$	5065\pm 21	$	&	$	4996\pm 18	$	\\
$\Omega_{\rm 1}$=$\Omega_{\rm 2}$        	&	$	4.402\pm 0.039	$	&	$	2.537\pm 0.007	$	&	$	3.120\pm 0.026	$	&	$	2.431 \pm 0.004	$	&	$	2.990\pm 0.031	$	\\
$q_{ph}$  	&	$	1.468\pm 0.281	$	&	$	0.346\pm 0.003	$	&	$	0.658\pm 0.033	$	&	$	0.280\pm 0.001	$	&	$	0.584\pm 0.036	$	\\
$L_{1}~(B)$             	&	$	5.729\pm 0.373	$	&	$	7.474\pm 0.036	$	&	$	7.618\pm 0.241	$	&	$	9.303\pm 0.052	$	&	$	8.192\pm 0.152	$	\\
$L_{1}~(V)$             	&	$	5.619\pm 0.370	$	&	$	7.763\pm 0.035	$	&	$	6.596\pm 0.239	$	&	$	9.393\pm 0.043	$	&	$	7.907\pm 0.159	$	\\
$L_{1}~(R)$             	&	$	5.438\pm 0.350	$	&	$	7.975\pm 0.061	$	&	$	6.525\pm 0.238	$	&	$	9.508\pm 0.039	$	&	$	7.818\pm 0.159	$	\\
$L_{1}~(I)$             	&	$	5.224\pm 0.337	$	&	$	8.133\pm 0.055	$	&	$	6.086\pm 0.231	$	&	$	9.535\pm 0.029	$	&	$	7.408\pm 0.149	$	\\
$L_{2}~(B)$         	&	$	5.747^{a}	$	&	$	4.035^{a}	$	&	$	3.756^{a}	$	&	$	2.339^{a}	$	&	$	3.597^{a}	$	\\
 $L_{2}~(V)$         	&	$	5.895^{a}	$	&	$	4.013^{a}	$	&	$	3.364^{a}	$	&	$	2.422^{a}	$	&	$	3.620^{a}	$	\\
 $L_{2}~(R)$         	&	$	5.911^{a}	$	&	$	3.982^{a}	$	&	$	3.429^{a}	$	&	$	2.503^{a}	$	&	$	3.703^{a}	$	\\
 $L_{2}~(I)$         	&	$	5.986^{a}	$	&	$	3.845^{a}	$	&	$	3.368^{a}	$	&	$	2.593^{a}	$	&	$	3.703^{a}	$	\\
$l_{3}~(B)$             	&	$	0.000\pm 0.002	$	&	$	0	$	&	$	0.006\pm 0.001	$	&	$	0	$	&	$	0.000\pm 0.001	$	\\
$l_{3}~(V)$             	&	$	0.012\pm 0.002	$	&	$	0	$	&	$	0.122\pm 0.027	$	&	$	0	$	&	$	0.021\pm 0.012	$	\\
$l_{3}~(R)$             	&	$	0.023\pm 0.003	$	&	$	0	$	&	$	0.132\pm 0.027	$	&	$	0	$	&	$	0.040\pm 0.012	$	\\
$l_{3}~(I)$             	&	$	0.041\pm 0.003	$	&	$	0	$	&	$	0.182\pm 0.026	$	&	$	0	$	&	$	0.069\pm 0.009	$	\\
\hline																					
r$_{1}~^{side}$         	&	$	0.4199	$	&	$	0.4835	$	&	$	0.4227	$	&	$	0.4943	$	&	$	0.4337	$	\\
r$_{2}~^{side}$         	&	$	0.3489	$	&	$	0.2902	$	&	$	0.3455	$	&	$	0.2646	$	&	$	0.3346	$	\\
\hline																					
co-latitude             	&	$	-	$	&	$	-	$	&	$	121.2\pm 10.9	$	&	$	127.8\pm 6.4	$	&	$	172.3\pm 2.7	$	\\
longitude               	&	$	-	$	&	$	-	$	&	$	281.6\pm 6.5	$	&	$	254.5\pm 3.1	$	&	$	318.1\pm 6.6	$	\\
radius                  	&	$	-	$	&	$	-	$	&	$	23.4\pm 6.5	$	&	$	16.5\pm 0.3	$	&	$	53.6 \pm 2.6	$	\\
temp. factor            	&	$	-	$	&	$	-	$	&	$	0.838\pm 0.057	$	&	$	0.704\pm 0.023	$	&	$	0.738 \pm 0.068	$	\\

\hline
\multicolumn{6}{l}{(*): Fixed parameter, (a): Calculated according to L$_{1}$} \\
\hline
\end{tabular}
\end{small}
\end{flushleft}
\end{table*}

\begin{table*}
\centering
\caption{Results derived from light curve modeling (continued).}
\label{Tab8.res5}
\begin{flushleft}
\begin{small}
\begin{tabular}{lcccccc}
\hline

Parameters	&		1SWASP		&		1SWASP		&		2MASS		&		2MASS		&		1SWASP		\\
	&		J180947.64+490255.0		&		J195900.31-252723.1		&		J21031997+0209339		&		J21042404+0731381		&		J212454.61+203030.8		\\
\hline																					
Fill-out factor	&	$	16 \pm 5\%	$	&	$	18 \pm 5 \%	$	&	$	13 \pm 4\%	$	&	$	3 \pm 2\%	$	&	$	5 \pm 5\%	$	\\
$i$ [deg]               	&	$	82.3\pm 0.2	$	&	$	88.9\pm 2.3	$	&	$	82.4\pm 0.5	$	&	$	78.9\pm 0.7	$	&	$	89.0\pm 2.1	$	\\
$T_{\rm 1}$ [K] 	&	$	4050^{*}	$	&	$	5400^{*}	$	&	$	4400^{*}	$	&	$	4800^{*}	$	&	$	5250^{*}	$	\\
$T_{\rm 2}$ [K] 	&	$	3833\pm 24	$	&	$	5244\pm 96	$	&	$	4179\pm 20	$	&	$	4772\pm 24	$	&	$	5264\pm 70	$	\\
$\Omega_{\rm 1}$=$\Omega_{\rm 2}$        	&	$	3.614\pm 0.027	$	&	$	4.915\pm 0.031	$	&	$	3.666\pm 0.020	$	&	$	2.471\pm 0.003	$	&	$	2.721\pm 0.014	$	\\
$q_{ph}$  	&	$	0.967\pm 0.016	$	&	$	1.836\pm 0.022	$	&	$	0.992\pm 0.011	$	&	$	0.306\pm 0.041	$	&	$	0.428\pm 0.007	$	\\
$L_{1}~(B)$             	&	$	7.940\pm 0.039	$	&	$	4.557\pm 0.037	$	&	$	6.939\pm 0.037	$	&	$	8.886\pm 0.139	$	&	$	8.167\pm 0.406	$	\\
$L_{1}~(V)$             	&	$	6.940\pm 0.042	$	&	$	4.666\pm 0.036	$	&	$	6.770\pm 0.036	$	&	$	8.475\pm 0.189	$	&	$	8.049\pm 0.409	$	\\
$L_{1}~(R)$             	&	$	6.481\pm 0.043	$	&	$	4.581\pm 0.034	$	&	$	6.696\pm 0.035	$	&	$	8.154\pm 0.179	$	&	$	7.992\pm 0.406	$	\\
$L_{1}~(I)$             	&	$	6.395\pm 0.044	$	&	$	4.573\pm 0.031	$	&	$	6.591\pm 0.034	$	&	$	7.764\pm 0.139	$	&	$	8.082\pm 0.421	$	\\
$L_{2}~(B)$         	&	$	3.531^{a}	$	&	$	6.693^{a}	$	&	$	4.492^{a}	$	&	$	2.921^{a}	$	&	$	3.812^{a}	$	\\
 $L_{2}~(V)$         	&	$	4.575^{a}	$	&	$	6.994^{a}	$	&	$	4.684^{a}	$	&	$	2.788^{a}	$	&	$	3.749^{a}	$	\\
 $L_{2}~(R)$         	&	$	5.077^{a}	$	&	$	6.981^{a}	$	&	$	4.864^{a}	$	&	$	2.694^{a}	$	&	$	3.716^{a}	$	\\
 $L_{2}~(I)$         	&	$	5.383^{a}	$	&	$	7.151^{a}	$	&	$	5.104^{a}	$	&	$	2.573^{a}	$	&	$	3.748^{a}	$	\\
$l_{3}~(B)$             	&	$	0	$	&	$	0.035\pm 0.001	$	&	$	0	$	&	$	0.009\pm 0.008	$	&	$	0.001\pm 0.001	$	\\
$l_{3}~(V)$             	&	$	0	$	&	$	0.041\pm 0.002	$	&	$	0	$	&	$	0.061\pm 0.014	$	&	$	0.017\pm 0.002	$	\\
$l_{3}~(R)$             	&	$	0	$	&	$	0.046\pm 0.003	$	&	$	0	$	&	$	0.098\pm 0.014	$	&	$	0.036\pm 0.002	$	\\
$l_{3}~(I)$             	&	$	0	$	&	$	0.046\pm 0.003	$	&	$	0	$	&	$	0.141\pm 0.008	$	&	$	0.030\pm 0.002	$	\\
\hline																					
r$_{1}~^{side}$         	&	$	0.3901	$	&	$	0.4436	$	&	$	0.3858	$	&	$	0.4899	$	&	$	0.4581	$	\\
r$_{2}~^{side}$         	&	$	0.3839	$	&	$	0.3316	$	&	$	0.3844	$	&	$	0.2743	$	&	$	0.3027	$	\\
\hline																					
co-latitude             	&	$	-	$	&	$	-	$	&	$	-	$	&	$	136.9\pm 10.2	$	&	$	88.8\pm 9.6	$	\\
longitude               	&	$	-	$	&	$	-	$	&	$	-	$	&	$	16.6\pm 4.4	$	&	$	354.9\pm 0.7	$	\\
radius                  	&	$	-	$	&	$	-	$	&	$	-	$	&	$	26.6\pm 5.9	$	&	$	40.0\pm 0.7	$	\\
temp. factor            	&	$	-	$	&	$	-	$	&	$	-	$	&	$	0.672\pm 0.098	$	&	$	0.963\pm 0.015	$	\\
\hline
\multicolumn{6}{l}{(*): Fixed parameter, (a): Calculated according to L$_{1}$} \\
\hline
\end{tabular}
\end{small}
\end{flushleft}
\end{table*}

\begin{table*}
\centering
\caption{Results derived from light curve modeling (continued).}
\label{Tab9.res6}
\begin{flushleft}
\begin{small}
\begin{tabular}{lcccccc}
\hline

Parameters	&		1SWASP		&		1SWASP		&		1SWASP		&		1SWASP		&		1SWASP		\\
	&		J212808.86+151622.0		&		J220734.47+265528.6		&		J221058.82+251123.4		&		J224747.20-351849.3		&		J232610.13-294146.6		\\
\hline																					
Fill-out factor	&	$	16 \pm 8\%	$	&	$	0 \pm 1 \%	$	&	$	17 \pm 10\%	$	&	$	0 \pm 5\%	$	&	$	8 \pm 7 \%	$	\\
$i$ [deg]               	&	$	75.4\pm 0.3	$	&	$	70.5\pm 0.2	$	&	$	78.8\pm 0.2	$	&	$	58.4\pm 1.1	$	&	$	75.9\pm 0.9	$	\\
$T_{\rm 1}$ [K] 	&	$	4700^{*}	$	&	$	4900^{*}	$	&	$	4950^{*}	$	&	$	4300^{*}	$	&	$	4850^{*}	$	\\
$T_{\rm 2}$ [K] 	&	$	4342\pm 48	$	&	$	4862\pm 20	$	&	$	4662\pm 28	$	&	$	3884\pm 33	$	&	$	4496\pm 72	$	\\
$\Omega_{\rm 1}$=$\Omega_{\rm 2}$        	&	$	5.305\pm 0.061	$	&	$	4.574 \pm 0.003	$	&	$	5.532\pm 0.059	$	&	$	3.344 \pm 0.020	$	&	$	6.519\pm 0.052	$	\\
$q_{ph}$  	&	$	2.108\pm 0.047	$	&	$	1.508\pm 0.004	$	&	$	2.276\pm 0.045	$	&	$	0.697\pm 0.013	$	&	$	2.965\pm 0.037	$	\\
$L_{1}~(B)$             	&	$	5.258\pm 0.091	$	&	$	4.868\pm 0.040	$	&	$	4.728\pm 0.050	$	&	$	9.648\pm 0.065	$	&	$	4.714\pm 0.052	$	\\
$L_{1}~(V)$             	&	$	4.893\pm 0.084	$	&	$	4.902\pm 0.036	$	&	$	4.578\pm 0.050	$	&	$	8.700\pm 0.059	$	&	$	4.617\pm 0.051	$	\\
$L_{1}~(R)$             	&	$	4.728\pm 0.083	$	&	$	4.922\pm 0.033	$	&	$	4.420\pm 0.050	$	&	$	8.151\pm 0.058	$	&	$	4.477\pm 0.050	$	\\
$L_{1}~(I)$             	&	$	4.519\pm 0.047	$	&	$	4.884\pm 0.027	$	&	$	4.029 \pm 0.048	$	&	$	7.845\pm 0.056	$	&	$	4.219\pm 0.047	$	\\
$L_{2}~(B)$         	&	$	5.833^{a}	$	&	$	6.833^{a}	$	&	$	6.816^{a}	$	&	$	2.010^{a}	$	&	$	7.742^{a}	$	\\
 $L_{2}~(V)$         	&	$	5.803^{a}	$	&	$	6.905^{a}	$	&	$	6.818^{a}	$	&	$	2.966^{a}	$	&	$	7.946^{a}	$	\\
 $L_{2}~(R)$         	&	$	5.927^{a}	$	&	$	6.962^{a}	$	&	$	6.796^{a}	$	&	$	3.526^{a}	$	&	$	8.045^{a}	$	\\
 $L_{2}~(I)$         	&	$	6.186^{a}	$	&	$	6.959^{a}	$	&	$	6.575^{a}	$	&	$	3.874^{a}	$	&	$	8.199^{a}	$	\\
$l_{3}~(B)$             	&	$	0.042\pm 0.009	$	&	$	0	$	&	$	0.003\pm 0.003	$	&	$	0	$	&	$	0	$	\\
$l_{3}~(V)$             	&	$	0.074\pm 0.008	$	&	$	0	$	&	$	0.035\pm 0.003	$	&	$	0	$	&	$	0	$	\\
$l_{3}~(R)$             	&	$	0.084\pm 0.008	$	&	$	0	$	&	$	0.056\pm 0.003	$	&	$	0	$	&	$	0	$	\\
$l_{3}~(I)$             	&	$	0.081\pm 0.008	$	&	$	0	$	&	$	0.113\pm 0.003	$	&	$	0	$	&	$	0	$	\\
\hline																					
r$_{1}~^{side}$         	&	$	0.4556	$	&	$	0.4084	$	&	$	0.4634	$	&	$	0.3995	$	&	$	0.483	$	\\
r$_{2}~^{side}$         	&	$	0.3186	$	&	$	0.3329	$	&	$	0.3125	$	&	$	0.3475	$	&	$	0.2852	$	\\
\hline																					
co-latitude             	&	$	137.1\pm 2.3	$	&	$	-	$	&	$	170.3\pm 0.6	$	&	$	-	$	&	$	166.1\pm 6.2	$	\\
longitude               	&	$	155.4\pm 1.4	$	&	$	-	$	&	$	50.5\pm 1.7	$	&	$	-	$	&	$	219.3\pm 7.4	$	\\
radius                  	&	$	35.7 \pm 1.7	$	&	$	-	$	&	$	49.3\pm 0.2	$	&	$	-	$	&	$	86.6\pm 7.9	$	\\
temp. factor            	&	$	0.853\pm 0.004	$	&	$	-	$	&	$	0.643\pm 0.021	$	&	$	-	$	&	$	0.813\pm 0.099	$	\\
\hline
\multicolumn{6}{l}{(*): Fixed parameter, (a): Calculated according to L$_{1}$} \\
\hline
\end{tabular}
\end{small}
\end{flushleft}
\end{table*}


\bsp	
\label{lastpage}
\end{document}